\documentstyle[12pt,epsf,epsfig]{article}
\newcount\timecount
\newcount\hours \newcount\minutes  \newcount\temp \newcount\pmhours

\hours = \time
\divide\hours by 60
\temp = \hours
\multiply\temp by 60
\minutes = \time
\advance\minutes by -\temp
\def\hour{\the\hours}
\def\minute{\ifnum\minutes<10 0\the\minutes
            \else\the\minutes\fi}
\def\clock{
\ifnum\hours=0 12:\minute\ AM
\else\ifnum\hours<12 \hour:\minute\ AM
      \else\ifnum\hours=12 12:\minute\ PM
            \else\ifnum\hours>12
                 \pmhours=\hours
                 \advance\pmhours by -12
                 \the\pmhours:\minute\ PM
                 \fi
            \fi
      \fi
\fi
}

\def\monthname{\relax\ifcase\month 0/\or January\or February\or
   March\or April\or May\or June\or July\or August\or September\or
   October\or November\or December\else\number\month/\fi}

\def\bold#1{\setbox0=\hbox{$#1$}%
     \kern-.025em\copy0\kern-\wd0
     \kern.05em\copy0\kern-\wd0
     \kern-.025em\raise.0433em\box0 }

\def\ga{\mathrel{\raise.3ex\hbox{$>$\kern-.75em\lower1ex\hbox{$\sim$}}}}
\def\la{\mathrel{\raise.3ex\hbox{$<$\kern-.75em\lower1ex\hbox{$\sim$}}}}
\def\gev{{\rm \, Ge\kern-0.125em V}}
\def\tev{{\rm \, Te\kern-0.125em V}}
\def\beq{\begin{equation}}
\def\eeq{\end{equation}}
\def\st{\scriptstyle}
\def\ss{\scriptscriptstyle}

\def\mst{m_{\tilde t}}

\def\mchi{m_{\chi}}
\def\ohsq{\Omega_{\chi} h^2}

\def\m12{m_{1\!/2}}

\def\ga{\mathrel{\raise.3ex\hbox{$>$\kern-.75em\lower1ex\hbox{$\sim$}}}}
\def\la{\mathrel{\raise.3ex\hbox{$<$\kern-.75em\lower1ex\hbox{$\sim$}}}}
\def\gyr{{\rm \, G\kern-0.125em yr}}
\def\gev{{\rm \, Ge\kern-0.125em V}}
\def\tev{{\rm \, Te\kern-0.125em V}}
\def\beq{\begin{equation}}
\def\eeq{\end{equation}}
\def\ss{\scriptscriptstyle}

\def\mst{m_{\tilde\tau_R}}
\def\mstop{m_{\tilde t_1}}
\def\msl{m_{\tilde{\ell}_1}}
\def\stau{\tilde \tau}
\def\stop{\tilde t}
\def\sbot{\tilde b}
\def\mchi{m_{\tilde \chi}}
\def\mxi{m_{\tilde{\chi}_i^0}}
\def\mxj{m_{\tilde{\chi}_j^0}}
\def\mchari{m_{\tilde{\chi}_i^+}}
\def\mcharj{m_{\tilde{\chi}_j^+}}
\def\mgluino{m_{\tilde g}}

\def\m12{m_{1\!/2}}

\def\ohsq{\Omega_{\widetilde\chi}\, h^2}
\def\ch{{\widetilde \chi}} 
\def\st{{\widetilde \tau}_{\scriptscriptstyle\rm 1}}
\def\sm{{\widetilde \mu}_{\scriptscriptstyle\rm R}}
\def\sel{{\widetilde e}_{\scriptscriptstyle\rm R}}

\def\tsq{|{\cal T}|^2}
\def\tcm{\theta_{\rm\scriptscriptstyle CM}}

\def\neq{n_{\rm eq}}
\def\qeq{q_{\rm eq}}
\def\mw{m_W}
\def\mz{m_Z}

\newcommand\f[1]{f_#1}
\def\nl{\hfill\nonumber\\&&}

\begin{document}
\begin{titlepage}
\pagestyle{empty}
\baselineskip=21pt
\rightline{hep-ph/0112113}
\rightline{CERN--TH/2001-339}
\rightline{UMN--TH--2032/01}
\rightline{TPI--MINN--01/50}
\vskip 0.2in
\begin{center}
{\large{\bf Calculations of Neutralino-Stop Coannihilation in the CMSSM}}
\end{center}
\begin{center}
\vskip 0.2in
{{\bf John Ellis}$^1$, {\bf Keith
A.~Olive}$^{2}$ and {\bf Yudi Santoso}$^{2}$}\\
\vskip 0.1in
{\it
$^1${TH Division, CERN, Geneva, Switzerland}\\
$^2${Theoretical Physics Institute,
University of Minnesota, Minneapolis, MN 55455, USA}}\\
\vskip 0.2in
{\bf Abstract}
\end{center}
\baselineskip=18pt \noindent

We present detailed calculations of the $\ch {\tilde t_1}$ coannihilation
channels that have the largest impact on the relic $\ch$ density in the
constrained minimal supersymmetric extension of the Standard Model
(CMSSM), in which scalar masses $m_0$, gaugino masses $m_{1/2}$ and the
trilinear soft supersymmetry-breaking parameters $A_0$ are each assumed to
be universal at some input grand unification scale. The most important
${\tilde t_1} {\tilde t_1}^*$ and ${\tilde t_1} {\tilde t_1}$ annihilation
channels are also calculated, as well as ${\tilde t_1} {\tilde \ell}$
coannihilation channels.  We illustrate the importance of these new
coannihilation calculations when $A_0$ is relatively large. While they do
not increase the range of $m_{1/2}$ and hence $m_\chi$
allowed by cosmology, these coannihilation channels do open up new `tails'
of parameter space extending to larger values of $m_0$. 

\vfill
\leftline{CERN--TH/2001-339}
\leftline{November 2001}
\end{titlepage}
\baselineskip=18pt

\section{Introduction}

A favoured candidate for cold dark matter is the lightest supersymmetric
particle (LSP), which is generally thought to be the lightest neutralino
$\ch$~\cite{EHNOS} in the minimal supersymmetric extension of the
Standard Model (MSSM).  It is common to focus attention on the constrained
MSSM (CMSSM), in which all the soft supersymmetry-breaking scalar masses
$m_0$ are required to be equal at an input superysmmetric GUT scale, as
are the gaugino masses $m_{1/2}$ and the trilinear soft
supersymmetry-breaking parameters $A_0$. These assumptions yield
well-defined relations between the various sparticle masses, and
correspondingly more definite predictions for the relic abundance
$\Omega_\ch h^2$ and observable signatures. This paper is devoted to
relic-abundance calculations including
coannihilations of the lightest neutralino $\ch$ with ${\tilde t}_1$, the
lighter supersymmetric partner of the top quark~\cite{stopco}.

The range $0.1 < \Omega_\ch h^2 < 0.3$ is generally thought to be
preferred by astrophysics and cosmology~\cite{omegah2}. Lower values of
$\Omega_\ch h^2$ might be possible if there is some other source of cold
dark matter, but higher values are incompatible with observation. The
regions of the $m_{1/2}, m_0$ plane where the relic density falls within
the preferred range $0.1 < \Omega_\ch h^2 < 0.3$ have generally been
divided into four generic parts.
There is a
`bulk' region at moderate $m_{1/2}$ and
$m_0$~\cite{EHNOS}. Then, extending to larger $m_{1/2}$, there is a `tail'
of the parameter space where the LSP $\ch$ is almost degenerate with the
next-to-lightest supersymmetric particle (NLSP), which is in this region
the ${\tilde \tau}_1$, the lighter supersymmetric partner of the $\tau$
lepton. Along this `tail', efficient coannihilations~\cite{gs,oldcoann,eg}
keep
$\Omega_\ch h^2$ down in the preferred range, even for larger values of
$m_\ch$~\cite{efo,efosi,coann,ADS}. At larger $m_0$, close to the boundary
where electroweak symmetry breaking is no longer possible, there is the
`focus-point' region where the LSP has a larger Higgsino component and
$m_\ch$ is small enough for $\Omega_\ch h^2$ to be
acceptable~\cite{focus}. Finally, extending to larger $m_{1/2}$ and $m_0$
at intermediate values of $m_{1/2} / m_0$, there may be a `funnel' of
CMSSM parameter space where rapid direct-channel annihilations via the
poles of the heavier Higgs bosons $A$ and $H$ keep $\Omega_\ch h^2$
in the preferred range~\cite{EFGOSi,funnel}.

In this paper, we emphasize the significance of coannihilation of the LSP
$\ch$ with ${\tilde t}_1$, the lighter supersymmetric
partner of the
$t$ quark \cite{stopco}. This mechanism opens up another `tail' of
parameter space, this time extending to larger values of $m_0$. It is not
relevant for the small values of $A_0$ considered in previous
coannihilation calculations~\cite{efosi,coann,ADS}, but may be important
for large
$A_0$, as we demonstrate in this paper. Coannihilations of $\ch$ with
${\tilde t}_1$ are important when the latter is the NLSP, just as $\ch
{\tilde
\tau}_1$ coannihilations are important when the ${\tilde \tau}_1$ is the
NLSP. In the latter case, one must also consider coannihilations with the
${\tilde e}_1$ and ${\tilde \mu}_1$, which are not much heavier than the
${\tilde \tau}_1$~\cite{efo,efosi,coann,ADS}. There are also regions of
CMSSM parameter space where
both the ${\tilde t}_1$ and ${\tilde \tau}_1$ are close in mass to the LSP
$\ch$, and ${\tilde t}_1 {\tilde \tau}_1$ coannihilations must also be
considered. We present here detailed calculations of the matrix elements
and cross sections for all the leading $\ch {\tilde t}_1$ and ${\tilde
t}_1 {\tilde \ell}$ coannihilation processes, and illustrate their
importance for $\Omega_\ch h^2$ in some instances in the CMSSM when $A_0
\ne 0$.

The structure of the paper is as follows. In Section 2 we recall some
important features of LSP relic-density calculations in general, and
coannihilations in particular. Then, in Section 3 we compare the relative
magnitudes of the $\ch \ch$, $\ch {\tilde t}_1$, ${\tilde t}_1 
{\tilde t}^{(*)}_1$ 
and ${\tilde t}_1 {\tilde \ell}^{(*)}$ processes for some specific
choices of the CMSSM parameters. Section 4 provides an overview of the
implications of $\ch {\tilde t}_1$ coannihilation and related
processes for the regions of the $m_{1/2}, m_0$ plane allowed by the
constraint $0.1 < \Omega_\ch h^2 < 0.3$ for various choices of the other
CMSSM parameters. Relevant details of our calculations of the matrix
elements are contained in an Appendix.

\section{Formalism for Annihilation and Coannihilation}

The density of neutralino relics left over from the early Universe may be
determined relatively simply in terms of relevant annihilation cross
sections, using the Boltzmann rate equation to determine a freeze-out
density. The relic density subsequently scales with the inverse of the
comoving volume, and hence with the entropy density. In the MSSM framework
discussed here, since neutralinos are Majorana fermions, the $S$-wave
annihilation cross sections into fermion-antifermion pairs are suppressed
by the masses of the final-state fermions, and it is therefore necessary
to compute $P$-wave contributions to the cross sections~\cite{EHNOS}.

The rate equation for a stable particle with density $n$ is
\beq
{dn \over dt} = -3 {\dot R \over R} n - \langle \sigma v_{\rm rel}
\rangle (n^2 - \neq^2) \;,
\label{rate}
\eeq
where $\neq$ is the equilibrium number density and $\langle \sigma v_{\rm 
rel}
\rangle$ is the thermally averaged product of the annihilation cross
section $\sigma$ and the relative velocity $v_{\rm rel}$.
In the early Universe, we can write $\dot R/R = (8\pi G_N \rho/3)^{1/2}$,
where $\rho = \pi^2 g(T) T^4/30$ is the energy density in radiation and
$g(T)$ is the number of relativistic degrees of freedom.
Conservation of the entropy density $s = 2 \pi^2 h(T) T^4/45$ implies that
$\dot R/R = - \dot T/T - h'\dot{T}/3h$ where 
$h' \equiv dh/dT$.  Generally, we have $h(T)
\approx g(T)$. Defining $x \equiv T/m$ and $q \equiv n/T^3h$, we can
write
\beq
{dq \over dx} = m \left({\textstyle{4\pi^3\over 45}} G_N g\right)^{-1/2}
                \left(h + {\textstyle{1 \over 3}}mxh'\right)
                \langle\sigma v_{\rm rel} \rangle
                (q^2 - \qeq^2) \; .
\label{rate2}
\eeq
The effect of the $h'$ term was discussed in detail in~\cite{swo}, and is
most important when the mass $m$ is between 2 and 10 GeV.  Since we
only consider neutralinos that are significantly more massive, we neglect
it below (though it is not neglected in our calculations). In the case of
the MSSM, freeze-out occurs when
$x
\sim 1/20$,  and the final relic density is determined by integrating
(\ref{rate}) down to
$x = 0$, and is given by
\beq
\rho_\ch = m q(0)h(0)T_0^3.
\eeq

When coannihilations are important, there are several relevant
particle species $i$, each with different mass, number density $n_i$ and
equilibrium number density $n_{{\rm eq},i}$. Even
in such a situation~\cite{gs}, the rate equation (\ref{rate}) still 
applies, provided $n$ is interpreted as the total number density,
\beq
n \equiv \sum_i n_i \;,
\label{n}
\eeq
$\neq$ as the total equilibrium number density,
\beq
\neq \equiv  \sum_i n_{{\rm eq},i} \;,
\label{neq}
\eeq
and the effective annihilation cross section as
\beq
\langle\sigma_{\rm eff} v_{\rm rel}\rangle \equiv
\sum_{ij}{ n_{{\rm eq},i} n_{{\rm eq},j} \over \neq^2}
\langle\sigma_{ij} v_{\rm rel}\rangle \;.
\label{sv2} 
\eeq
In (\ref{rate2}),  $m$ is now understood as the mass of the lightest
particle under consideration. For $T \la m_i$,
the equilibrium number density of each species is given by \cite{swo,gg}
\begin{eqnarray}
n_{{\rm eq},i} &=& g_i\int {d^3p\over(2\pi)^3} \; e^{-E/T}
\nonumber \\
               &=& {g_i m_i^2 T \over 2\pi^2} K_2(m_i/T) \;,
\nonumber \\
               &=& g_i \left({m_i T \over 2\pi}\right)^{3/2} \exp(-m_i/T)
                       \left(1 + {15 T\over 8m_i}+ \ldots \right) \;,
\label{neqi}
\end{eqnarray}
where $g_i$ is a spin and color degeneracy factor and
$K_2(x)$ is a modified Bessel function.
We make the approximation of
Boltzmann statistics for the annihilating particles, which is
excellent in practice.

We now recall how to compute
$\langle\sigma_{12} v_{\rm rel}\rangle$
for the process $1+2\to 3+4$ in an efficient manner.
Suppose we have determined the squared transition matrix element $\tsq$
(summed over final spins and averaged over initial spins) and expressed 
it as a function of the Mandelstam variables $s$, $t$, $u$.
We then express $\tsq$ in terms of $s$ and the
scattering angle $\tcm$ in the center-of-mass frame, as described 
in~\cite{efosi}. We now define
\begin{eqnarray}
w(s) &\equiv& {1\over4}
              \int {d^3 p_3\over(2\pi)^3 E_3}\,{d^3 p_4\over(2\pi)^3 E_4}
              \,(2\pi)^4\delta^4(p_1+p_2-p_3-p_4)\; \tsq
\nonumber \\
&=& {1\over32\pi}\,{p_3(s)\over s^{1/2}} \int_{-1}^{+1}d\cos\tcm\,\tsq \;.
\label{w}
\end{eqnarray}  
In terms of $w(s)$, the total annihilation cross section
$\sigma_{12}(s)$ is given by
$\sigma_{12}(s) = w(s)/s^{1/2}p_1(s)$~\footnote{Our $w(s)$
is also the same as $w(s)$ in~\cite{swo,fkosi,efo},
which is written as $W/4$ in~\cite{eg}.}.

The above analysis is exact.  To reproduce the usual partial wave 
expansion,
we expand $\tsq$ in powers of $p_1(s)/m_1$.
The odd powers vanish upon
integration over $\tcm$, while the zeroth- and second-order terms
correspond to the usual $S$ and $P$ waves, respectively.
Each factor of $p_1(s)$ is accompanied by a factor of $\cos\tcm$,
so we have
\begin{equation}
\int_{-1}^{+1}d\cos\tcm\,\tsq =
 \left(\tsq_{\cos\tcm\,\to\, +1/\sqrt3} \; + \;
       \tsq_{\cos\tcm\,\to\, -1/\sqrt3}\right) + {\cal O}(p_1^4)\;.
\label{itsq}
\end{equation}
We can therefore evaluate the $S$- and $P$-wave contributions to
$w(s)$ simply by evaluating $\tsq$
at two different values of $\cos\tcm$;
no integrations are required.

The proper procedure for thermal averaging has been discussed
in~\cite{swo,gg} for the case of $m_1=m_2$, and
in~\cite{fkosi,eg} for the case of $m_1\ne m_2$, so we do not discuss
it in detail here. One finds
\begin{equation}
\langle\sigma_{12} v_{\rm rel}\rangle
= a_{12} + b_{12} \, x + {\cal O}(x^2) \;,
\label{sv3}
\end{equation}
where $x \equiv T/m_1$ (assuming $m_1<m_2$).
In our case, we extract $a_{12}$ and $b_{12}$
from the transition amplitudes
listed in the Appendix for each final state.   We set
$x=0$ to get $a_{12}$, and then compute $b_{12}$ by setting $x$
to a numerical value small enough to render the ${\cal O}(x^2)$ terms 
negligible.
We then compute
$a_{\rm eff}$ and $b_{\rm eff}$ by performing the sum over initial
states as in (\ref{sv2}), and integrate the rate equation (\ref{rate2}) 
numerically to obtain the relic
LSP abundance. To a fair approximation, the relic density can simply be
written as~\cite{EHNOS,gs}
\begin{equation}
  \label{eq:ohsq}
  \Omega h^2 \approx
{10^9 \gev^{-1} \over g_{\ss f}^{1/2} M_{\rm pl}(a_{\rm eff}+
b_{\rm eff} x_{\ss f}/2)x_{\ss f}},
\end{equation} 
where the freeze-out temperature $T_{\!f}\sim m_\ch/20$, and $g_{\ss f}$
is the number of relativistic degrees of freedom at $T_{\!f}$. 

This implies that the
ratio of relic densities computed with and without coannihilations is
approximately
\begin{equation}
  \label{eq:R}
 R\equiv{\Omega^0\over\Omega}
 \approx \left({\hat\sigma_{\rm eff}\over\hat\sigma_0}\right)
\left({x_{\!\ss f}\over x_{\!\ss f}^{0}}\right),
\end{equation}
where $\hat\sigma\equiv a + b x/2$ and sub- and superscripts 0 denote
quantities computed ignoring coannihilations.  The ratio ${x_{\!\ss
    f}^0 / x_{\!\ss f}}\approx 1+x_{\!\ss f}^0 \ln (g_{\rm
  eff}\sigma_{\rm eff}/g_1\sigma_0)$, where
$g_{\rm eff}\equiv\sum_i g_i (m_i/m_1)^{3/2}e^{-(m_i-m_1)/T}$.
For the case where the ${\tilde t}_1$ and $\ch$ are almost degenerate, 
$g_{\rm eff} \approx \sum_i g_i =8$ and
${x_{\!\ss f}^0 / x_{\!\ss f}}\approx 1.2$.

\section{Coannihilation Rates for ${\tilde t}_1$ in the MSSM}

We now use the above formalism to estimate the relative importance of the
${\tilde t}_1 \ch$ coannihilation processes, ${\tilde t}_1 {\tilde
t}_1^*$ and ${\tilde t}_1 {\tilde t}_1$ annihilations calculated in the
Appendix.  We also take into account the $\ch {\tilde \ell}$
coannihilations calculated previously~\cite{efo,efosi} and, for
completeness, include the
${\tilde t}_1 {\tilde \ell}$ and ${\tilde t}_1 {\tilde \ell}^*$
coannihilations also calculated in the Appendix.

To compute the effective annihilation cross sections for light
sparticles in the MSSM, we allow the index $i$ in (\ref{n}) to run
over ${\tilde t}_1, {\tilde t}_1^*, \st, \st^*, \sel, \sel^*, \sm$ and 
$\sm^*$, as well as $\ch$. The following is the change in $\sigma_{\rm 
eff}$ compared with~\cite{efosi}, where
49 of the $\sigma_{ij}$ in (\ref{sv2}) were already included:
\begin{equation}
\Delta \sigma_{\rm eff} =  2\, (\sigma_{\stop_1\stop_1}+
\sigma_{\stop_1\stop_1^*})r_{\stop_1}^2 + 4\, \sigma_{\ch {\tilde
t}_1} r_{\ch}r_{{\tilde t}_1} + 
8\, (\sigma_{\stop_1 {\tilde e_R}}+\sigma_{\stop_1
{\tilde e_R}^*})r_{\stop_1}r_{\tilde e_R} + 4\, (\sigma_{\stop_1 {\tilde
\tau}_1}+\sigma_{\stop_1 {\tilde \tau}_1^*}) r_{\st}  r_{{\tilde t}_1}
\end{equation}
where $r_i\equiv n_{{\rm eq},i}/n_{\rm eq}$.
We have taken the $\sel$ and $\sm$ (but not the $\stau$) to be degenerate 
in mass, thus accounting for the 81 possible initial state combinations.
Note that we have summed over color states in the cross sections
amplitudes listed in the Appendix, and we have taken the stop degeneracy
factor
$g_{\stop} = 3$. 
We list in Table~\ref{table:states} the sets of initial and final states
for which we compute the annihilation cross sections, using the transition
amplitudes given in the Appendix. We use $q$ to denote the four light
quarks, which we have taken to be massless.

\begin{table}[htb]\caption{Initial and Final States for 
${\tilde t}_1$ Annihilation and Coannihilation Processes}
\begin{center}
\begin{tabular}{|c|l|}
\hline
Initial State & Final States\\ 
\hline
~ & ~ \\
${\tilde t}_1 {\tilde t}^*_1$ & $g g, \gamma g, Z g, t {\bar t}, b {\bar
b}, q {\bar q}, g h, g H, Z h, Z H, Z A, W^\pm H^\mp$, \\
~ & $h h, h H, H H, A A, h A, H A, H^+ H^-$ \\
${\tilde t}_1 {\tilde t}_1$ & $t t$ \\
$\ch {\tilde t}_1$ & $t g, t Z, b W^+, t H, t h, t A, b H^+$ \\
${\tilde t}_1 {\tilde \ell}$ & $t \ell, b \nu $ \\
${\tilde t}_1 {\tilde \ell}^*$ & $t {\bar \ell}$ \\
\hline
\end{tabular}
\label{table:states}
\end{center}
\end{table}
\vspace{-.5cm}

In the CMSSM, the diagonal entries of the squark mass matrix tend to pick
up large contributions from the gaugino masses, $m_{LL,RR}^2 \ni {\cal O}
(6) m_{1/2}^2$,
thus making the squarks heavier than the neutralinos.  The off-diagonal
entry for an up-type
squark~\footnote{Note here our sign convention for $A_q$.} 
\beq
m_{LR}^2 = -m_q (A_q + \mu \cot \beta)
\label{offdiag}
\eeq
can, however, be large, particularly for the stops, or for sbottoms at
large
$\tan \beta$~\footnote{For down-type squarks, the factor $\cot \beta$ in
(\ref{offdiag}) is replaced by $\tan \beta$.}. When $A_t$ is sufficiently
large, the lighter
stop, ${\tilde t_1}$, can become degenerate with (or lighter than) the
neutralino.
Thus, it is when $A_0$ is large that we expect $\ch \stop_1$
coannihilations to become important.  

It is important to distinguish
between the effective low-energy parameters $A_t$, etc., and the 
high-energy input parameter
$A_0$, which are related through the running of the renormalization-group
equations. For example, for $\tan \beta = 10$ and $m_0 = 300\gev$,
$\stop_1 \ch$ coannihilations are important when $m_{1/2} = 200,
450$,and $670 \gev$ and
$A_0 = 1000, 2000$, and $3000\gev$, but these values correspond to $A_t
\simeq 565, 1200$, and $1700
\gev$ respectively. Furthermore, the values of $A$ for the light squarks
are different and typically larger than $A_t$.

\begin{figure}
\begin{minipage}[b]{8in}
\epsfig{file=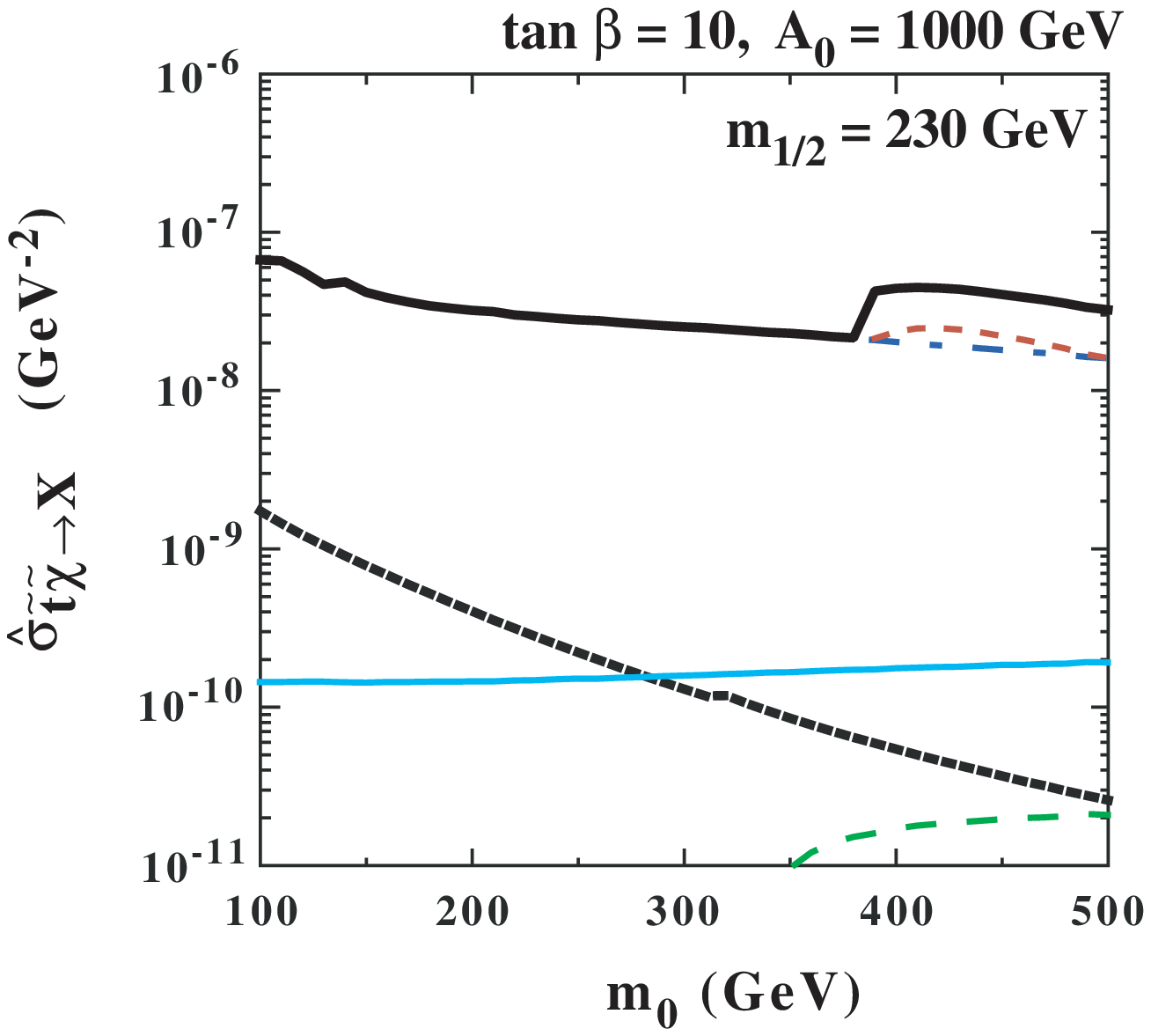,height=2.5in}
\epsfig{file=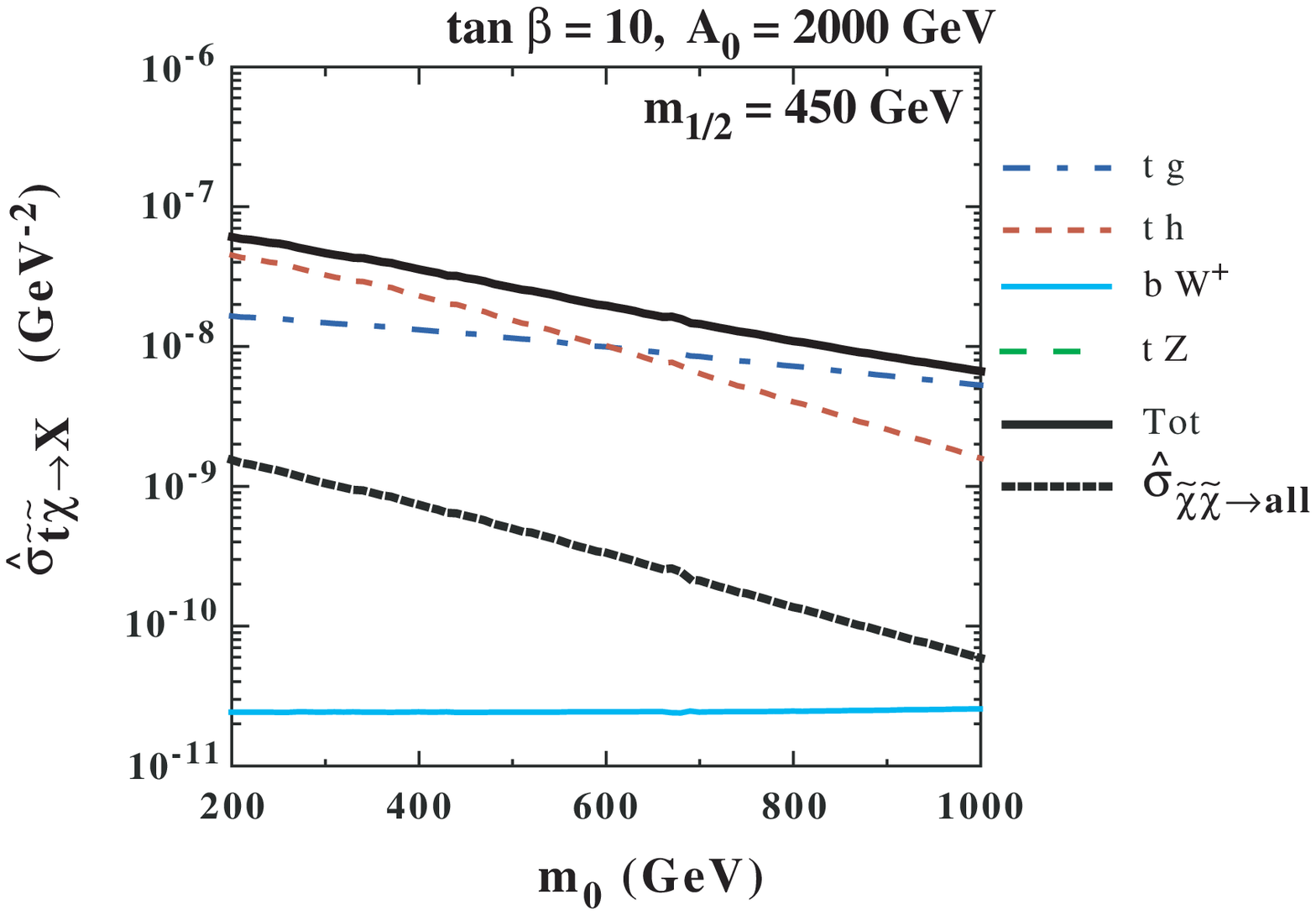,height=2.5in} \hfill
\end{minipage}
\hspace*{1.6in}
\caption{{\it 
The separate contributions to the $\ch {\tilde t_1}$ coannihilation cross 
sections $\hat\sigma\equiv
a+{1\over2}b x$ for $x=T/m_\ch=1/23$ as functions of
$m_0$ for (a) $\m12 = 230\gev$, $A_0 = 1000\gev$ and (b) $\m12 = 450\gev$,
$A_0 = 2000\gev$. Also shown are the total cross section and , for
comparison, the much smaller total cross  section for $\ch \ch$ annihilation.
}
\label{fig:sigmahatchit}}
\end{figure}

We display in Fig.~\ref{fig:sigmahatchit} numerical values of the
contributions to $\hat\sigma\equiv a + b x/2$ (see (\ref{eq:ohsq})) in
$\ch {\tilde t_1}$ coannihilation, for the representative values $x=1/23,
\tan\beta=10, \mu>0$ and (a) $\m12 =
230\gev, A_0 = 1000\gev$ and (b) $\m12 =
450\gev, A_0 = 2000\gev$ as functions of $m_0$. 
For comparison, the total cross section for $\ch
\ch$ annihilation to all final states is also shown, as a thick dotted
line. We see that the
$\ch {\tilde t_1}
\to  t g$ and $t h$ coannihilation cross sections dominate by large
factors over the total $\ch \ch$ annihilation cross section, suggesting
that they may have a greater importance than that suggested by simply
comparing Boltzmann suppression factors. The feature in
Fig.~\ref{fig:sigmahatchit}(a) at $m_0 \sim 400 \gev$ is due to the 
threshold for the production of $t h$ final states.  At smaller values of
$m_0$, this final state is kinematically forbidden.

\begin{figure}
\begin{minipage}[b]{8in}
\epsfig{file=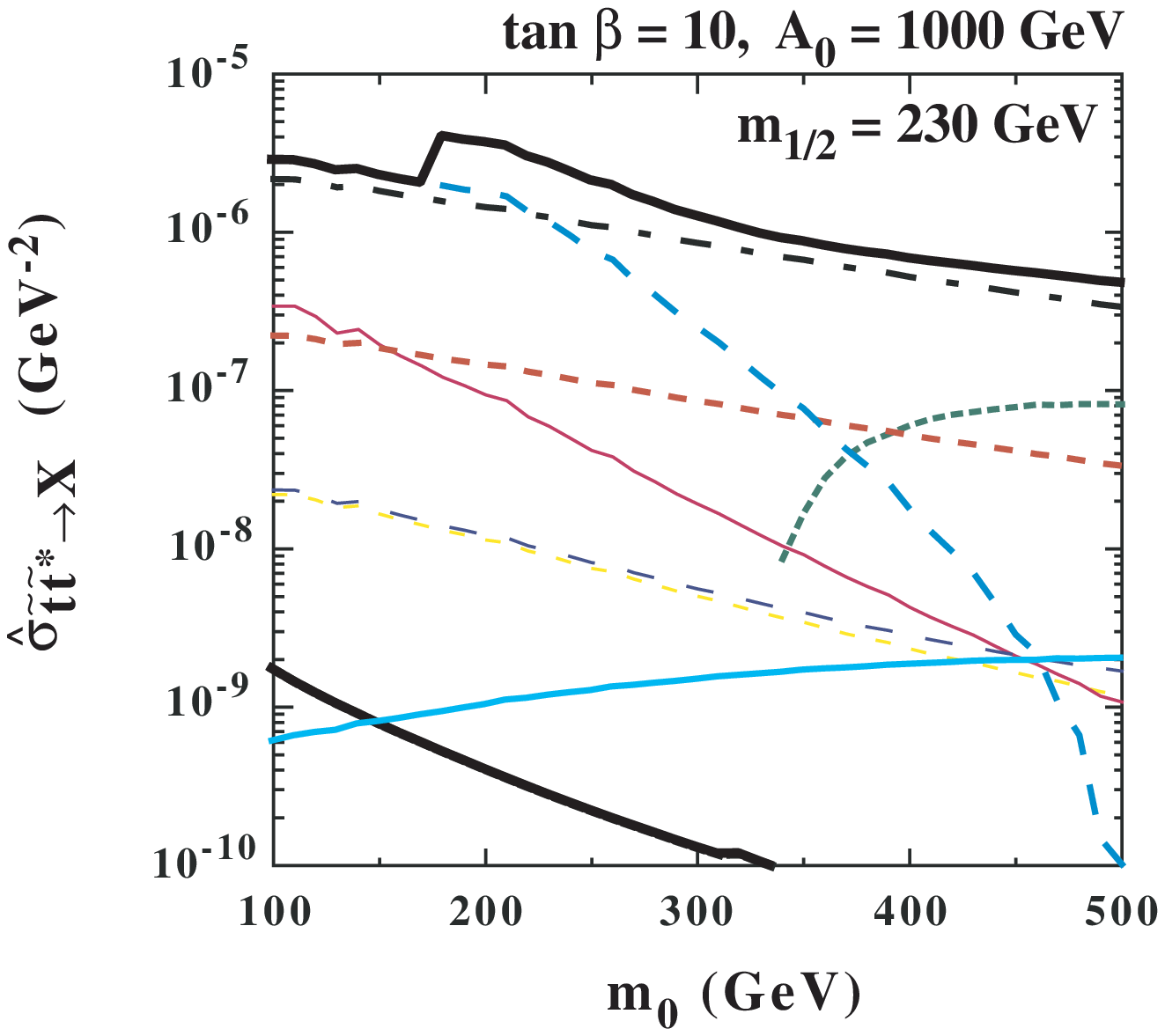,height=2.5in}
\epsfig{file=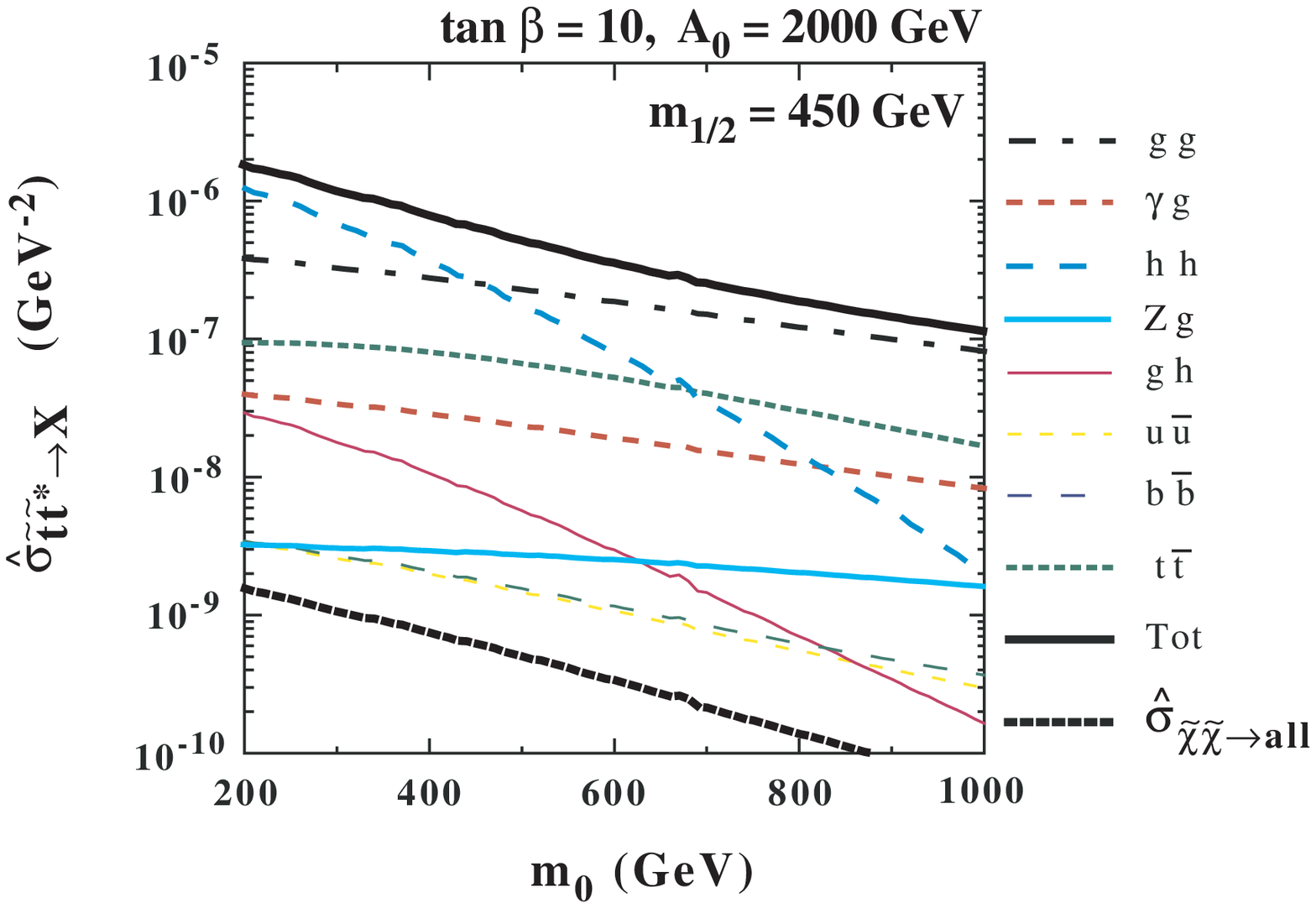,height=2.5in} \hfill
\end{minipage}
\hspace*{1.6in}
\caption{{\it
The separate contributions to the ${\tilde t_1} {\tilde t_1}^*$ 
annihilation cross
sections $\hat\sigma\equiv
a+{1\over2}b x$ for $x=T/m_\ch=1/23$, as functions of  
$m_0$ for (a) $\m12 = 230\gev$,  $A_0 = 1000\gev$ and (b) $\m12 = 450\gev$,
$A_0 =
2000\gev$. Also shown are the total cross section and, for comparison,
the much smaller total cross section for $\ch \ch$ annihilation.
}
\label{fig:sigmahatttbar}}
\end{figure}

Fig.~\ref{fig:sigmahatttbar} displays similar plots for ${\tilde t_1}
{\tilde t_1}^*$ annihilation, for the same parameter choices as in
Fig.~\ref{fig:sigmahatchit}.
In this case, the dominant ${\tilde t_1} {\tilde t_1}^*$ annihilation
cross sections are into $g g$ and $h h$, and even subdominant cross
sections such as $\gamma g$, $Z g$, $g h$ and the various quark-antiquark
channels are far larger than the total $\ch \ch$ annihilation cross 
section. Once again, when $A_0 = 1000\gev$ we see
thresholds, in this case corresponding to $h h$ production at $m_0 \sim
180 \gev$ and $t \bar t$ production at $m_0 \sim 330\gev$. 

\begin{figure}
\vspace*{-0.5in}
\begin{minipage}[b]{8in}
\epsfig{file=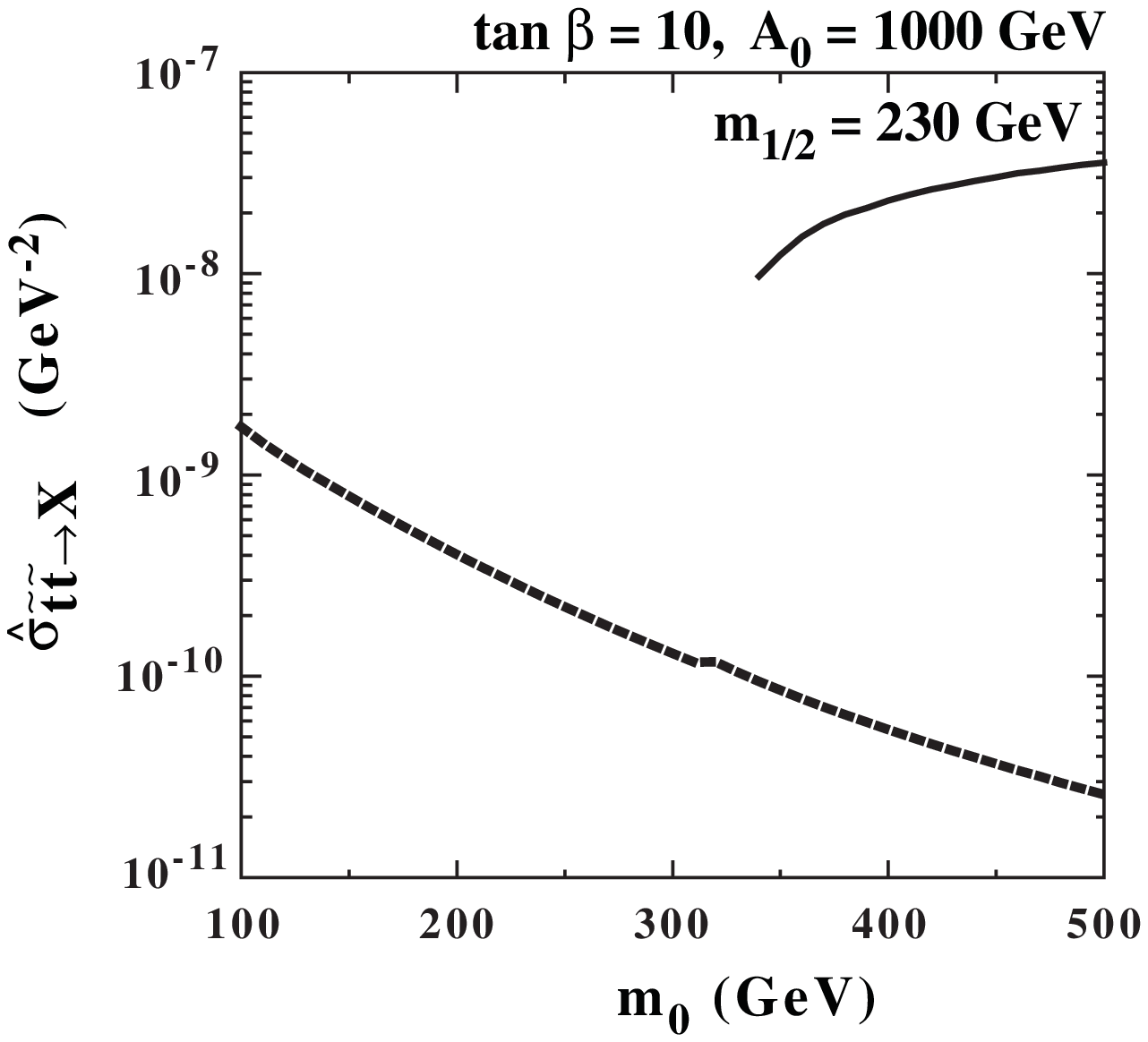,height=2.5in}
\epsfig{file=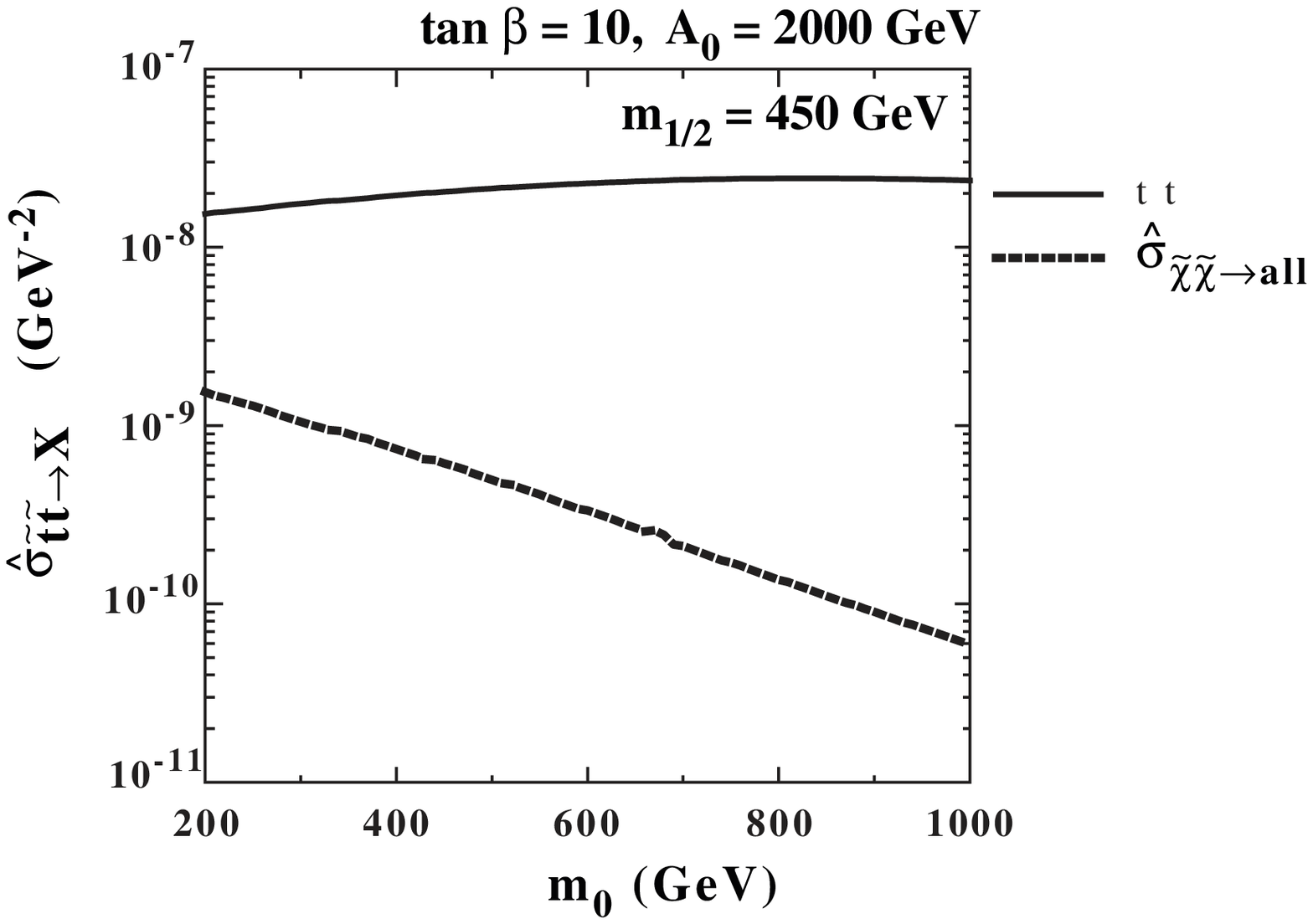,height=2.5in} \hfill
\end{minipage}
\hspace*{1.6in}
\caption{{\it
The ${\tilde t_1} {\tilde t_1} \to t t$ 
annihilation cross
sections $\hat\sigma\equiv
a+{1\over2}b x$ for $x=T/m_\ch=1/23$, as functions of  
$m_0$ for (a)  $\m12 = 230\gev$, $A_0 = 1000\gev$ and (b) $\m12 = 450\gev$, 
$A_0 = 2000\gev$. Also shown, for comparison, is the much smaller total
cross section for $\ch \ch$ annihilation.
}
\label{fig:sigmahattt}}
\end{figure}

The ${\tilde t_1} {\tilde t_1}$ annihilation cross sections shown in 
Fig.~\ref{fig:sigmahattt} show that the cross section for annihilation 
into the $t t$ final state, when it is kinematically open, is also far
larger  than the total $\ch \ch$ annihilation cross section. 

\begin{figure}
\vspace*{-0.5in}
\begin{minipage}[b]{8in}
\epsfig{file=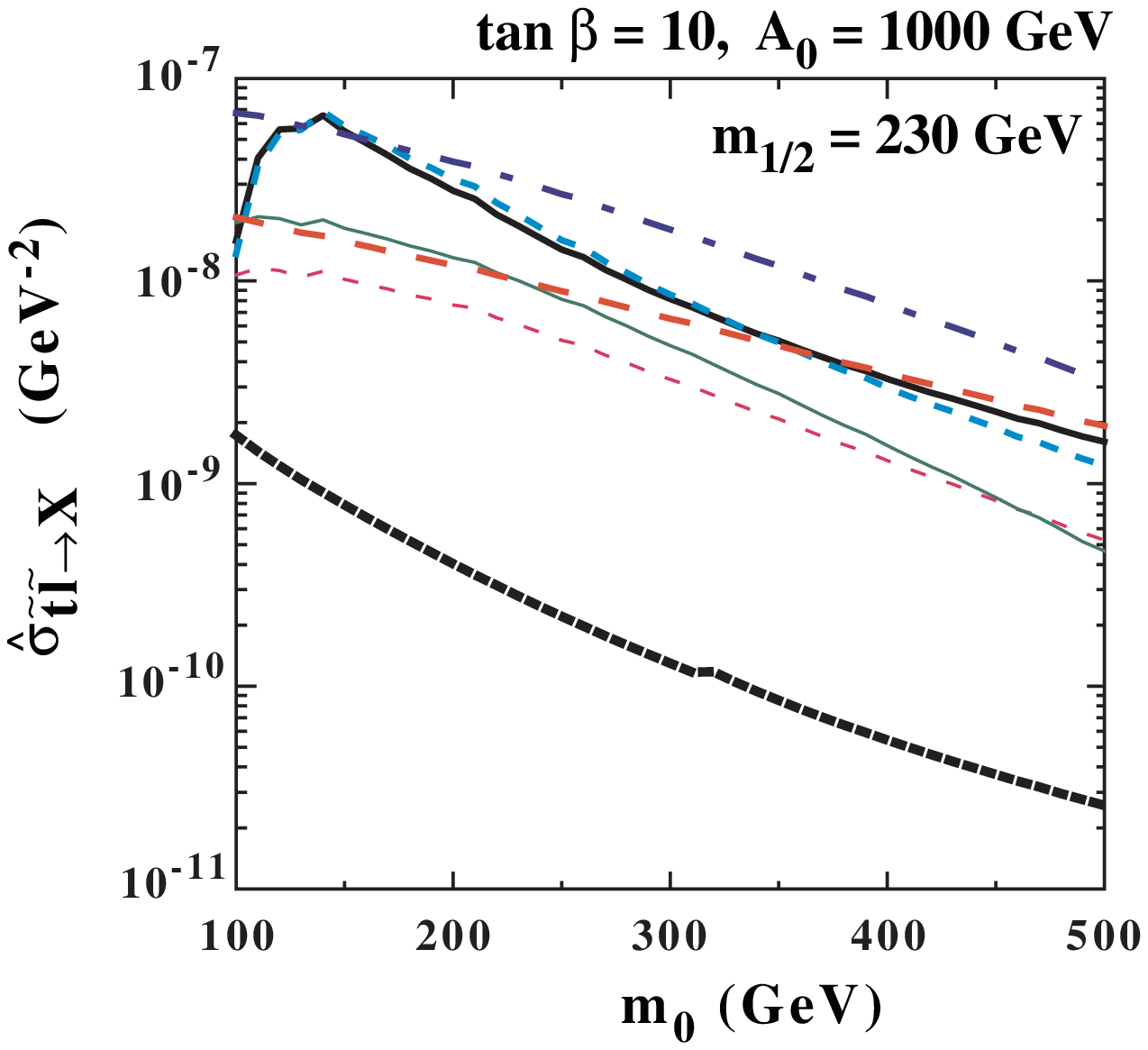,height=2.5in}
\epsfig{file=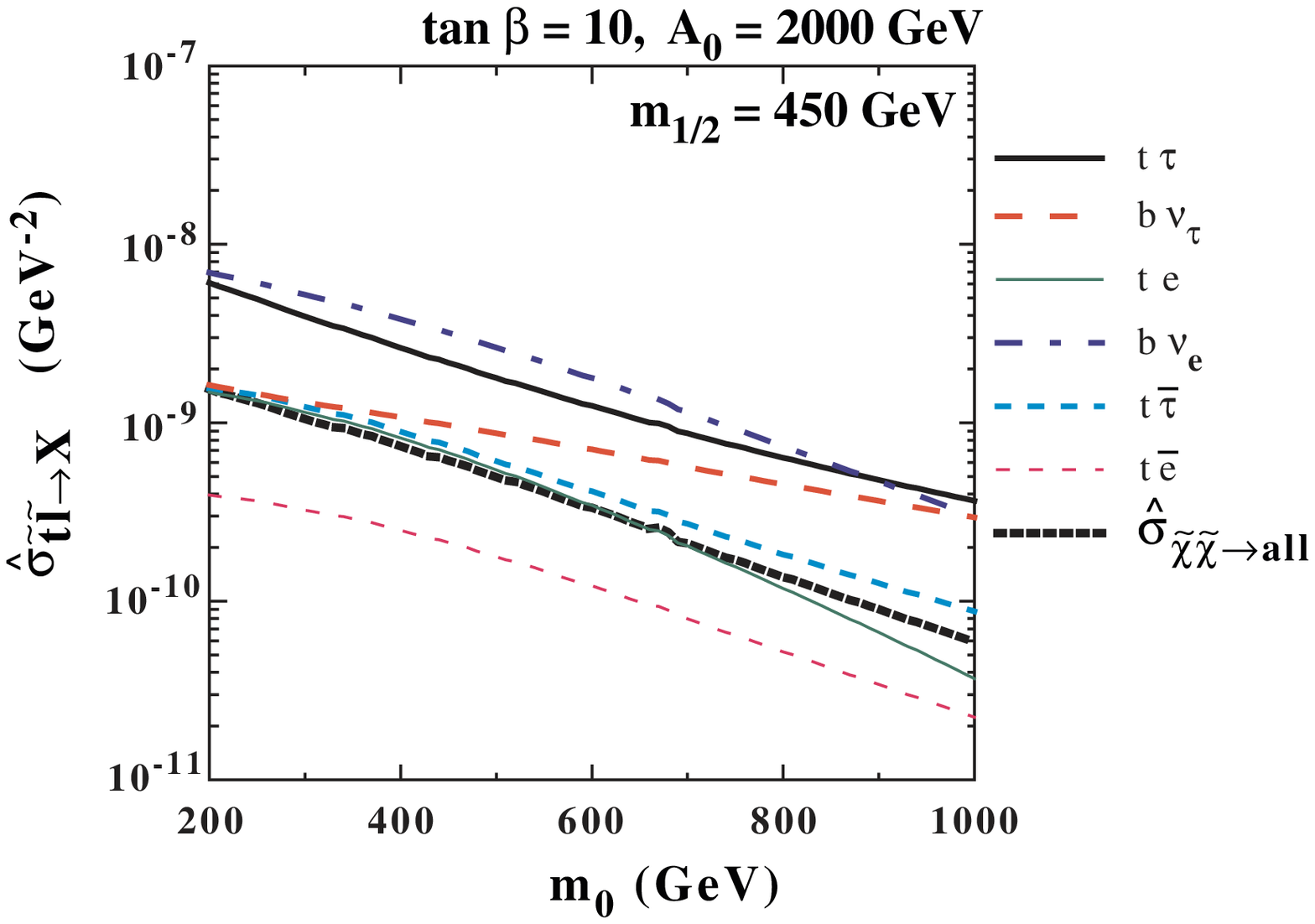,height=2.5in} \hfill
\end{minipage}
\hspace*{1.6in}
\caption{{\it
The separate contributions to the ${\tilde \ell_1} {\tilde t_1}$ 
coannihilation cross
sections $\hat\sigma\equiv
a+{1\over2}b x$ for $x=T/m_\ch=1/23$, as functions of  
$m_0$ for (a)  $\m12 = 230\gev$,  $A_0 = 1000\gev$ and (b)   $\m12 = 450\gev$,
$A_0 = 2000\gev$. Also shown, for comparison, is the much smaller total
cross section for $\ch \ch$ annihilation.
}
\label{fig:sigmahattl}}
\end{figure}

A complete study of coannihilation effects must include not only the $\ch
{\tilde \ell}$ processes considered previously~\cite{efo,efosi}, and the
$\ch {\tilde t_1}$
processes considered above, but also ${\tilde \ell_1} {\tilde t_1}$
coannihilations. Accordingly, the final set of coannihilation cross
sections we present are those for ${\tilde \ell_1} {\tilde t_1}$ and
${\tilde \ell_1^*} {\tilde t_1}$, shown in Fig.~\ref{fig:sigmahattl}. We
see that, when (a) $A_0 = 1000\gev$, the $t \tau$, $t {\bar \tau}$ and $b
\nu_e$ final states are the most important, followed by $t e$, $b
\nu_\tau$ and $t {\bar e}$, whereas (b) the $t {\bar \tau}$ and $t {\bar
e}$
final states are relatively much less important when $A_0 = 2000\gev$.
In all panels of Fig.~\ref{fig:sigmahattl}, there are coannihilation cross 
sections much greater than the total $\ch \ch$ annihilation cross section, 
which is also plotted.

The basic reason for the relatively small magnitude of the $\ch \ch$
annihilation cross section is that it is dominated by the $P$-wave
suppressed cross sections for $\ch \ch$ annihilation to fermion pairs.
This was also the basic reason why $\ch {\tilde \ell}$ coannihilation
processes were previously found to be so
important~\cite{efo,efosi,coann,ADS}.

\begin{figure}
\begin{minipage}{8in}
\epsfig{file=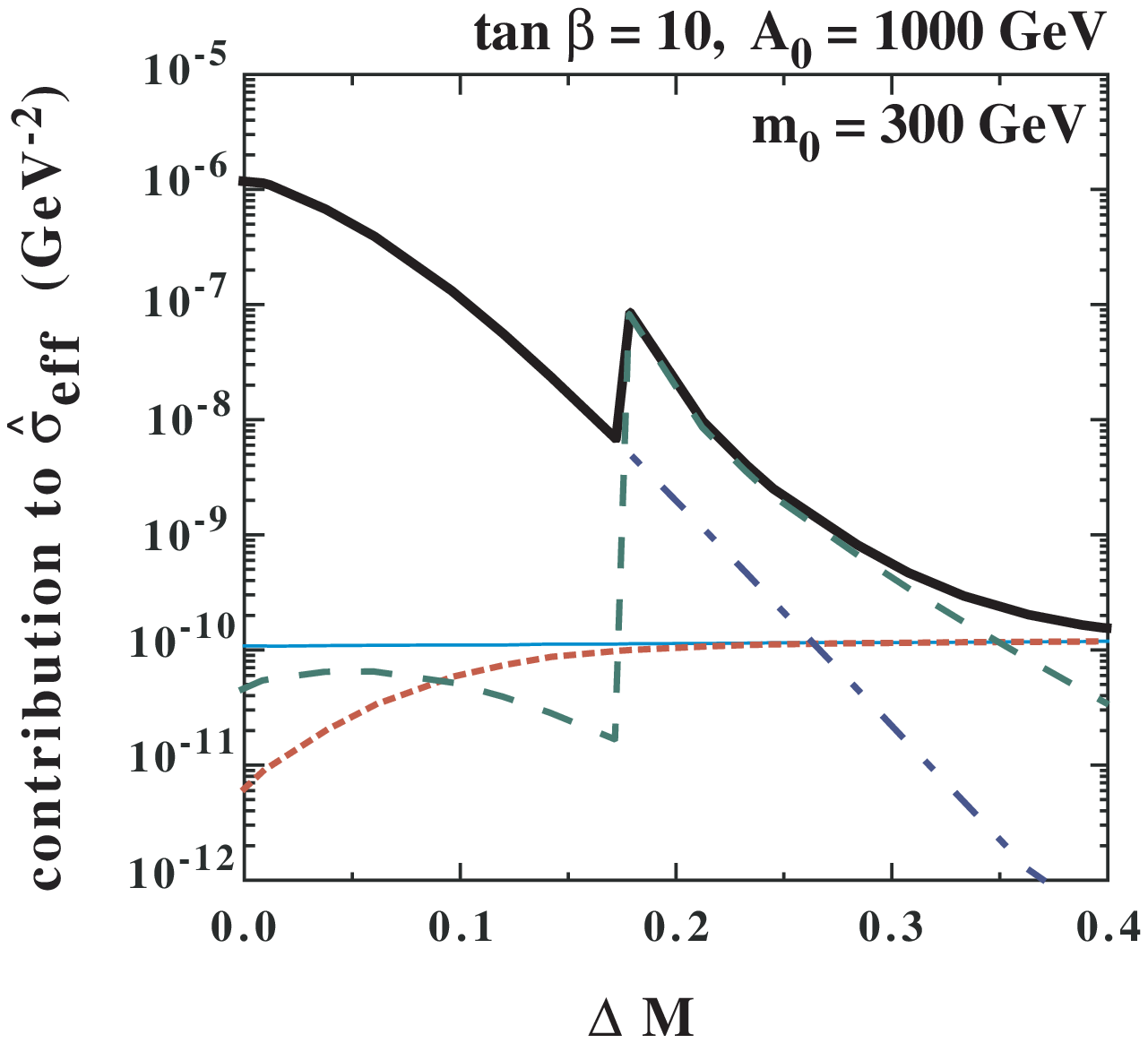,height=2.5in}
\epsfig{file=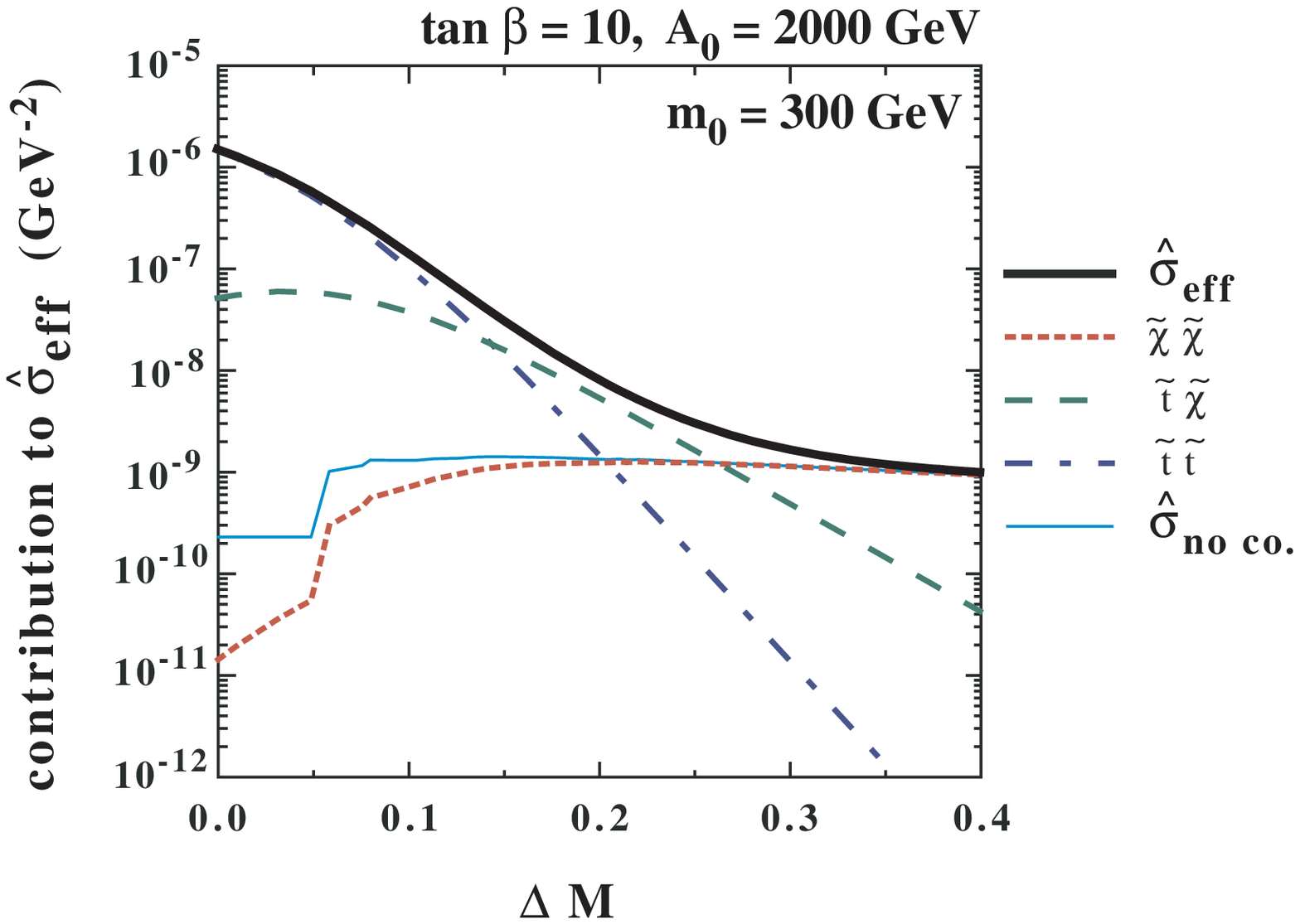,height=2.5in} \hfill
\end{minipage}
\caption{\label{fig:svdm}
{\it 
The separate contributions to the total effective cross sections
$\hat\sigma_{\rm eff}$ for $x=T/m_\ch=1/23$, as functions
of $\Delta M\equiv(\mst- m_\ch)/m_\ch$, obtained by varying $m_{1/2}$, 
with (a) $A_0 = 1000\gev$ 
and (b) $A_0 = 2000\gev$, both for $\tan \beta = 10, \mu > 0$ and 
$m_0 = 300\gev$.
}}
\end{figure}

The contributions of the various annihilation channels to $\sigma_{\rm
eff}$ are weighted by the relative abundances of the $\ch$, $\tilde t_1$
and $\tilde \ell_1$.  For a stop degenerate with the $\ch$, ${\tilde t_1}
{\tilde t_1}^*$ annihilation, ${\tilde t_1} {\tilde t_1}$ annihilation and
$\ch {\tilde t_1}$ coannihilation are clearly the dominant contributions
to $\Delta \sigma_{\rm eff}$, and hence to $\sigma_{\rm eff}$ in
(\ref{sv2}), and the final neutralino relic density is greatly reduced.  
As the stops become heavier than the neutralinos, their number densities
are exponentially suppressed and the stop contributions to $\sigma_{\rm
eff}$ become less important. Fig.~\ref{fig:svdm} shows the sizes of the
separate contributions to $\hat\sigma_{\rm eff}$ from $\ch \ch$
annihilation, ${\tilde t_1} \ch$ coannihilation and ${\tilde t_1} {\tilde
t_1}^*$, ${\tilde t_1} {\tilde t_1}$ annihilations (combined), as
functions of the mass difference between the ${\tilde t_1}$ and $\ch$.  In
Fig.~\ref{fig:svdm}, we have fixed $m_0=300\gev$, $\tan\beta=10$, $\mu >
0$,
$A_0=$ (a) 1000 and (b)  2000~GeV, and computed $\hat\sigma_{\rm eff}$ for
varying $m_{1/2}$, which amounts to varying the fractional mass difference
$\Delta M \equiv (m_{\tilde t_1} - m_\ch)/ m_\ch$. For these choices, the
stau mass, $m_{\stau_1}$, is much larger than $\mchi$. The thin dotted
lined is the value of
$\hat\sigma$ that one would compute if one ignored all coannihilation
contributions~\footnote{This differs from
${\hat \sigma}_{\ch \ch}$ because of the number-density weighting
factor.}.
Note that, in the case of close degeneracy between the
$\ch$ and
${\tilde t_1}$, the
${\tilde t_1} {\tilde t_1}$ and ${\tilde t_1} {\tilde t_1}^*$
annihilations dominate
$\hat\sigma_{\rm eff}$. However, since these contributions are suppressed
by two powers of $n_{\rm eq,{\tilde t_1}}$, they drop rapidly with $\Delta
M$, and neutralino-stop coannihilation takes over. For $A_0 = 2000
\gev$, this occurs at
$\Delta M\ga 0.18$.  This contribution in turn falls with one power of
$n_{\rm eq,{\tilde t_1}}$, and $\ch \ch$ annihilation re-emerges as the
dominant reaction for $\Delta M\ga0.25$. When $\Delta M\ga 0.35$,
${\tilde t_1}$ coannihilation effects can be neglected. In
We see the presence of kinematic thresholds also in Fig.~\ref{fig:svdm}.
In panel (a), we see the threshold for producing a single top
quark in $\stop \ch$ coannihilations, and in panel (b) we see
the threshold for producing a $t {\bar t}$ pair in $\ch \ch$
annihilations.

\section{Implications of ${\tilde t_1}$ Coannihilations for the 
Region of CMSSM Parameter Space Favoured by Cosmology}

\begin{figure}
\hspace*{-.70in}
\begin{minipage}{8in}
\epsfig{file=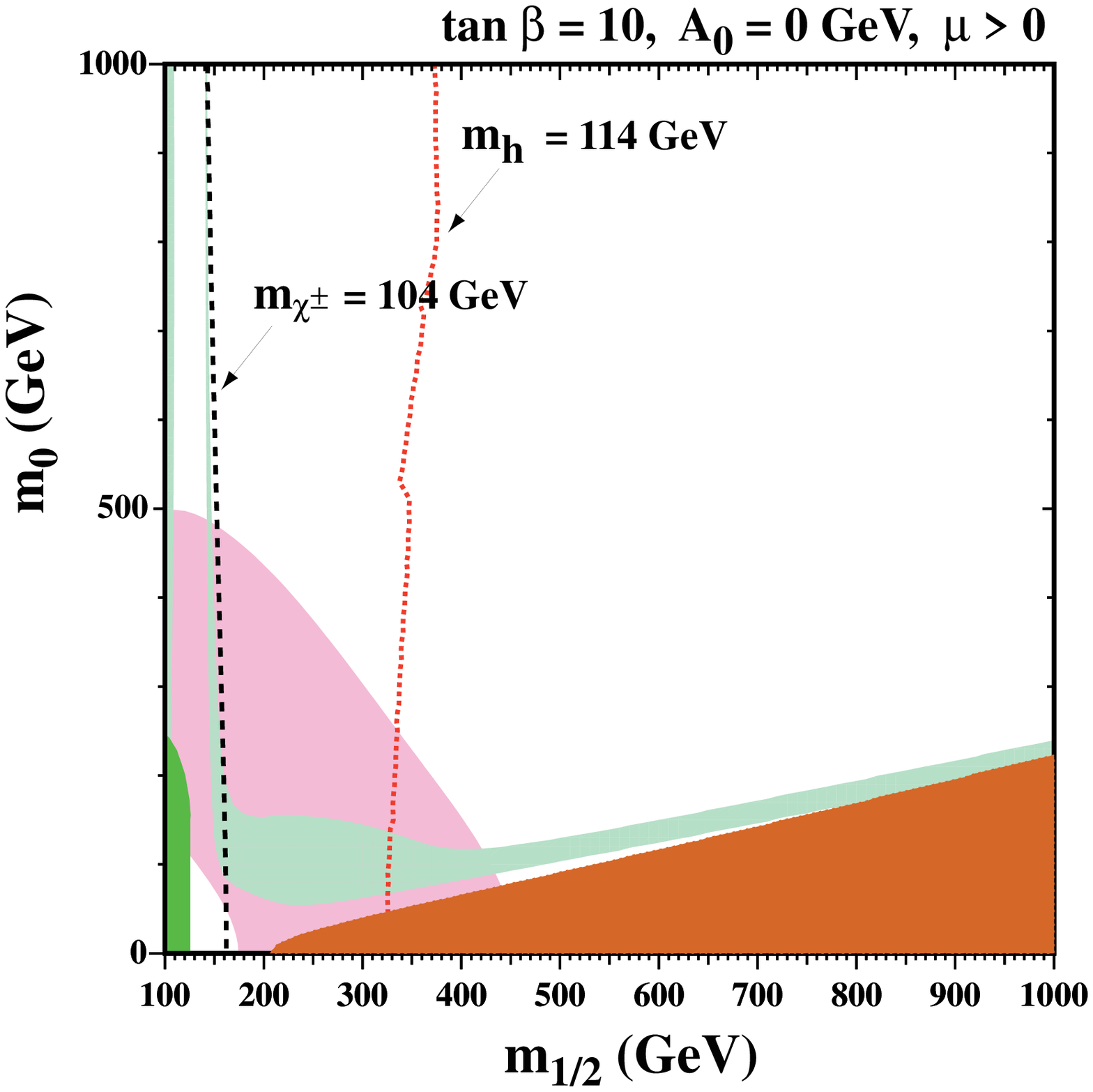,height=3.5in} 
\epsfig{file=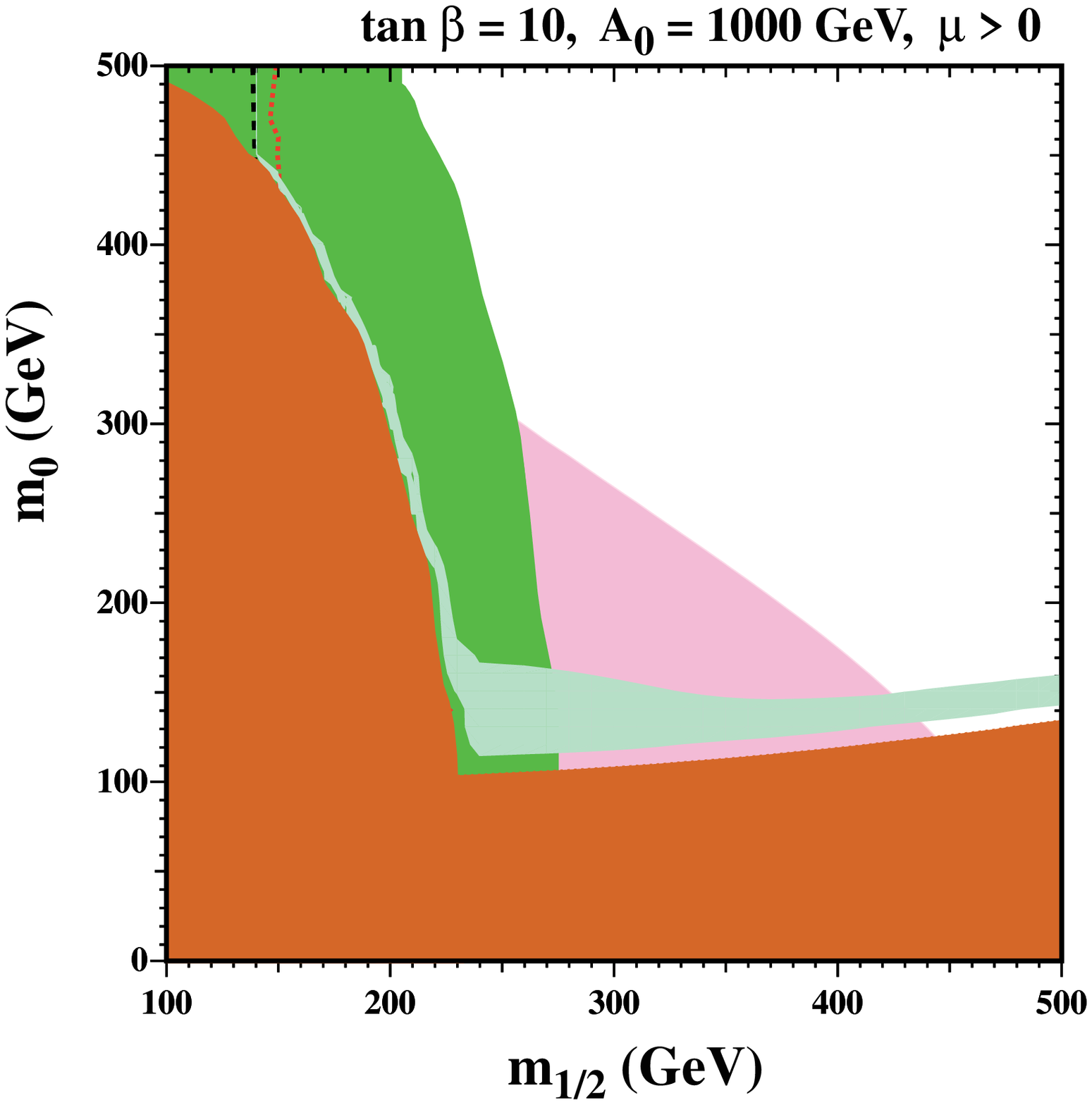,height=3.5in} \hfill
\end{minipage}
\begin{minipage}{8in}
\hspace*{-.70in}
\epsfig{file=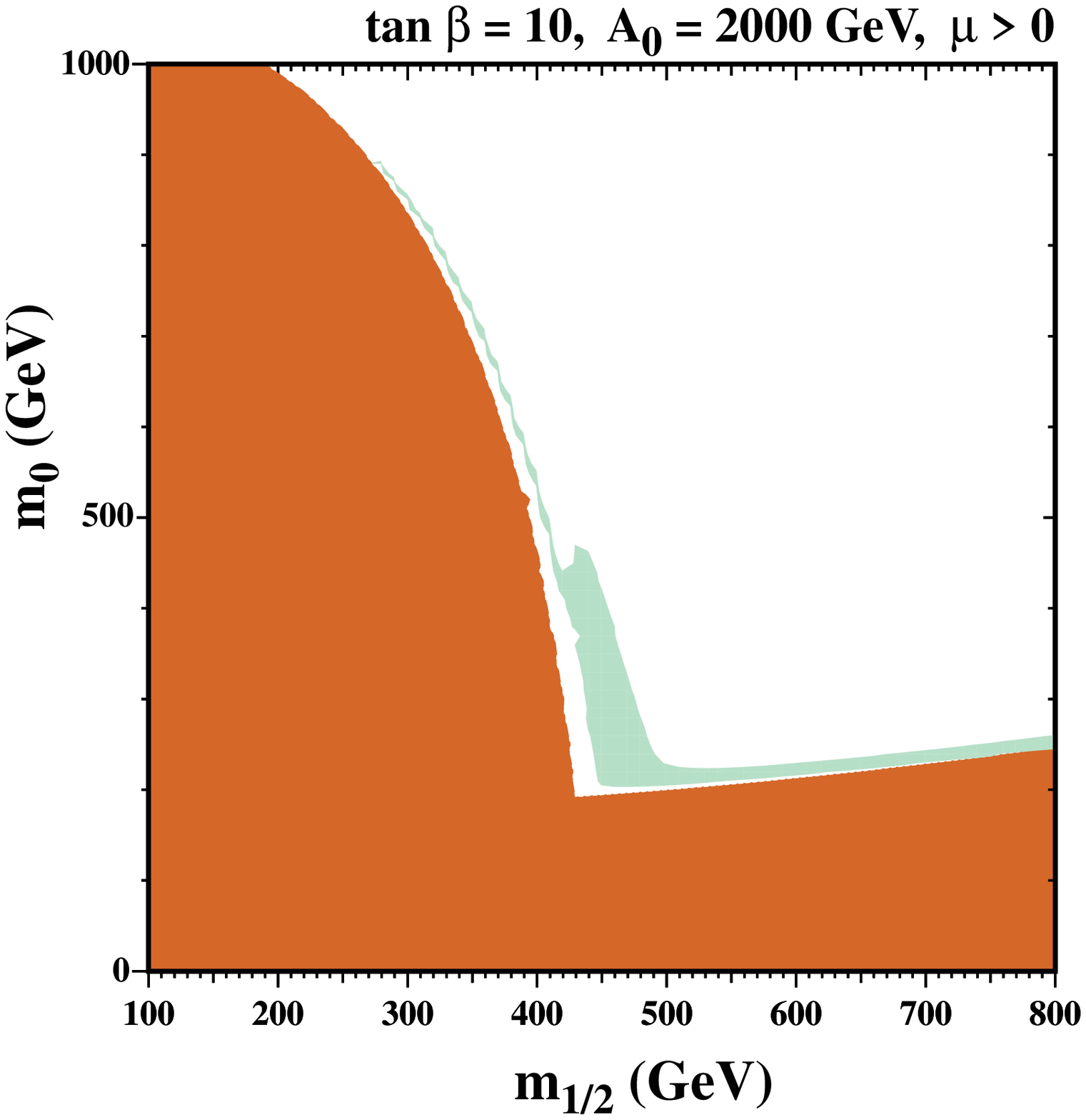,height=3.5in} 
\epsfig{file=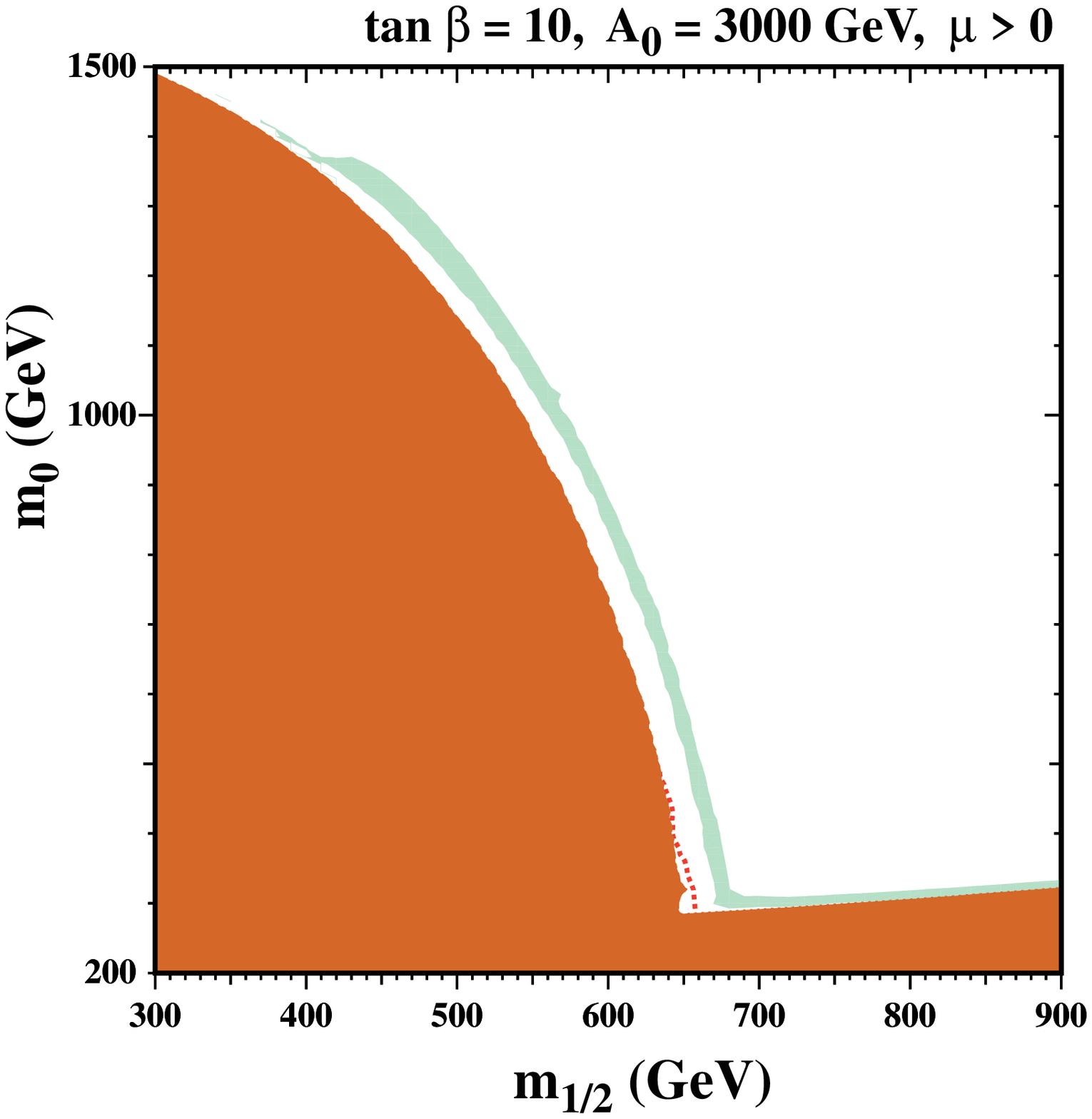,height=3.5in} \hfill
\end{minipage}
\caption{\label{fig:plane10}
{\it
The $(m_{1/2}, m_0)$ planes for $\tan \beta = 10, \mu > 0$ and $A_0 =$ 
(a) 0, (b) 1000, (c) 2000 and (d) 3000~GeV.
The very dark (red) shaded regions are excluded because the LSP is 
the ${\tilde t_1}$ or the ${\tilde \tau_1}$. The dark shaded regions 
are excluded by $b \to s \gamma$. The intermediate shaded regions are 
favoured by $g_\mu - 2$ at the 2-$\sigma$ level. The light 
(turquoise) shaded regions are those favoured by cosmology,
with \protect\mbox{$0.1\leq\ohsq\leq 0.3$} after the inclusion of 
coannihilation effects. The (black) dashed lines are the contours 
$m_{\ch^\pm} = 104$~GeV, and the (red) dotted lines are the contours $m_h 
= 114$~GeV.
}}
\end{figure}

We now explore the consequences of ${\tilde t_1}$ coannihilations for the
region of CMSSM parameter space in which $0.1 < \Omega_\ch h^2 < 0.3$, as
favoured by cosmology. We display in Fig.~\ref{fig:plane10} the
$(\m12, m_0)$ planes for $\tan\beta = 10$ and $\mu > 0$, for the different
values of $A_0 =$ (a) 0, (b) 1000, (c) 2000 and (d) 3000~GeV. The very
dark
(red) shaded regions have $m_{\tilde{\tau}_1}$ or $m_{\tilde t_1} < m_\ch$, and
hence are excluded by the very stringent bounds on charged dark
matter~\cite{EHNOS}. The light (turquoise) shaded regions correspond to
the
preferred relic-density range \mbox{$0.1<\ohsq<0.3$}. The dark (green)
shaded regions are those excluded by measurements of $b \to s \gamma$. The
intermediate (pink) shaded regions in panels (a) and (b) are those
favoured by the BNL measurement of $g_\mu - 2$ at the 2-$\sigma$
level~\cite{g-2}.  
The (black) dashed line in panel (a)  is the contour $m_{\ch^\pm} =
104$~GeV, corresponding to the kinematic range of LEP~2, and the (red)
dotted line is the contour $m_h = 114$~GeV, as evaluated using {\tt
FeynHiggs}~\cite{FeynHiggs}, corresponding to the LEP lower limit on the
mass of the Higgs
boson. These contours are also glimpsed in the other panels.

Panel (a) of Fig.~\ref{fig:plane10} is indistinguishable from analogous
plots shown previously~\cite{efosi,EFGOSi}: ${\tilde t_1}$ coannihilations
are not important
when $A_0 = 0$. Panel (b) shows how ${\tilde t_1}$ coannihilation
generates a `tail' of allowed CMSSM parameter space, extending the `bulk'
region of the $(m_{1/2}, m_0)$ plane up as far as $m_0 \sim 450$~GeV. 
The boundary of the region where $m_{\stop_1} < m_\ch$, and hence
the $\stop_1 \ch$ coannihilation region, slopes to the left as $m_0$ is
increased, since we have fixed $A_0$ over the plane.  As $m_0$ is
increased, the impact of the $A$-dependent off-diagonal term is diminished
and the light stop is heavier. To compensate for this, one must decrease
$m_{1/2}$ to obtain the necessary degree of degeneracy between
$\stop_1$ and $\ch$. This effect is also seen in panels (c) and (d) of
Fig.~\ref{fig:plane10}.

In the particular cae of $A_0 = 1000\gev$ shown in panel (b) of 
Fig.~\ref{fig:plane10}, the ${\tilde t_1}$ coannihilation tail
happens to be excluded by $b \to s \gamma$. However, this is not the case
for (c) $A_0 = 2000$~GeV and (d) $A_0 = 3000$~GeV, where the ${\tilde
t_1}$ coannihilation tail extends up to $m_0 \sim 900$ and beyond
1350~GeV, respectively. For (b) $A_0 = 2000$~GeV, the region
favoured by $g_\mu -
2$ is hidden on the left, inside the excluded region.
However, around
$A_0 = 1500$~GeV there is a $\ch \stop_1$ coannihilation region that
satisfies both the $b \to s \gamma$ and $g_\mu -
2$ constraints. 

The cosmologically-favoured region is broadened significantly when
$t {\bar t}$ production is kinematically allowed in $\ch \ch$
annihilation The t-channel stop exchange contribution to this process
is significantly enhanced when
$\stop_1$ is relatively light: $m_{\stop_1} \simeq
m_\ch$.
This feature can be seen in Fig.~\ref{fig:plane10}(c) and
Fig.~\ref{fig:plane10}(d). The  
$\stop_1 {\tilde \ell}$ coannihilations are only important in the
corner area near the instep of the dark
(red) shaded region where $m_{\tilde{\tau}_1} \simeq m_{\tilde t_1}
\simeq
m_\ch$. 

\begin{figure}
\hspace*{-.70in}
\begin{minipage}{8in}
\epsfig{file=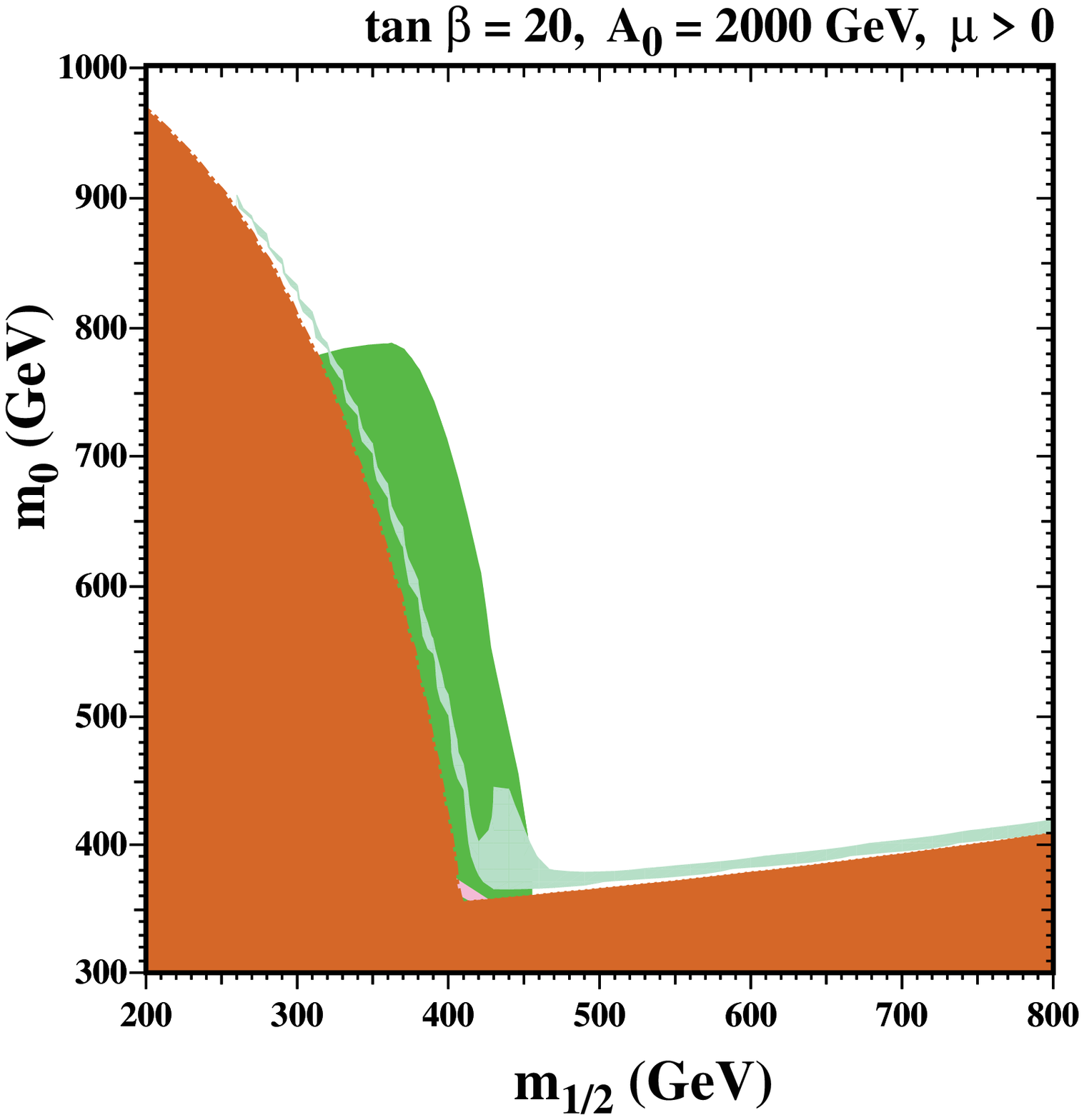,height=3.5in}
\epsfig{file=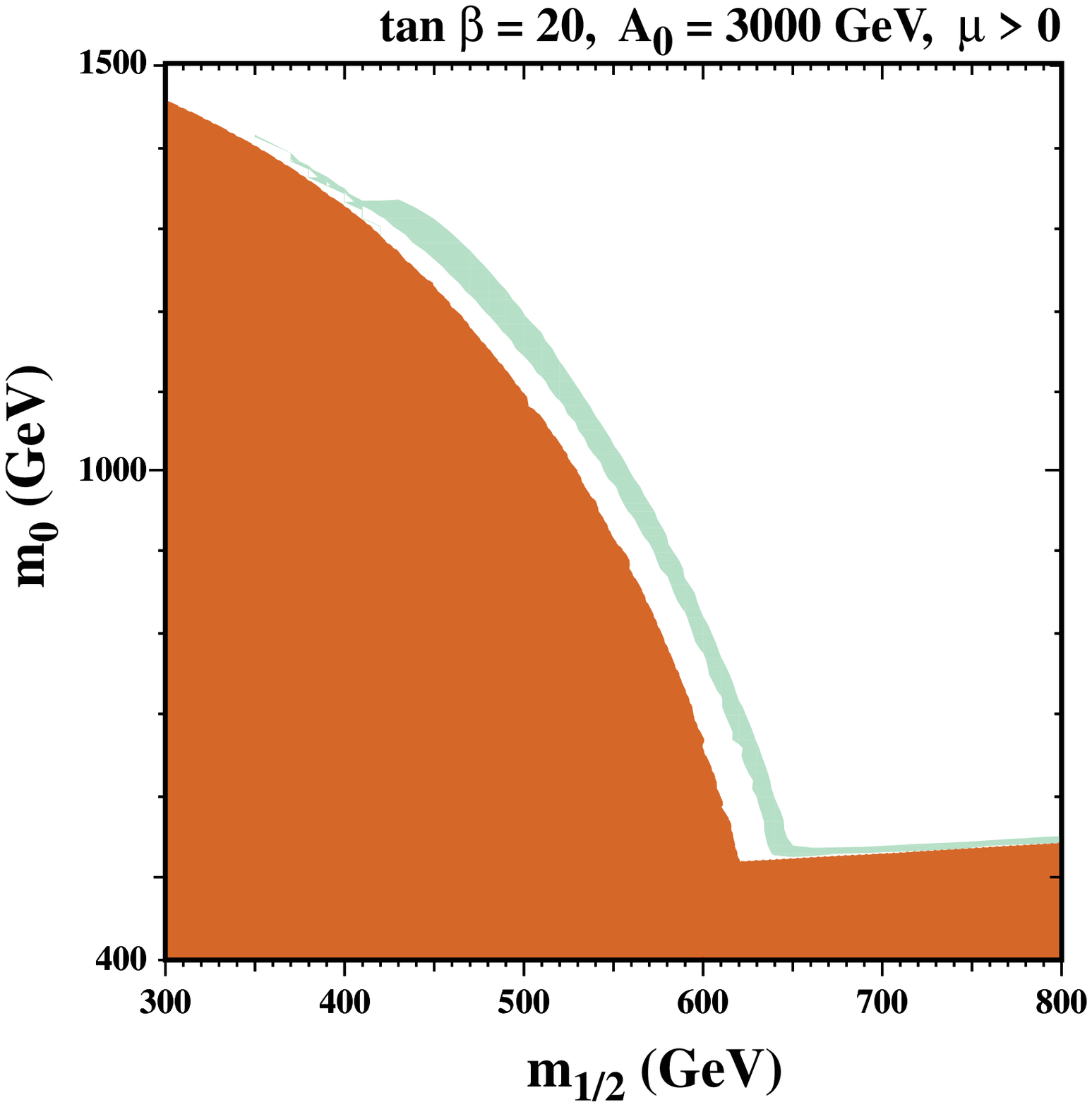,height=3.5in} \hfill
\end{minipage}
\begin{minipage}{8in}
\hspace*{-.70in}
\epsfig{file=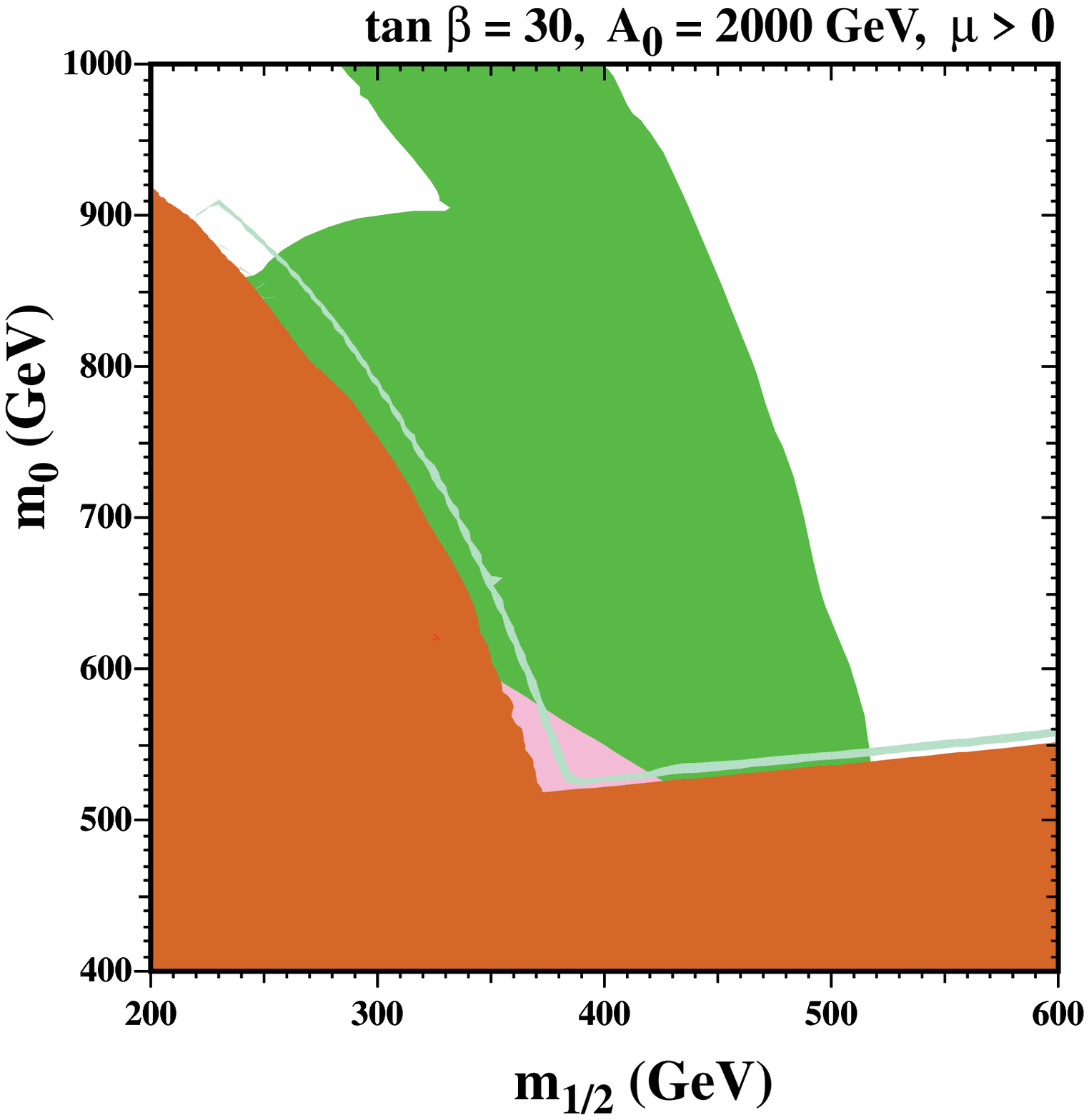,height=3.5in}
\epsfig{file=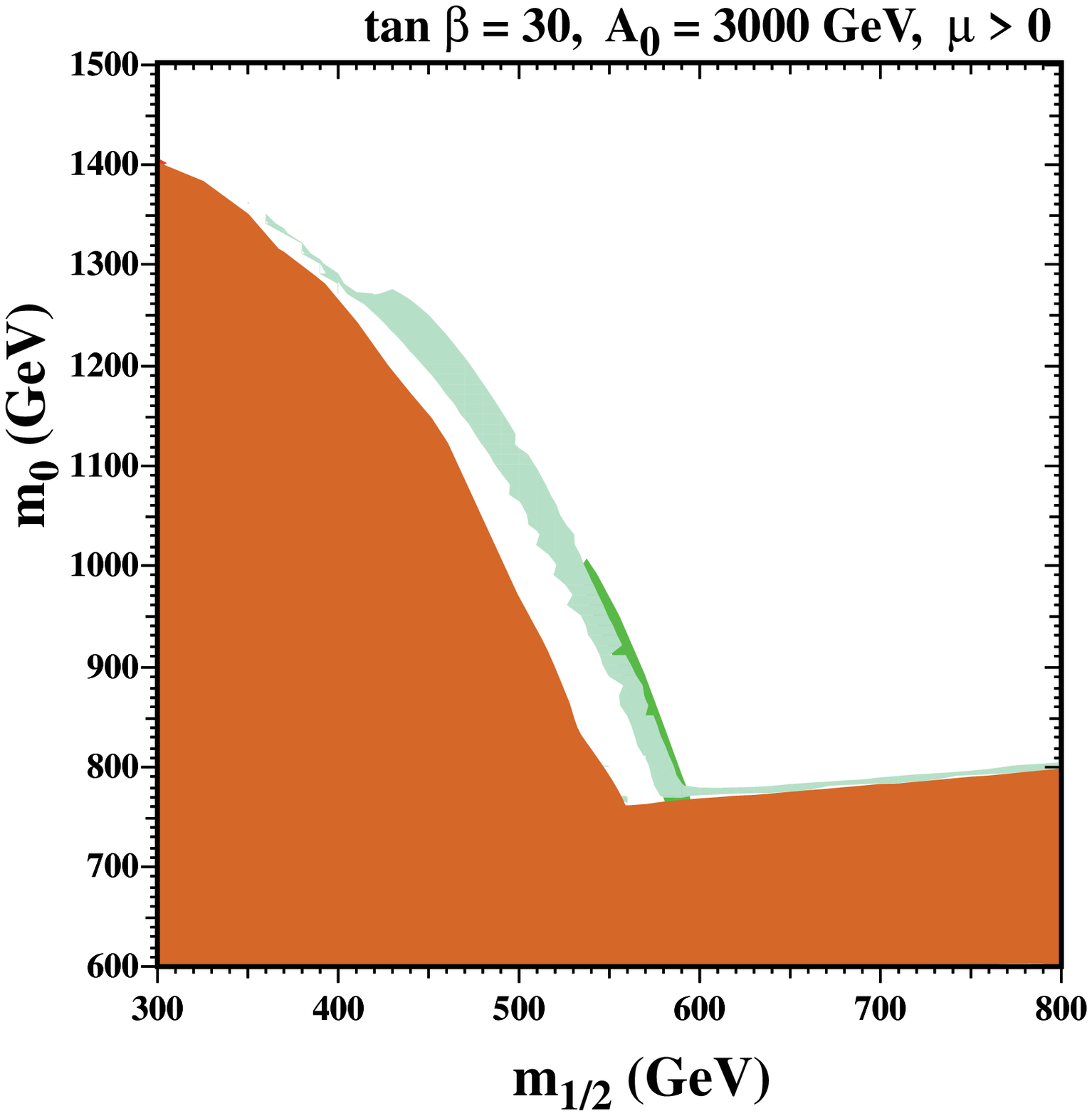,height=3.5in} \hfill
\end{minipage}
\caption{\label{fig:plane2030}
{\it 
Notation as for Fig.~\ref{fig:plane10}. Panels (a,b) are for $\tan \beta = 
20, \mu > 0$, and panels (c,d) are for $\tan \beta = 30, \mu > 0$. Panels 
(a,c) are for $A_0 = 2000$~GeV, and panels (b,d) for $A_0 = 3000$~GeV.
}}
\end{figure}

Fig.~\ref{fig:plane2030} shows analogous $(m_{1/2}, m_0)$ planes for the
larger values $\tan \beta = 20$ and 30, both again for $\mu > 0$. We see
in panel (a) that the ${\tilde t_1}$ coannihilation tail extends up to
$m_0 \sim 900$~GeV when $\tan \beta = 20$ and $A_0 = 2000$~GeV, beyond the
region forbidden by $b \to s \gamma$. We then see in panel (b) that the
${\tilde t_1}$ coannihilation tail extends beyond $m_0 \sim 1400$~GeV when
$\tan \beta = 20$ and $A_0 = 3000$~GeV. A similar portion of the ${\tilde
t_1}$ coannihilation tail that is consistent with $b \to s \gamma$ is also
visible in panel (c), for $\tan \beta = 20$ and $A_0 = 2000$~GeV, where
$m_0 \sim 900$~GeV. For fixed $A_0$, as $\tan \beta$ is increased the $b \to s
\gamma$ constraint becomes more severe.

The
${\tilde t_1}$ coannihilation tails do not increase the allowed range of
$m_{1/2}$, as was the case for the
${\tilde
\ell}$ coannihilation tail~\cite{efo,efosi,coann,ADS}, but they do add a
significant filament to
the CMSSM region allowed by experiment and favoured by cosmology.

In the above illustrative examples, we have chosen the sign of $A_0$ to be
the same as that of $\mu$, with the effect of maximizing the stop
off-diagonal mass terms, and hence minimizing $m_{\stop_1}$. If 
the sign
of $A_0$ is switched while keeping $\mu > 0$, an analogous $\ch
\stop_1$
coannihilation region is found only at even larger negative $A_0$.
However, the $m_h$ constraint is more severe for
$A_0 < 0$, and also 
for $\mu < 0$, excluding the region of  parameter space of interest in the
$\ch \stop_1$
coannihilation context.    

We have not discussed in this paper the potential constraint imposed by
the absence of
colour and charge breaking (CCB) vacua, or at least the suppression of
transitions to CCB vacua. This constraint would restrict
the value of $A_0$ relative to $m_0$~\cite{AF}. If the CCB constraint is
imposed, 
some of the parameter space for small $m_0$ will be excluded, but
there will still be regions at high $m_0$ where $\ch {\tilde t_1}$
coannihilation is crucial. The issue of compatibity with the
potential $g_\mu
-2$ constraint would then arise. Since raising $m_0$ suppresses the
neutralino-proton elastic
scatttering cross
section, $\ch \stop_1$ coannihilation would tend to lower the neutralino
direct
detection rate for the same range of $m_\ch$.   

\section{Conclusions and Open Issues}

We have documented in this paper the potential importance of $\ch {\tilde
t_1}$
coannihilation in delineating the preferred domain of CMSSM parameter
space for $A_0 \ne 0$. In this paper, we have only sctratched the surface
of this subject, whose higher-dimensional parameter space merits more
detailed exploration. The Appendix provides details of the diagrammatic
calculations that should be sufficient for our results to be verified and
used by other authors. Although applied in the context of the CMSSM, our
results may also be used in more general MSSM contexts.  However, other
coannihilation processes are also important in other regions of the
general MSSM parameter space. For example, in the CMSSM the sbottom mass
is
generally larger than the stop mass even for large $\tan \beta$. However, if one
allow non-degeneracy in the scalar soft breaking mass term, a sbottom NLSP
becomes
possible~\cite{ADS}. A complete calculation of the LSP relic
density in the MSSM requires a careful discussion of all such
coannihilation possibilities.

\
\vskip 0.5in
\vbox{
\noindent{ {\bf Acknowledgments} } \\
\noindent 
We thank Toby Falk for many related discussions. 
The work of K.A.O. and Y.S. was supported in part by DOE grant
DE--FG02--94ER--40823.}

\newpage

\baselineskip=14pt
\appendix
\setcounter{equation}{0}
\renewcommand{\theequation}{A\arabic{equation}}
\section*{Appendix}

We
calculate the tree contributions to the coannihilation and annihilation
amplitudes that are leading order in $\alpha_s$, $\alpha_t \equiv
\lambda_t^2 / 4 \pi$ and/or $\alpha_b \equiv \lambda_b^2 / 4 \pi$. We keep terms
that are of the same order in $\sqrt{\alpha_s \alpha_{em}}$,
$\sqrt{\alpha_t \alpha_{em}}$,
$\sqrt{\alpha_b \alpha_{em}}$, $\sqrt{\alpha_s \alpha_W}$,
$\sqrt{\alpha_t \alpha_W}$ and/or 
$\sqrt{\alpha_b \alpha_W}$, but 
neglect most terms in $\alpha_{em}$ and/or
$\alpha_W$. This approximation is motivated by the relatively large
magnitudes of $\alpha_s$ and $\alpha_t$, and the fact that $\alpha_b \sim
\alpha_t$ at large $\tan \beta$. We have however included s-channel $Z$,
and $\gamma$ exchange for $\stop_1 \stop_1^* \rightarrow t \bar t$ and checked
that they make negligible contribution to the total cross section and therefore
neglect these channels in the remaining amplitudes.   

\subsection*{Couplings}

Here we list couplings used in the calculation. Factors not written, such
as $i$, $\gamma$'s, QCD generators and momentum , are taken into account in the
calculation of amplitudes squared, which are listed subsequently. 

\baselineskip=20pt
\begin{eqnarray*}
C_{g-g-g} &=& -g_s \\
C_{\stop_1-\stop_1-g} &=& -g_s \\
C_{\stop_1-\stop_1-g-g} &=& g_s^2 \\
C_{\stop_1-\stop_1-\gamma} &=& -2\, e/3 \\
C_{\stop_1-\stop_1-\gamma-g} &=& 4\, e\, g_s/3 \\
C_{\stop_1-\stop_1-Z} &=& \frac{-g_2}{\cos \theta_{\ss W}} ((-2/3) \sin^2
\theta_{\ss W}   + \cos^2 \theta_t /2 ) \\
C_{\stop_1-\stop_2-Z} &=& \frac{g_2}{\cos \theta_{\ss W}} (1/2) \cos \theta_t
\sin \theta_t \\
C_{\stop_1-\stop_1-Z-g} &=& \frac{2 \, g_s \, g_2}{\cos \theta_{\ss W}} ((1/2)
\cos^2
\theta_t - (2/3)      \sin^2 \theta_{\ss W}) \\
C_{q-q-g} &=& -g_s \\
C_{\stop_1-t-\tilde{g}}^L &=&  \sqrt{2} \; g_s \sin \theta_t \\
C_{\stop_1-t-\tilde{g}}^R &=&  -\sqrt{2} \; g_s \cos \theta_t \\
C_{\stop_1-\stop_1-h} &=& \frac{g_2 \, \mz}{\cos \theta_{\ss W}} \left( \left(
\frac{1}{2} -
\frac{2}{3} \sin^2\! \theta_{\ss W} \right) \cos^2\! \theta_t  + \frac{2}{3}
\sin^2\! \theta_{\ss W} \sin^2\! \theta_t \right) \sin (\alpha
+ \beta) \nl - \, \frac{g_2\,  m_t^2 \cos \alpha}{\mw \sin \beta} - 
\frac{g_2 \, m_t\, (\mu  \sin \alpha - A_t \cos \alpha)}{\mw \sin \beta}
\sin \theta_t \cos \theta_t \\
C_{\stop_1-\stop_2-h} &=& \frac{g_2 \, \mz}{\cos \theta_{\ss W}} \left(
\frac{2}{3} \sin^2\! \theta_{\ss W} - \left(
\frac{1}{2} - \frac{2}{3} \sin^2\! \theta_{\ss W} \right)
 \right) \cos \theta_t  \sin \theta_t
\sin(\alpha+\beta) \nl - \frac{g_2\, m_t \, (\mu \sin \alpha - A_t \cos
\alpha)}{2\, \mw \sin \beta}( \cos^2\! \theta_t - \sin^2\! \theta_t)  \\
C_{\stop_1-\stop_1-H} &=& - \,\frac{g_2 \, \mz}{\cos \theta_{\ss W}} \left( \left(
\frac{1}{2} -
\frac{2}{3} \sin^2\! \theta_{\ss W} \right)
\cos^2\! \theta_t
+ \frac{2}{3} \sin^2\! \theta_{\ss W} \sin^2\! \theta_t \right) \cos
(\alpha+\beta)  \nl  - \, \frac{g_2\, m_t^2 \sin \alpha}{\mw  \sin \beta}
 + \frac{g_2 \, m_t\,  (\mu \cos \alpha + A_t \sin \alpha)}{\mw \sin \beta}
\sin \theta_t \cos \theta_t \\
C_{\stop_1-\stop_2-H} &=& - \, \frac{g_2 \, \mz}{ \cos \theta_{\ss W}} \left(
\frac{2}{3} \sin^2 \! \theta_{\ss W} - \left(
\frac{1}{2} - \frac{2}{3} \sin^2\! \theta_{\ss W} \right) \right)
 \sin \theta_t \cos \theta_t 
\cos(\alpha+\beta) \nl + \, \frac{g_2 \, m_t\, (\mu \cos \alpha + A_t \sin
\alpha)}{2 \, \mw \sin
\beta} ( \cos^2\! \theta_t - \sin^2 \!  \theta_t) \\
C_{\stop_1-\stop_1-A} &=& 0 \\
C_{\stop_1-\stop_2-A} &=&  \frac{g_2 \, m_t}{2 \, \mw}(\mu-A_t \cot \beta) \\
C_{t-t-\gamma} &=& -2\, e/3 \\
C_{f-f-Z}^L &=&  \frac{g_2}{\cos \theta_{\ss W}} (Q_f \sin^2 \theta_{\ss W} -
T_{3f}) \\
C_{f-f-Z}^R &=&  \frac{g_2}{\cos \theta_{\ss W}} \, Q_f \sin^2 \theta_{\ss W}
\\
C_{t-t-h} &=& - \, 
\frac{g_2 \, m_t \cos \alpha}{2 \, \mw \sin
\beta} \\
C_{t-t-H} &=& - \, \frac{g_2\, m_t \sin \alpha}{2\, \mw \sin \beta} \\
C_{t-t-A} &=&    \frac{g_2\, m_t \cot \beta}{2\, \mw} \\
C_{b-b-h} &=& \frac{g_2 \, m_b \sin \alpha}{2\, \mw \cos \beta} \\
C_{b-b-H} &=& - \, \frac{g_2 \, m_b \cos \alpha}{2 \, \mw \cos \beta} \\
C_{\stop_1-b-\chi^+_i}^{L} &=&  \frac{g_2\, m_b}{ \sqrt{2} \, \mw \cos \beta}
\, U_{i2} \cos \theta_t \\
C_{\stop_1-b-\chi^+_i}^{R} &=& -\,g_2 \, V_{i1} \cos \theta_t + 
\frac{g_2\, m_t}{ \sqrt{2} \, \mw \sin \beta} \, V_{i2} \sin \theta_t \\
C_{\tilde{\ell}_1-\nu-\chi^+_i}^{L} &=&  0 \\
C_{\tilde{\ell}_1-\nu-\chi^+_i}^{R} &=&  \frac{g_2\, m_{\ell}}{\sqrt{2}\, \mw
\cos \beta} \, U_{i2} \sin \theta_{\ell} - g_2\, U_{i1} \cos \theta_{\ell} \\
C_{h-h-h} &=&  -\, \frac{3\, g_2\, \mz}{2 \cos \theta_{\ss W}} \cos 2 \alpha
\sin(\beta+\alpha) \\
C_{h-h-H} &=&  - \, \frac{g_2 \, \mz}{2 \cos \theta_{\ss W}} (2 \sin 2 \alpha
\sin (\beta+\alpha) - \cos(\beta+\alpha) \cos 2 \alpha) \\
C_{h-H-H} &=& \frac{g_2 \, \mz}{2 \cos \theta_{\ss W}} (2 \sin 2 \alpha
\cos (\beta+\alpha)+  \sin (\beta+\alpha) \cos 2 \alpha) \\
C_{H-H-H} &=& -\, \frac{3\, g_2 \, \mz}{2 \cos \theta_{\ss W}} \cos 2 \alpha
\cos(\beta+\alpha) \\
C_{h-A-A} &=&  -\, \frac{g_2\, \mz}{2 \cos \theta_{\ss W}} \cos 2 \beta
\sin(\beta+\alpha)   \\
C_{H-A-A} &=&  \frac{g_2 \, \mz}{2 \cos \theta_{\ss W}} \cos 2 \beta
\cos(\beta+ \alpha)    \\
C_{h-H^+-H^-} &=& -g_2 (\mw \sin (\beta-\alpha) + \mz/(2 \cos \theta_{\ss W})
\cos (2 \beta)
             \sin (\beta + \alpha)) \\
C_{H-H^+-H^-} &=& -g_2 (\mw \cos(\beta-\alpha) - \mz/(2 \cos \theta_{\ss W})
\cos (2 \beta)
   \cos (\beta+\alpha)) \\
C_{h-Z-A} &=& g_2 \cos (\alpha-\beta)/(2 \cos \theta_{\ss W})\\
C_{H-Z-A} &=& g_2 \sin (\alpha-\beta)/(2 \cos \theta_{\ss W})\\
C_{h-W^+-H^-} &=& \frac{g_2}{2} \cos (\alpha - \beta) \\
C_{H-W^+-H^-} &=& \frac{g_2}{2} \sin (\alpha - \beta) \\
C_{\stop_1- \stop_1-h-h} &=& \frac{g_2^2}{2} \left(
\frac{\cos 2\alpha }{\cos^2\! \theta_{\ss W}}
\left( \left( \frac{1}{2} - \frac{2}{3} \sin^2\! \theta_{\ss W} \right)
\cos^2\! \theta_t + \frac{2}{3} \sin^2\! \theta_{\ss W} \sin^2 \theta_t \right)
- \frac{m_t^2}{\mw^2} \frac{\cos^2\! \alpha}{ \sin^2\! \beta} \right) \\
C_{\stop_1- \stop_1-h-H} &=& \frac{g_2^2 \sin 2 \alpha}{2} \left( \left(
\frac{1}{2} - \frac{2}{3} \sin^2\! \theta_{\ss W} \right) \sec^2\! \theta_{\ss
W}  \cos^2 \! \theta_t + \frac{2}{3} \tan^2\!
\theta_{\ss W} \sin^2 \! \theta_t  \right. \nl \left. - \frac{m_t^2}{2\, \mw^2
\sin^2\! \beta}
\right) \\
C_{\stop_1- \stop_1-H-H} &=& \frac{g_2^2}{2} \left( \frac{-\cos 2
\alpha}{\cos^2\! \theta_{\ss W}} \left( \left( \frac{1}{2} - \frac{2}{3}
\sin^2 \theta_{\ss W} \right) \cos^2\! \theta_t + \frac{2}{3} \sin^2 \!
\theta_{\ss W} \sin^2 \! \theta_t \right)  -
\frac{m_t^2}{\mw^2}
\frac{\sin^2 \! \alpha}{ \sin^2\! \beta} \right) \\
C_{\stop_1- \stop_1-A-A} &=& \frac{g_2^2}{2} \left( \frac{\cos 2
\beta}{\cos^2\! \theta_{\ss W}} \left( \left( \frac{1}{2} - \frac{2}{3} \sin^2
\! \theta_{\ss W} \right)
\cos^2\! \theta_t + \frac{2}{3} \sin^2\! \theta_{\ss W} \sin^2 \theta_t \right) 
 - \frac{m_t^2}{\mw^2 \tan^2\! \beta} \right) \\
C_{\stop_1- \stop_1-H^+-H^-} &=& \frac{g_2^2 \cos 2 \beta }{2} \left( \left( -1
+ \left( \frac{1}{2} 
- \frac{2}{3} \sin^2\! \theta_{\ss W} \right)
 \sec^2\! \theta_{\ss W} \right) \cos^2\! \theta_t  +
 \frac{2}{3} \tan^2\!
 \theta_{\ss W} \sin^2\! \theta_t \right) \nl  
- \frac{g_2^2}{2\, \mw^2}(m_b^2 \tan^2 \! \beta \cos^2\! \theta_t + 
m_t^2 \cot^2 \! \beta \sin^2\! \theta_t) \\
C_{\stop_1- \sbot_1-H^+} &=& - \, \frac{g_2\, \mw}{\sqrt{2}} \left( \sin 2 \beta
- \frac{m_b^2 \tan \beta + m_t^2 \cot \beta}{\mw^2} \right) \cos \theta_t \cos
\theta_b + \, \frac{g_2}{\sqrt{2}\, \mw} \nl \times \, 
\left( m_t \, m_b \, (\tan \beta + \cot \beta)
\sin \theta_t \sin \theta_b 
- m_b \, (A_b \tan \beta- \mu) \cos \theta_t \sin
\theta_b \right. \nl \left.
- m_t \, (A_t \cot \beta - \mu) \sin \theta_t \cos
\theta_b \right) \\
C_{\stop_1- \sbot_2-H^+} &=& \frac{g_2\, \mw}{\sqrt{2}} \left( \sin 2
\beta - \frac{m_b^2
\tan \beta + m_t^2  \cot \beta}{\mw^2} \right) \cos \theta_t \sin \theta_b 
+ \, \frac{g_2}{\sqrt{2} \,\mw} \nl \times \, 
\left( m_t\, m_b (\tan \beta + \cot \beta)
\sin \theta_t \cos \theta_b 
- m_b (A_b \tan \beta - \mu) \cos \theta_t \cos \theta_b \right. \nl \left.
+ m_t (A_t \cot \beta - \mu) \sin \theta_t \sin \theta_b \right) \\
C_{\stop_1- \sbot_1-W^+} &=& \frac{- g_2}{\sqrt{2}} \cos \theta_t \cos \theta_b
\\
C_{\stop_1- \sbot_2-W^+} &=&  \frac{g_2}{\sqrt{2}} \cos \theta_t \sin \theta_b
\\
C_{\stop_1-t-\tilde{\chi}^0_i}^{L} &=& \frac{- g_2}{\sqrt{2}} \left( \cos
\theta_t \,  ( N_{i2} + \frac{\tan \theta_{\ss W}}{3} N_{i1} ) + \sin \theta_t
\,
\frac{m_t
\, N_{i4}}{\mw \sin \beta} \right) \\
C_{\stop_1-t-\tilde{\chi}^0_i}^{R} &=& \frac{- g_2}{\sqrt{2}} \left( \cos
\theta_t \, \frac{m_t\,
N_{i4}}{\mw \sin \beta} - \frac{4}{3} \sin \theta_t \,
  \tan \theta_{\ss W} N_{i1} \right) {\rm sign}(m_{\tilde{\chi}^0_i}) \\
C_{\stop_2-t-\tilde{\chi}^0_i}^{L} &=& \frac{- g_2}{\sqrt{2}} \left( -\sin
\theta_t \,( N_{i2}
+ \frac{\tan \theta_{\ss W}}{3} N_{i1} )  + \cos \theta_t \,\frac{  m_t \,
N_{i4}}{\mw \sin \beta} \right) \\
C_{\stop_2-t-\tilde{\chi}^0_i}^{R} &=& \frac{- g_2}{\sqrt{2}} \left( -\sin
\theta_t \,\frac{ m_t \,
N_{i4}}{\mw \sin \beta} - \frac{4}{3} \cos \theta_t \,
 \tan \theta_{\ss W} N_{i1} \right) {\rm sign}(m_{\tilde{\chi}^0_i})
\\
C_{\sbot_1-b-\tilde{\chi}^0_i}^{L} &=& \frac{- g_2}{\sqrt{2}} \left( -\cos
\theta_b \, 
(N_{i2} - \frac{\tan \theta_{\ss W}}{3} N_{i1}) + \sin \theta_b \,\frac{m_b \, 
N_{i3}}{\mw \cos \beta} \right) \\
C_{\sbot_1-b-\tilde{\chi}^0_i}^{R} &=& \frac{- g_2}{\sqrt{2}} \left( \cos
\theta_b \, \frac{m_b \, N_{i3}}{\mw \cos \beta}  + \frac{2}{3}\sin \theta_b \,
  \tan \theta_{\ss W} N_{i1} \right) {\rm
sign}(m_{\tilde{\chi}^0_i})  \\
C_{\sbot_2-b-\tilde{\chi}^0_i}^{L} &=& \frac{- g_2}{\sqrt{2}} \left( \sin
\theta_b \, 
(N_{i2} - \frac{\tan \theta_{\ss W}}{3} N_{i1}) + \cos \theta_b \,\frac{ m_b \,
N_{i3}}{\mw \cos \beta} \right) \\
C_{\sbot_2-b-\tilde{\chi}^0_i}^{R} &=& \frac{- g_2}{\sqrt{2}} \left( -\sin
\theta_b \, \frac{ m_b\,
N_{i3}}{\mw  \cos \beta}  + \frac{2}{3} \cos \theta_b \,
 \tan \theta_{\ss W} N_{i1} \right) {\rm
sign}(m_{\tilde{\chi}^0_i})  \\
C_{\tilde{\ell}_1-\ell-\tilde{\chi}^0_i}^{L} &=& \frac{- g_2}{\sqrt{2}} \left(
- \cos \theta_{\ell} 
(N_{i2} + \tan \theta_{\ss W} N_{i1}) + \sin \theta_{\ell} \, 
\frac{ m_{\ell}\, N_{i3}}{\mw \cos \beta} \right)  \\
C_{\tilde{\ell}_1-\ell-\tilde{\chi}^0_i}^{R} &=& \frac{- g_2}{\sqrt{2}} \left(
 \cos \theta_{\ell} \, \frac{ m_{\ell}\, N_{i3}}{\mw \cos \beta} +  2 \sin
 \theta_{\ell}  \tan \theta_{\ss W} N_{i1} \right) {\rm
     sign}(m_{\tilde{\chi}^0_i})  \\     
C_{\tilde{\chi}^0_i - \tilde{\chi}^0_j - h}^{L} &=&  \frac{g_2}{2} [
(N_{i3} (N_{j2} -
N_{j1} \tan \theta_{\ss W}) + N_{j3} (N_{i2} - N_{i1} \tan \theta_{\ss W}))
\, {\rm sign}(m_{\tilde{\chi}^0_i})  \sin \alpha \nl    
 + \, (N_{i4} (N_{j2} - N_{j1} \tan
\theta_{\ss W}) + N_{j4} (N_{i2} - N_{i1} \tan \theta_{\ss W})) \cos
\alpha ] \\
C_{\tilde{\chi}^0_i - \tilde{\chi}^0_j - h}^{R} &=& \frac{g_2}{2} [
(N_{j3} (N_{i2} -
N_{i1} \tan \theta_{\ss W}) + N_{i3} (N_{j2} - N_{j1} \tan \theta_{\ss W})) \,
 {\rm sign}(m_{\tilde{\chi}^0_j})   \sin \alpha  \nl + \, (N_{j4}
(N_{i2} - N_{i1} \tan
\theta_{\ss W}) + N_{i4} (N_{j2} - N_{j1} \tan \theta_{\ss W})) \cos \alpha
] \\
C_{\tilde{\chi}^0_i - \tilde{\chi}^0_j - H}^{L} &=& -\, \frac{g_2}{2} [
(N_{i3} (N_{j2} -
N_{j1} \tan \theta_{\ss W}) + N_{j3} (N_{i2} - N_{i1} \tan \theta_{\ss W}))
\, {\rm sign}(m_{\tilde{\chi}^0_i}) \nl \times \, \cos \alpha   - \, (N_{i4}
(N_{j2} - N_{j1} \tan
\theta_{\ss W}) + N_{j4} (N_{i2} - N_{i1} \tan \theta_{\ss W})) \sin \alpha
]  \\
C_{\tilde{\chi}^0_i - \tilde{\chi}^0_j - H}^{R} &=& - \, \frac{g_2}{2} [
(N_{j3} (N_{i2} -
N_{i1} \tan \theta_{\ss W}) + N_{i3} (N_{j2} - N_{j1} \tan \theta_{\ss W}))
\, {\rm sign}(m_{\tilde{\chi}^0_j}) \nl \times \, \cos \alpha   - \, (N_{j4}
(N_{i2} - N_{i1} \tan
\theta_{\ss W}) + N_{i4} (N_{j2} - N_{j1} \tan \theta_{\ss W})) \sin \alpha
]  \\
C_{\tilde{\chi}^0_i - \tilde{\chi}^0_j - A}^{L} &=&  \frac{g_2}{2} [
(N_{i3} (N_{j2} -
N_{j1} \tan \theta_{\ss W}) + N_{j3} (N_{i2} - N_{i1}  \tan \theta_{\ss W}))
\, {\rm sign}(m_{\tilde{\chi}^0_i}) \nl \times \, \sin \beta   - \, (N_{i4}
(N_{j2} - N_{j1} \tan
\theta_{\ss W}) + N_{j4} (N_{i2} - N_{i1} \tan \theta_{\ss W})) \cos \beta
] \\
C_{\tilde{\chi}^0_i - \tilde{\chi}^0_j - A}^{R} &=&  -\frac{g_2}{2} [
((N_{j3} (N_{i2}
- N_{i1} \tan \theta_{\ss W}) + N_{i3} (N_{j2} - N_{j1} \tan \theta_{\ss W}))
\, {\rm sign}(m_{\tilde{\chi}^0_j}) ) \nl \times \, \sin \beta   - \, (N_{j4}
(N_{i2} - N_{i1} \tan
\theta_{\ss W}) + N_{i4} (N_{j2} - N_{j1} \tan \theta_{\ss W})) \cos \beta
]  \\
C_{t - b - H^+}^{L} &=&   \frac{g_2}{ \sqrt{2} \, \mw} \, m_b \tan \beta   \\
C_{t - b - H^+}^{R} &=&  \frac{g_2}{ \sqrt{2}\, \mw} \, m_t \cot \beta  \\
C_{\tilde{\chi}^0_i - \tilde{\chi}^-_j - H^+}^{L} &=&  -g_2 \left( N_{i3} \,
U_{j1} - (N_{i2} + N_{i1} \tan \theta_{\ss W}) U_{j2}/\sqrt{2} \right)
{\rm sign}(m_{\tilde{\chi}^0_i})  \sin \beta  \\
C_{\tilde{\chi}^0_i - \tilde{\chi}^-_j - H^+}^{R} &=&  -g_2 \left( N_{i4}\,
V_{j1} + (N_{i2} + N_{i1} \tan \theta_{\ss W}) V_{j2}/\sqrt{2} \right) \cos
\beta  \\
C_{\tilde{\chi}^0_i - \tilde{\chi}^0_j - Z}^{L} &=& \frac{g_2}{4 \cos
\theta_{\ss W}} {\rm sign}(m_{\tilde{\chi}^0_j})
         (N_{i4} \, N_{j4} - N_{i3} \, N_{j3}) \\
C_{\tilde{\chi}^0_i - \tilde{\chi}^0_j - Z}^{R} &=& - C_{\tilde{\chi}^0_i -
\tilde{\chi}^0_j - Z}^{L}\\
C_{\tilde{\chi}^0_i - \tilde{\chi}^-_j - W^+}^{L} &=& g_2 (N_{i2} \, V_{j1} -
N_{i4} \, V_{j2}   / \sqrt{2} )\\
C_{\tilde{\chi}^0_i - \tilde{\chi}^-_j - W^+}^{R} &=& g_2 (N_{i2} \, U_{j1} +
N_{i3} \, U_{j2}  / \sqrt{2} )
\end{eqnarray*}

\subsection*{Squared Amplitudes}
Below is the list of the amplitudes squared.
Note that, for identical-particle final states,
one needs to divide them by two when performing the 
momentum integrations.

\subsection*{$\stop_1\stop_1^*\longrightarrow gg$}
 I. s-channel gluon annihilation \hfill\\
 II. t-channel $\stop_1$ exchange \hfill\\
 III. u-channel $\stop_1$ exchange \hfill\\
 IV.  point interaction\hfill\\
\begin{eqnarray}
\f1 &=& C_{\stop_1-\stop_1-g} \, C_{g-g-g}    \nonumber \\
\f2 &=& (C_{\stop_1-\stop_1-g})^2    \nonumber \\
\f3 &=& (C_{\stop_1-\stop_1-g})^2  \nonumber \\
\f4 &=& C_{\stop_1-\stop_1-g-g}\nonumber \\
{\cal T}_{\rm I}\!\!\times\!\!{\cal T}_{\rm I} &=& (12/9)(4 (\mstop^2 - s) +
(5/2)(u-t)^2)/s^2   \nonumber \\
{\cal T}_{\rm II}\!\!\times\!\!{\cal T}_{\rm II} &=& (16/27)(2 \mstop^2 + 2
t)^2/(t - \mstop^2)^2    \nonumber \\
{\cal T}_{\rm III}\!\!\times\!\!{\cal T}_{\rm III} &=& (16/27)(2 \mstop^2 + 2
u)^2/(u - \mstop^2)^2    \nonumber \\
{\cal T}_{\rm IV}\!\!\times\!\!{\cal T}_{\rm IV} &=& (28/27)(4)    \nonumber \\
{\cal T}_{\rm I}\!\!\times\!\!{\cal T}_{\rm II} &=& 0     \nonumber \\
{\cal T}_{\rm I}\!\!\times\!\!{\cal T}_{\rm III} &=& 0     \nonumber \\
{\cal T}_{\rm I}\!\!\times\!\!{\cal T}_{\rm IV} &=& 0     \nonumber \\
{\cal T}_{\rm II}\!\!\times\!\!{\cal T}_{\rm III} &=& (-2/27)(4 \mstop^2 -
s)^2/((u-\mstop^2)(t-\mstop^2))     \nonumber \\  
{\cal T}_{\rm II}\!\!\times\!\!{\cal T}_{\rm IV} &=& (-14/27)(6 \mstop^2 - 2 u
- (5/2) s)/(t - \mstop^2)     \nonumber \\
{\cal T}_{\rm III}\!\!\times\!\!{\cal T}_{\rm IV} &=& (-14/27)(6 \mstop^2 - 2 t
- (5/2) s)/(u - \mstop^2)     \nonumber \\ 
\tsq &=&   \f1^2 {\cal T}_{\rm I}\!\!\times\!\!{\cal T}_{\rm I} 
+  \f2^2 {\cal T}_{\rm II}\!\!\times\!\!{\cal T}_{\rm II} 
+ \f3^2 {\cal T}_{\rm III}\!\!\times\!\!{\cal T}_{\rm III} 
+ \f4^2 {\cal T}_{\rm IV}\!\!\times\!\!{\cal T}_{\rm IV}
+ 2 \f1 \f2 {\cal T}_{\rm I}\!\!\times\!\!{\cal T}_{\rm II} 
+ \nl   2 \f1 \f3 {\cal T}_{\rm I}\!\!\times\!\!{\cal T}_{\rm III}
+ 2 \f1 \f4 {\cal T}_{\rm I}\!\!\times\!\!{\cal T}_{\rm IV} 
+ 2 \f2 \f3 {\cal T}_{\rm II}\!\!\times\!\!{\cal T}_{\rm III}
+   2 \f2 \f4 {\cal T}_{\rm II}\!\!\times\!\!{\cal T}_{\rm IV}
 + \nl 2 \f3 \f4 {\cal T}_{\rm III}\!\!\times\!\!{\cal T}_{\rm IV}   
\end{eqnarray}

\subsection*{$\stop_1\stop_1^*\longrightarrow \gamma g$}
 I. t-channel $\stop_1$ exchange \hfill\\
 II. u-channel $\stop_1$ exchange \hfill\\
 III.  point interaction\hfill\\
\begin{eqnarray}
\f1 &=& C_{\stop_1-\stop_1-g} \, C_{\stop_1-\stop_1-\gamma}    
\nonumber \\
\f2 &=& C_{\stop_1-\stop_1-g} \, C_{\stop_1-\stop_1-\gamma}    
\nonumber \\
\f3 &=& C_{\stop_1-\stop_1-\gamma-g}     \nonumber \\
{\cal T}_{\rm I}\!\!\times\!\!{\cal T}_{\rm I} &=& (4/9)(2 \mstop^2 + 2 t)^2/(t
- \mstop^2)^2   \nonumber \\
{\cal T}_{\rm II}\!\!\times\!\!{\cal T}_{\rm II} &=&  (4/9)(2 \mstop^2 + 2
u)^2/(u - \mstop^2)^2   \nonumber \\
{\cal T}_{\rm III}\!\!\times\!\!{\cal T}_{\rm III} &=&   (4/9)(4)  \nonumber \\
{\cal T}_{\rm I}\!\!\times\!\!{\cal T}_{\rm II} &=&  (4/9)(4 \mstop^2 -
s)^2/((t - \mstop^2)(u - \mstop^2))     \nonumber \\
{\cal T}_{\rm I}\!\!\times\!\!{\cal T}_{\rm III} &=&   (-4/9)(6 \mstop^2 - (5/2)
s - 2 u)/(t - \mstop^2)   \nonumber \\
{\cal T}_{\rm II}\!\!\times\!\!{\cal T}_{\rm III} &=&  ( -4/9)(6 \mstop^2 -
(5/2) s - 2 t)/(u - \mstop^2)    \nonumber \\   
\tsq &=&   \f1^2 {\cal T}_{\rm I}\!\!\times\!\!{\cal T}_{\rm I} 
+  \f2^2 {\cal T}_{\rm II}\!\!\times\!\!{\cal T}_{\rm II} 
+ \f3^2 {\cal T}_{\rm III}\!\!\times\!\!{\cal T}_{\rm III} 
+ 2 \f1 \f2 {\cal T}_{\rm I}\!\!\times\!\!{\cal T}_{\rm II} 
+ 2 \f1 \f3 {\cal T}_{\rm I}\!\!\times\!\!{\cal T}_{\rm III} 
+ \nl 2 \f2 \f3 {\cal T}_{\rm II}\!\!\times\!\!{\cal T}_{\rm III}   
\end{eqnarray}

\subsection*{$\stop_1\stop_1^*\longrightarrow Z g$}
 I. t-channel $\stop_1$ exchange \hfill \\
 II. u-channel $\stop_1$ exchange \hfill \\ 
 III. point interaction \hfill \\
\begin{eqnarray}
\f1 &=&  C_{\stop_1-\stop_1-Z} \;  C_{\stop_1-\stop_1-g} \nonumber \\
\f2 &=&  C_{\stop_1-\stop_1-Z} \;  C_{\stop_1-\stop_1-g}  \nonumber \\
\f3 &=&  C_{\stop_1-\stop_1-Z-g}  \nonumber \\
{\cal T}_{\rm I}\!\!\times\!\!{\cal T}_{\rm I} &=&  (4/9)(2 \mstop^2 - \mz^2 +
2 t - (\mstop^2-t)^2/\mz^2)(2 \mstop^2 + 2 t)/(t - \mstop^2)^2  \nonumber \\
{\cal T}_{\rm II}\!\!\times\!\!{\cal T}_{\rm II} &=&  (4/9)(2 \mstop^2 - \mz^2
+ 2 u - (u-\mstop^2)^2
     /\mz^2)(2 \mstop^2 + 2 u)/(u - \mstop^2)^2  \nonumber \\
{\cal T}_{\rm III}\!\!\times\!\!{\cal T}_{\rm III} &=&  12/9  \nonumber \\
{\cal T}_{\rm I}\!\!\times\!\!{\cal T}_{\rm II} &=&  (4/9)(4 \mstop^2 - s -
(u-\mstop^2)
         (\mstop^2-t)/\mz^2)(4 \mstop^2-\mz^2-s)
        \nl   /((t - \mstop^2)(u - \mstop^2))  \nonumber \\
{\cal T}_{\rm I}\!\!\times\!\!{\cal T}_{\rm III} &=&  -(4/9) (6 \mstop^2 + (3/2)
\mz^2 - 2 u- (5/2) s 
            \nl -(\mstop^2-t)((1/2)s-(3/2)\mz^2-\mstop^2+u)/\mz^2) 
	     /(t - \mstop^2)  \nonumber \\
{\cal T}_{\rm II}\!\!\times\!\!{\cal T}_{\rm III} &=&  -(4/9) (6 \mstop^2 +
(3/2) \mz^2 - 2 t- (5/2) s 
            \nl -(\mstop^2-u)((1/2)s-(3/2)\mz^2-\mstop^2+t)/\mz^2) 
	     /(u - \mstop^2)  \nonumber \\
\tsq &=&   \f1^2 {\cal T}_{\rm I}\!\!\times\!\!{\cal T}_{\rm I}
+  \f2^2 {\cal T}_{\rm II}\!\!\times\!\!{\cal T}_{\rm II}
+  \f3^2 {\cal T}_{\rm III}\!\!\times\!\!{\cal T}_{\rm III}
+  2 \f1 \f2 {\cal T}_{\rm I}\!\!\times\!\!{\cal T}_{\rm II}
+  2 \f1 \f3 {\cal T}_{\rm I}\!\!\times\!\!{\cal T}_{\rm III}
+  \nl 2 \f2 \f3 {\cal T}_{\rm II}\!\!\times\!\!{\cal T}_{\rm III}
\end{eqnarray}

\subsection*{$\stop_1\stop_1^*\longrightarrow t \bar t$}
 I. s-channel gluon annihilation \hfill\\
 II. t-channel gluino exchange \hfill\\
 III. s-channel $h$ annihilation \hfill\\
 IV. s-channel $H$ annihilation \hfill\\ 
 V. t-channel $\tilde{\chi}^0_{(1,2,3,4)}$ exchange \hfill\\ 
 VI. s-channel $Z$ annihilation  \hfill\\
 VII. s-channel $\gamma$ annihilation   \hfill\\
\begin{eqnarray}
\f1 &=& C_{\stop_1-\stop_1 - g} \, C_{t-t-g}     \nonumber \\
f_{2L} &=&  \left( C_{\stop_1-t-\tilde{g}}^{L} \right)^2 \nonumber \\
f_{2R} &=& \left( C_{\stop_1-t-\tilde{g}}^{R} \right)^2 \nonumber \\
f_3 &=& C_{\stop_1-\stop_1-h} \, C_{t-t-h} \nonumber \\
f_4 &=& C_{\stop_1-\stop_1-H} \, C_{t-t-H} \nonumber \\
J_i &=& (C_{\stop_1-t-\tilde{\chi}_i^0}^{L} +
C_{\stop_1-t-\tilde{\chi}_i^0}^{R} )/2 \nonumber \\ 
K_i &=& (C_{\stop_1-t-\tilde{\chi}_i^0}^{L} -
C_{\stop_1-t-\tilde{\chi}_i^0}^{R} )/2  \nonumber \\
f_{6c} &=& C_{\stop_1-\stop_1- Z} \, (C_{t-t - Z}^R + C_{t-t - Z}^L)/2
\nonumber \\
f_{6d} &=& C_{\stop_1-\stop_1- Z} \, (C_{t-t - Z}^R - C_{t-t - Z}^L)/2\nonumber
\\
f_{7c} &=& C_{\stop_1-\stop_1 - \gamma} \, C_{t-t- \gamma} \nonumber \\
f_{7d} &=& 0 \nonumber \\
{\cal T}_{\rm I}\!\!\times\!\!{\cal T}_{\rm I} &=&  (2/9) (2 s (s - 4 \mstop^2)
- 2 (u - t)^2)/s^2  \nonumber \\
{\cal T}_{\rm II}\!\!\times\!\!{\cal T}_{\rm II} &=&   (2/9) ((f_{2L}^2 +
f_{2R}^2) ((2 m_t^2 - s) t - (\mstop^2 - m_t^2 - t)^2)  - \, 4 \, \sqrt{f_{2L}
f_{2R}} \, (f_{2L}+f_{2R}) \nl \times \,  
m_t \, \mgluino \, (\mstop^2 - m_t^2 - t)  + \,
          2 f_{2L} f_{2R}\, (\mgluino^2 (s - 4 m_t^2) - 2 m_t^2 t))
	   / (t - \mgluino^2)^2  \nonumber \\
{\cal T}_{\rm III}\!\!\times\!\!{\cal T}_{\rm III} &=&  (2 (s - 4 m_t^2))/(s -
m_h^2)^2   \nonumber \\
{\cal T}_{\rm IV}\!\!\times\!\!{\cal T}_{\rm IV} &=&   (2 (s - 4 m_t^2))/(s -
m_H^2)^2  \nonumber \\
{\cal T}_{\rm V}\!\!\times\!\!{\cal T}_{\rm V} &=&  2(-\mstop^4 (4\, J_i\, J_j\,
K_i \,
K_j +  J_i^2 (J_j^2 + K_j^2) +   K_i^2 (J_j^2 + K_j^2)) + 
4  \mxi \, \mxj \, (J_i^2 - K_i^2) \nl \times \, (J_j^2 - K_j^2) \, m_t^2 - 
2 \mstop^2 (4 \, J_i \, J_j \, K_i \, K_j +  J_i^2 (J_j^2 + K_j^2) + 
    K_i^2 (J_j^2 + K_j^2)) \, m_t^2 \nl + 
4  \mxj \, (J_i^2 + K_i^2) (J_j^2 - K_j^2) \, m_t^3  + 
          4  \mxi (J_i^2 - K_i^2) (J_j^2 + K_j^2) \, m_t^3 \nl + 
          3 (4 \, J_i\,  J_j \, K_i \, K_j + J_i^2 (J_j^2 + K_j^2)   + 
                K_i^2 (J_j^2 + 
                      K_j^2)) m_t^4 - \mxi \, \mxj \,
(J_i^2 - K_i^2) \nl \times \, (J_j^2 - 
                K_j^2) s  + \mstop^2 (4 \, J_i \, J_j \, K_i \, 
K_j + J_i^2 (J_j^2 + K_j^2) + 
                K_i^2 (J_j^2 + 
                      K_j^2)) s \nl - \mxj (J_i^2 + 
                K_i^2) (J_j^2 - 
                K_j^2) m_t s  - \mxi (J_i^2 - 
                K_i^2) (J_j^2 + 
                K_j^2) \, m_t \, s \nl - (4 \, J_i \, J_j \, K_i \, K_j + 
                J_i^2 (J_j^2 + K_j^2) + 
                K_i^2 (J_j^2 + 
                      K_j^2)) \, m_t^2 \, s  + \mstop^2 (4 \, 
J_i \, J_j \, K_i \, K_j \nl + J_i^2 (J_j^2 + K_j^2) + 
                K_i^2 (J_j^2 + 
                      K_j^2)) \, t  + \mxj (J_i^2 + 
                K_i^2) (J_j^2 - 
                K_j^2) \, m_t \, t \nl + \mxi (J_i^2 - 
                K_i^2) (J_j^2 + K_j^2) \, m_t \, t + 
          2 (-4 \, J_i \, J_j \, K_i \, K_j + J_i^2 (J_j^2 + K_j^2) \nl + 
                K_i^2 (J_j^2 + 
                      K_j^2)) \, m_t^2 \, t  + (4 \, J_i \, J_j \, K_i \, K_j + 
                J_i^2 (J_j^2 + K_j^2) + 
                K_i^2 (J_j^2 + 
                      K_j^2)) \, m_t^2 \, t \nl + \mstop^2 (4 \, 
J_i \, J_j \, K_i \, K_j + J_i^2 (J_j^2 + K_j^2) + 
                K_i^2 (J_j^2 + 
                      K_j^2)) \, u - \mxj (J_i^2 + 
                K_i^2) \nl \times \, (J_j^2 - 
                K_j^2) \, m_t \, u  - \mxi (J_i^2 - 
                K_i^2) (J_j^2 + 
                K_j^2) m_t u - (4 \, J_i \, J_j \, K_i \, K_j \nl + 
                J_i^2 (J_j^2 + K_j^2)  + 
                K_i^2 (J_j^2 + 
                      K_j^2)) \, m_t^2 \, u - (4 \, J_i \, J_j \, K_i \, K_j + 
                J_i^2 (J_j^2 + K_j^2) \nl + 
                K_i^2 (J_j^2 + K_j^2)) \, t \, u)
	       /
       ((\mxi^2 - t) (-\mxj^2 + t))
          \nonumber \\
{\cal T}_{\rm VI}\!\!\times\!\!{\cal T}_{\rm VI} &=&  (2(16 f_{6d}^2 \mstop^2
m_t^2 - 4 f_{6c}^2 \mstop^2 s - 
     4 f_{6d}^2 \mstop^2 s - 4 f_{6d}^2 m_t^2 s + 
           f_{6c}^2 s^2 + f_{6d}^2 s^2 - f_{6c}^2 t^2 \nl - 
           f_{6d}^2 t^2 + 2 f_{6c}^2 t u + 2 f_{6d}^2 t u - 
           f_{6c}^2 u^2 - f_{6d}^2 u^2))/(\mz^2 - s)^2 \nonumber \\
{\cal T}_{\rm VII}\!\!\times\!\!{\cal T}_{\rm VII} &=& (2 (16 f_{7d}^2 \mstop^2
m_t^2 - 4 f_{7c}^2 \mstop^2 s - 
           4 f_{7d}^2 \mstop^2 s - 4 f_{7d}^2 m_t^2 s + 
           f_{7c}^2 s^2 + f_{7d}^2 s^2 - f_{7c}^2 t^2 \nl - 
           f_{7d}^2 t^2 + 2 f_{7c}^2 t u + 2 f_{7d}^2 t u - 
           f_{7c}^2 u^2 - f_{7d}^2 u^2))/s^2 \nonumber \\
{\cal T}_{\rm I}\!\!\times\!\!{\cal T}_{\rm II} &=&    (2/27) ((f_{2L}+f_{2R})
((t-u)
           (\mstop^2 - m_t^2 - t) - (1/2) s
           (4 \mstop^2 - s + t - u)) \nl + 4 \sqrt{f_{2L} f_{2R}} \, \mgluino \,
	    m_t \, (t-u))/(s (t - \mgluino^2))   \nonumber \\
{\cal T}_{\rm I}\!\!\times\!\!{\cal T}_{\rm III} &=&   0   \nonumber \\
{\cal T}_{\rm I}\!\!\times\!\!{\cal T}_{\rm IV} &=&   0   \nonumber \\
{\cal T}_{\rm I}\!\!\times\!\!{\cal T}_{\rm V} &=&   0   \nonumber \\
{\cal T}_{\rm I}\!\!\times\!\!{\cal T}_{\rm VI} &=&   0   \nonumber \\
{\cal T}_{\rm I}\!\!\times\!\!{\cal T}_{\rm VII} &=&   0   \nonumber \\
{\cal T}_{\rm II}\!\!\times\!\!{\cal T}_{\rm III} &=&  0    \nonumber \\
{\cal T}_{\rm II}\!\!\times\!\!{\cal T}_{\rm IV} &=&   0   \nonumber \\
{\cal T}_{\rm II}\!\!\times\!\!{\cal T}_{\rm V} &=&   0   \nonumber \\ 
{\cal T}_{\rm II}\!\!\times\!\!{\cal T}_{\rm VI} &=&   0   \nonumber \\
{\cal T}_{\rm II}\!\!\times\!\!{\cal T}_{\rm VII} &=&   0   \nonumber \\
{\cal T}_{\rm III}\!\!\times\!\!{\cal T}_{\rm IV} &=&  2 \, (s - 4 m_t^2)/((s -
m_h^2) (s - m_H^2))    \nonumber \\
{\cal T}_{\rm III}\!\!\times\!\!{\cal T}_{\rm V} &=&   2\, f_3 \, ( ( J_i^2 + 
K_i^2) (-4 m_t^3 + m_t s - m_t t + m_t u) + ( J_i^2 -  K_i^2) ( - 4 m_t^2 \mxi
+ \mxi s )) \nl  /
       ((m_h^2 - s) (\mxi^2 - t))   \nonumber \\
{\cal T}_{\rm III}\!\!\times\!\!{\cal T}_{\rm VI} &=&  (4 (f_3 \, f_{6c} m_t t
- f_3 \, f_{6c} m_t u))/
       ((m_h^2 - s)(-\mz^2 + s))  \nonumber \\     
{\cal T}_{\rm III}\!\!\times\!\!{\cal T}_{\rm VII} &=&  (4 (f_3 \, f_{7c} m_t t
- f_3 \, f_{7c} m_t u))/((m_h^2 - s) s)   \nonumber \\      
{\cal T}_{\rm IV}\!\!\times\!\!{\cal T}_{\rm V} &=&  2 \, f_4 \, ( ( J_i^2 + 
K_i^2) (-4 m_t^3 + m_t s - m_t t + m_t u ) + 
           ( J_i^2 -  K_i^2) ( - 4 m_t^2 \mxi +  \mxi s  )) \nl/
       ((m_H^2 - s) (\mxi^2 - t))   \nonumber \\
{\cal T}_{\rm IV}\!\!\times\!\!{\cal T}_{\rm VI} &=&   (4 (f_4 \, f_{6c} m_t t -
f_4 \, f_{6c} m_t u))/
       ((m_H^2 - s)(-\mz^2 + s))  \nonumber \\   
{\cal T}_{\rm IV}\!\!\times\!\!{\cal T}_{\rm VII} &=&  (4 (f_4 \, f_{7c} m_t t -
f_4 \, f_{7c} m_t u))/((m_H^2 - s)s)   \nonumber \\
{\cal T}_{\rm V}\!\!\times\!\!{\cal T}_{\rm VI} &=&  (-16 f_{6d} \, K_i \, 
J_i \, \mstop^2 m_t^2 - 
         16 f_{6d} \, J_i \, K_i \, \mstop^2 m_t^2 - 
       4 f_{6c} \, J_i^2 \, \mstop^2 s + 4 f_{6d} \, K_i \, J_i \, \mstop^2 s
       \nl + 
     4 f_{6d} \, J_i \, K_i \, \mstop^2 s - 4 f_{6c} \, K_i^2 \, \mstop^2 s + 
      4 f_{6d} \, K_i \, J_i \, m_t^2 s + 4 f_{6d} \, J_i \, K_i \, m_t^2 s + 
      f_{6c} \, J_i^2 \, s^2 \nl - f_{6d} \, K_i \, J_i \, s^2 - 
      f_{6d} \, J_i \, K_i \, s^2 + f_{6c} \, K_i^2 \, s^2 - 
      4 f_{6c} \, J_i^2 \, m_t^2 t - 4 f_{6c} \, K_i^2 \, m_t^2 t \nl - 
      4 f_{6c} \, J_i^2 \, m_t \mxi \, t + 4 f_{6c} \, K_i^2 \, m_t \mxi \, t - 
      f_{6c} \, J_i^2 \, t^2 + f_{6d} \, K_i \, J_i \, t^2 + 
      f_{6d} \, J_i \, K_i \, t^2 \nl - f_{6c} \, K_i^2 \, t^2  + 
      4 f_{6c} \, J_i^2 \, m_t^2 u + 4 f_{6c} \, K_i^2 \, m_t^2 u + 
      4 f_{6c} \, J_i^2 \, m_t \mxi \, u \nl - 4 f_{6c} \, K_i^2 \, m_t \mxi \,
      u + 
      2 f_{6c} \, J_i^2 \, t u  - 2 f_{6d} \, K_i \, J_i \, t u - 
      2 f_{6d} \, J_i \, K_i \, t u + 2 f_{6c} \, K_i^2 \, t u \nl - 
      f_{6c} \, J_i^2 \, u^2  + f_{6d} \, K_i \, J_i \, u^2  + 
      f_{6d} \, J_i \, K_i \, u^2 - f_{6c} \, K_i^2 \, u^2)/
     ((\mz^2 - s)(\mxi^2 - t))  \nonumber \\
{\cal T}_{\rm V}\!\!\times\!\!{\cal T}_{\rm VII} &=&  -((-4 f_{7c} \, J_i^2
\mstop^2 s 
           - 4 f_{7c} \, K_i^2 \, \mstop^2 s + 
           f_{7c} \, J_i^2 \, s^2 + f_{7c} \, K_i^2 \, s^2 - 
           4 f_{7c} \, J_i^2 \, m_t^2 t \nl - 4 f_{7c} \, K_i^2 \, m_t^2 t  - 
           4 f_{7c} \, J_i^2 \, m_t \mxi \, t + 
           4 f_{7c} \, K_i^2 \, m_t \mxi \, t - f_{7c} \, J_i^2 \, t^2 - 
           f_{7c} \, K_i^2 \, t^2 \nl + 4 f_{7c} \, J_i^2 \, m_t^2 u  + 
           4 f_{7c} \, K_i^2 \, m_t^2 u + 
           4 f_{7c} \, J_i^2 \, m_t \mxi \, u - 
           4 f_{7c} \, K_i^2 \, m_t \mxi \, u \nl + 2 f_{7c} \, J_i^2 \, t u + 
           2 f_{7c} \, K_i^2 \, t u  - f_{7c} \, J_i^2 \, u^2 - 
           f_{7c} \, K_i^2 \, u^2)/(s (\mxi^2 - t)))  \nonumber \\
 {\cal T}_{\rm VI}\!\!\times\!\!{\cal T}_{\rm VII} &=&  (2 (-16 f_{6d} \, f_{7d}
 \, \mstop^2 m_t^2 + 
           4 f_{6c} \, f_{7c} \, \mstop^2 \, s + 4 f_{6d} \, f_{7d} \, \mstop^2
	   \, s + 
           4 f_{6d} \, f_{7d} \, m_t^2 \, s \nl - f_{6c} \, f_{7c} \, s^2  - 
           f_{6d} \, f_{7d} \, s^2 + f_{6c} \, f_{7c} \, t^2 + f_{6d} \, f_{7d}
	   \, t^2 - 
         2 f_{6c} \, f_{7c} \, t u - 2 f_{6d} \, f_{7d} \, t u \nl + f_{6c} \,
	 f_{7c} \, 
	   u^2 + 
           f_{6d} \, f_{7d} \, u^2))/((\mz^2 - s)s)  \nonumber \\  
\tsq &=&   \f1^2 {\cal T}_{\rm I}\!\!\times\!\!{\cal T}_{\rm I} 
+   {\cal T}_{\rm II}\!\!\times\!\!{\cal T}_{\rm II} 
+ \f3^2 {\cal T}_{\rm III}\!\!\times\!\!{\cal T}_{\rm III}
+ \f4^2 {\cal T}_{\rm IV}\!\!\times\!\!{\cal T}_{\rm IV}
+  \sum_{i,j=1}^4 {\cal T}_{\rm V}\!\!\times\!\!{\cal T}_{\rm V} 
+   {\cal T}_{\rm VI}\!\!\times\!\!{\cal T}_{\rm VI}
+ \nl  {\cal T}_{\rm VII}\!\!\times\!\!{\cal T}_{\rm VII}
+ 2 \f1  {\cal T}_{\rm I}\!\!\times\!\!{\cal T}_{\rm II} 
+  2 \f1 \f3 {\cal T}_{\rm I}\!\!\times\!\!{\cal T}_{\rm III}
+ 2 \f1 \f4 {\cal T}_{\rm I}\!\!\times\!\!{\cal T}_{\rm IV} 
+ 2 \f1 \sum_{i=1}^4 {\cal T}_{\rm I}\!\!\times\!\!{\cal T}_{\rm V}
+ \nl 2 \f1  {\cal T}_{\rm I}\!\!\times\!\!{\cal T}_{\rm VI}
+  2 \f1  {\cal T}_{\rm I}\!\!\times\!\!{\cal T}_{\rm VII} 
+  2  \f3 {\cal T}_{\rm II}\!\!\times\!\!{\cal T}_{\rm III}
+  2  \f4 {\cal T}_{\rm II}\!\!\times\!\!{\cal T}_{\rm IV}   
+   2 \sum_{i=1}^4 {\cal T}_{\rm II}\!\!\times\!\!{\cal T}_{\rm V}
+  \nl 2 {\cal T}_{\rm II}\!\!\times\!\!{\cal T}_{\rm VI} 
+   2 {\cal T}_{\rm II}\!\!\times\!\!{\cal T}_{\rm VII} 
+  2 \f3 \f4 {\cal T}_{\rm III}\!\!\times\!\!{\cal T}_{\rm IV}   
+  2 \f3 \sum_{i=1}^4 {\cal T}_{\rm III}\!\!\times\!\!{\cal T}_{\rm V}
+  2 {\cal T}_{\rm III}\!\!\times\!\!{\cal T}_{\rm VI} 
+  \nl 2 {\cal T}_{\rm III}\!\!\times\!\!{\cal T}_{\rm VII}
+   2 \f4 \sum_{i=1}^4 {\cal T}_{\rm IV}\!\!\times\!\!{\cal T}_{\rm V}
+  2 {\cal T}_{\rm IV}\!\!\times\!\!{\cal T}_{\rm VI} 
+  2 {\cal T}_{\rm IV}\!\!\times\!\!{\cal T}_{\rm VII}
+  2 {\cal T}_{\rm V}\!\!\times\!\!{\cal T}_{\rm VI}
+  \nl 2 {\cal T}_{\rm V}\!\!\times\!\!{\cal T}_{\rm VII}
+   2 {\cal T}_{\rm VI}\!\!\times\!\!{\cal T}_{\rm VII}
\end{eqnarray}

\subsection*{$\stop_1\stop_1^*\longrightarrow b \bar b$}
 I. s-channel $h$ annihilation \hfill\\
 II. s-channel $H$ annihilation \hfill\\ 
 III. t-channel $\tilde{\chi}^+_{(1,2)}$ exchange \hfill\\
 IV. s-channel gluon annihilation \hfill\\ 
\begin{eqnarray}
\f1 &=& C_{\stop_1 -\stop_1-h} \; C_{b-b-h} \nonumber \\
\f2 &=& C_{\stop_1-\stop_1-H} \; C_{b-b-H} \nonumber \\
f_{3LL}^i &=& C_{\stop_1-b-\tilde{\chi}^+_i}^{L} \;
C_{\stop_1-b-\tilde{\chi}^+_i}^{L} \nonumber \\
f_{3LR}^i &=& C_{\stop_1-b-\tilde{\chi}^+_i}^{L} \;
C_{\stop_1-b-\tilde{\chi}^+_i}^{R} \nonumber \\
f_{3RL}^i &=& C_{\stop_1-b-\tilde{\chi}^+_i}^{R} \;
C_{\stop_1-b-\tilde{\chi}^+_i}^{L} \nonumber \\
f_{3RR}^i &=& C_{\stop_1-b-\tilde{\chi}^+_i}^{R} \;
C_{\stop_1-b-\tilde{\chi}^+_i}^{R} \nonumber \\
\f4 &=& C_{\stop_1-\stop_1-g} \; C_{b-b-g} \nonumber \\
{\cal T}_{\rm I}\!\!\times\!\!{\cal T}_{\rm I} &=&  2 (s - 4 m_b^2)/(s -
m_h^2)^2  \nonumber \\
{\cal T}_{\rm II}\!\!\times\!\!{\cal T}_{\rm II} &=&  2 (s - 4 m_b^2)/(s -
m_H^2)^2  \nonumber \\
{\cal T}_{\rm III}\!\!\times\!\!{\cal T}_{\rm III} &=&  ((f_{3RR}^i
f_{3RR}^j+f_{3LL}^i f_{3LL}^j)
             (\, t \, (2 m_b^2-s)
            -(\mstop^2-m_b^2-t)^2) 
	    + (f_{3RL}^i f_{3LR}^j \nl
       +f_{3LR}^i f_{3RL}^j) \, \mchari \, \mcharj \, (s - 2 m_b^2) 
        - 
            (f_{3RR}^i f_{3LR}^j+f_{3LL}^i f_{3RL}^j+f_{3LL}^i f_{3LR}^j
            + \nl f_{3RR}^i f_{3RL}^j) \, \mcharj \, m_b \, (\mstop^2-m_b^2-t) 
	    - (f_{3RL}^i f_{3RR}^j+f_{3LR}^i f_{3LL}^j+f_{3LR}^i f_{3RR}^j
            + \nl f_{3RL}^i f_{3LL}^j) \, \mchari \, m_b \, (\mstop^2-m_b^2-t) 
           - (f_{3LL}^i f_{3RR}^j+f_{3RR}^i f_{3LL}^j) \,  
            m_b^2 \, 2 \, t 
	    \nl - (f_{3LR}^i f_{3LR}^j+f_{3RL}^i f_{3RL}^j) 2 \, m_b^2 \, 
             \mchari \, \mcharj) 
            /((t-\mchari^2) (t-\mcharj^2)) \nonumber \\
{\cal T}_{\rm IV}\!\!\times\!\!{\cal T}_{\rm IV} &=&  (2/9) (2 s (s - 4
\mstop^2) - 2 (u - t)^2)
        /s^2  \nonumber \\
{\cal T}_{\rm I}\!\!\times\!\!{\cal T}_{\rm II} &=&  2 (s - 4 m_b^2)/((s -
m_H^2) (s-m_h^2))  \nonumber \\
{\cal T}_{\rm I}\!\!\times\!\!{\cal T}_{\rm III} &=&  ((f_{3LL}^i+f_{3RR}^i)
2 m_b (t+m_b^2-\mstop^2)
     		+(f_{3RL}^i + f_{3LR}^i)  \mchari 
     		(s - 4 m_b^2)) \nl /((s-m_h^2) (t-\mchari^2))  \nonumber \\
{\cal T}_{\rm I}\!\!\times\!\!{\cal T}_{\rm IV} &=&  0  \nonumber \\
{\cal T}_{\rm II}\!\!\times\!\!{\cal T}_{\rm III} &=& ((f_{3LL}^i+f_{3RR}^i) 2
m_b (t+m_b^2-\mstop^2)
     		+(f_{3RL}^i + f_{3LR}^i)  \mchari 
     		(s - 4 m_b^2)) \nl /((s-m_H^2) (t-\mchari^2))   \nonumber \\
{\cal T}_{\rm II}\!\!\times\!\!{\cal T}_{\rm IV} &=&  0  \nonumber \\
{\cal T}_{\rm III}\!\!\times\!\!{\cal T}_{\rm IV} &=&  0  \nonumber \\
\tsq &=&   \f1^2 {\cal T}_{\rm I}\!\!\times\!\!{\cal T}_{\rm I} 
+  \f2^2 {\cal T}_{\rm II}\!\!\times\!\!{\cal T}_{\rm II} 
+ \sum_{i,j=1}^2 {\cal T}_{\rm III}\!\!\times\!\!{\cal T}_{\rm III}
 +  \f4^2 {\cal T}_{\rm IV}\!\!\times\!\!{\cal T}_{\rm IV}
+  2 \f1 \f2 {\cal T}_{\rm I}\!\!\times\!\!{\cal T}_{\rm II} 
+ \nl 2 \f1 \sum_{i=1}^2 {\cal T}_{\rm I}\!\!\times\!\!{\cal T}_{\rm III}
 + 2 \f1 \f4 {\cal T}_{\rm I}\!\!\times\!\!{\cal T}_{\rm IV}
+  2 \f2 \sum_{i=1}^2 {\cal T}_{\rm II}\!\!\times\!\!{\cal T}_{\rm III}
+  2 \f2 \f4 {\cal T}_{\rm II}\!\!\times\!\!{\cal T}_{\rm IV}
+  \nl 2 \f4 \sum_{i=1}^2 {\cal T}_{\rm III}\!\!\times\!\!{\cal T}_{\rm IV}
\end{eqnarray}

\subsection*{$\stop_1\stop_1^*\longrightarrow q \bar q$, \quad $ q = u,d,s,c$}
 I. s-channel gluon annihilation \hfill\\
\begin{eqnarray}
\f1 &=& C_{\stop_1-\stop_1-g} \; C_{q-q-g} \nonumber \\
{\cal T}_{\rm I}\!\!\times\!\!{\cal T}_{\rm I} &=&  (2/9)(2 s(s - 4
m_{\stop_1}^2) - 2(u - t)^2)/s^2  \nonumber \\
\tsq &=&   \f1^2 {\cal T}_{\rm I}\!\!\times\!\!{\cal T}_{\rm I}
\end{eqnarray} 

\subsection*{$\stop_1\stop_1^*\longrightarrow g H \quad [g h]$}
 I. t-channel $\stop_1$ exchange \hfill\\
 II.  u-channel $\stop_1$ exchange \hfill\\
\begin{eqnarray}
\f1 &=& C_{\stop_1 - \stop_1 -g} \; C_{\stop_1 - \stop_1 - H} \quad [C_{\stop_1
- \stop_1 -g} \; C_{\stop_1 - \stop_1 - h}] \nonumber \\
\f2 &=& C_{\stop_1 - \stop_1 -g} \; C_{\stop_1 - \stop_1 - H} \quad [C_{\stop_1
- \stop_1 -g} \; C_{\stop_1 - \stop_1 - h}] \nonumber \\
{\cal T}_{\rm I}\!\!\times\!\!{\cal T}_{\rm I} &=&  (-4/9) (2 (m_{\stop_1}^2 +
t))/(t - m_{\stop_1}^2)^2  \nonumber \\
{\cal T}_{\rm II}\!\!\times\!\!{\cal T}_{\rm II} &=&  (-4/9) (2 (m_{\stop_1}^2 +
u))/(u - m_{\stop_1}^2)^2  \nonumber \\
{\cal T}_{\rm I}\!\!\times\!\!{\cal T}_{\rm II} &=& (-4/9)(6 m_{\stop_1}^2
- u - t -2 s) / ((t - m_{\stop_1}^2)(u - m_{\stop_1}^2)) 
\nonumber \\
\tsq &=&   \f1^2 {\cal T}_{\rm I}\!\!\times\!\!{\cal T}_{\rm I} 
+ \f2^2 {\cal T}_{\rm II}\!\! \times\!\!{\cal T}_{\rm II} 
+ 2 \f1 \f2 {\cal T}_{\rm I}\!\!\times\! \!{\cal T}_{\rm II}
\end{eqnarray} 

\subsection*{$\stop_1\stop_1^*\longrightarrow Z h \quad [Z H]$}
 I. t-channel $\stop_{(1,2)}$ exchange \hfill\\
 II. u-channel $\stop_{(1,2)}$ exchange \hfill\\ 
\begin{eqnarray}
\f1^i &=& C_{\stop_1-\stop_i-Z} \; C_{\stop_1-\stop_i-h} \nonumber \\
\f2^i &=& C_{\stop_1-\stop_i-Z} \; C_{\stop_1-\stop_i-h} \nonumber \\
{\cal T}_{\rm I}\!\!\times\!\!{\cal T}_{\rm I} &=&  -(2 \mstop^2 + 2 t -\mz^2
            - (\mstop^2-t)^2/\mz^2)
            /((t-m_{\stop_i}^2)(t-m_{\stop_j}^2))  \nonumber \\
{\cal T}_{\rm II}\!\!\times\!\!{\cal T}_{\rm II} &=& -(2 \mstop^2 + 2 u - \mz^2
            - (\mstop^2-u)^2/\mz^2)
             /((u-m_{\stop_i}^2)(u-m_{\stop_j}^2))   \nonumber \\
{\cal T}_{\rm I}\!\!\times\!\!{\cal T}_{\rm II} &=&  -(4 \mstop^2 - m_h^2 - s -
(\mstop^2-t)
          (u-\mstop^2)/\mz^2)/((t-m_{\stop_i}^2)(u-m_{\stop_j}^2))  \nonumber \\
\tsq &=&   \sum_{i=1}^2 \sum_{j=1}^2 \left[ \f1^i \f1^j {\cal T}_{\rm
I}\!\!\times\!\!{\cal T}_{\rm I}
+ \f2^i \f2^j {\cal T}_{\rm II}\!\!\times\!\!{\cal T}_{\rm II}
+ 2 \f1^{i} \f2^j {\cal T}_{\rm I}\!\!\times\! \!{\cal T}_{\rm II} \right]
\end{eqnarray}

\subsection*{$\stop_1\stop_1^*\longrightarrow Z A$}
 I. s-channel $h$ annihilation \hfill\\
 II. s-channel $H$ annihilation \hfill\\
 III. t-channel $\stop_2$ exchange \hfill\\
 IV. u-channel $\stop_2$ exchange \hfill\\
\begin{eqnarray}
\f1 &=& C_{\stop_1-\stop_1-h} \; C_{h-Z-A} \nonumber \\
\f2 &=& C_{\stop_1-\stop_1-H} \; C_{H-Z-A} \nonumber \\
\f3 &=& C_{\stop_1-\stop_2-Z} \; C_{\stop_1-\stop_2-A} \nonumber \\
\f4 &=& C_{\stop_1-\stop_2-Z} \; C_{\stop_1-\stop_2-A} \nonumber \\
{\cal T}_{\rm I}\!\!\times\!\!{\cal T}_{\rm I} &=&  ((s-m_A^2)^2/\mz^2-(2
s-\mz^2 + 2 m_A^2))/(s-m_h^2)^2  \nonumber \\
{\cal T}_{\rm II}\!\!\times\!\!{\cal T}_{\rm II} &=&  ((s-m_A^2)^2/\mz^2-(2
s-\mz^2 + 2 m_A^2))/(s-m_H^2)^2  \nonumber \\
{\cal T}_{\rm III}\!\!\times\!\!{\cal T}_{\rm III} &=&  ((\mstop^2-t)^2/\mz^2-(2
\mstop^2 + 2 t))
             /(t-m_{\stop_2}^2)^2  \nonumber \\
{\cal T}_{\rm IV}\!\!\times\!\!{\cal T}_{\rm IV} &=&  ((\mstop^2-u)^2/\mz^2- (2
\mstop^2 + 2 u))
             /(u-m_{\stop_2}^2)^2  \nonumber \\
{\cal T}_{\rm I}\!\!\times\!\!{\cal T}_{\rm II} &=&  ((s-m_A^2)^2/\mz^2-(2
s-\mz^2 + 2 m_A^2))
          /((s-m_H^2)(s-m_h^2))  \nonumber \\
{\cal T}_{\rm I}\!\!\times\!\!{\cal T}_{\rm III} &=&
((s-m_A^2)(\mstop^2-t)/\mz^2 - \mstop^2 + 2 m_A^2-u)
         /((s-m_h^2)(t-m_{\stop_2}^2))   \nonumber \\
{\cal T}_{\rm I}\!\!\times\!\!{\cal T}_{\rm IV} &=&  ((s-m_A^2)(\mstop^2-u)
/\mz^2 - \mstop^2 + 2 m_A^2-t)
         /((s-m_h^2)(u-m_{\stop_2}^2))  \nonumber \\
{\cal T}_{\rm II}\!\!\times\!\!{\cal T}_{\rm III} &=&  ((s-m_A^2)(\mstop^2-t)
/\mz^2 - \mstop^2 + 2 m_A^2-u)
         /((s-m_H^2)(t-m_{\stop_2}^2))  \nonumber \\
{\cal T}_{\rm II}\!\!\times\!\!{\cal T}_{\rm IV} &=&  ((s-m_A^2)(\mstop^2-u)/
\mz^2 - \mstop^2 + 2 m_A^2 - t)
         /((s-m_H^2)(u-m_{\stop_2}^2))  \nonumber \\
{\cal T}_{\rm III}\!\!\times\!\!{\cal T}_{\rm IV} &=&  ((u-\mstop^2)(\mstop^2-t)
/\mz^2 - (4 \mstop^2 
         - m_A^2 -s))/((t-m_{\stop_2}^2)(u-m_{\stop_2}^2))  \nonumber \\
\tsq &=&   \f1^2 {\cal T}_{\rm I}\!\!\times\!\!{\cal T}_{\rm I}
+  \f2^2 {\cal T}_{\rm II}\!\!\times\!\!{\cal T}_{\rm II}
+  \f3^2 {\cal T}_{\rm III}\!\!\times\!\!{\cal T}_{\rm III}
+  \f4^2 {\cal T}_{\rm IV}\!\!\times\!\!{\cal T}_{\rm IV}
+  2 \f1 \f2 {\cal T}_{\rm I}\!\!\times\!\!{\cal T}_{\rm II}
+  \nl 2 \f1 \f3 {\cal T}_{\rm I}\!\!\times\!\!{\cal T}_{\rm III}
+  2 \f1 \f4 {\cal T}_{\rm I}\!\!\times\!\!{\cal T}_{\rm IV}
+  2 \f2 \f3 {\cal T}_{\rm II}\!\!\times\!\!{\cal T}_{\rm III}
+  2 \f2 \f4 {\cal T}_{\rm II}\!\!\times\!\!{\cal T}_{\rm IV}
+ \nl  2 \f3 \f4 {\cal T}_{\rm III}\!\!\times\!\!{\cal T}_{\rm IV}
\end{eqnarray}

\subsection*{$\stop_1\stop_1^*\longrightarrow W^+ H^- \quad [H^+ W^-]$}
 I. s-channel $h$ annihilation \hfill\\
 II. s-channel $H$ annihilation \hfill\\
 III. t-channel $\sbot_{(1,2)}$ exchange \hfill\\  
\begin{eqnarray}
\f1 &=& C_{\stop_1-\stop_1-h} \; C_{h - W^+ - H^-} \nonumber \\
\f2 &=& C_{\stop_1-\stop_1-H} \; C_{H - W^+ - H^-} \nonumber \\
\f3^i &=& C_{\stop_1-\sbot_i-W^+} \; C_{\stop_1 - \sbot_i - H^+} \nonumber \\
{\cal T}_{\rm I}\!\!\times\!\!{\cal T}_{\rm I} &=&  ((s-m_{H^+}^2)^2/\mw^2 - (2
s- \mw^2 + 2 m_{H^+}^2))/(s-m_h^2)^2  \nonumber \\
{\cal T}_{\rm II}\!\!\times\!\!{\cal T}_{\rm II} &=&  ((s-m_{H^+}^2)^2 / \mw^2 -
(2 s- \mw^2 + 2 m_{H^+}^2))
          /(s-m_H^2)^2  \nonumber \\
{\cal T}_{\rm III}\!\!\times\!\!{\cal T}_{\rm III} &=&  ((\mstop^2-t)^2/ \mw^2 -
2 (\mstop^2 + t))
            /((t-m_{\sbot_i}^2)(t-m_{\sbot_j}^2))  \nonumber \\
{\cal T}_{\rm I}\!\!\times\!\!{\cal T}_{\rm II} &=&  ((s-m_{H^+}^2)^2/ \mw^2 -
(2 s - \mw^2 + 2 m_{H^+}^2))
          /((s-m_H^2)(s-m_h^2))  \nonumber \\
{\cal T}_{\rm I}\!\!\times\!\!{\cal T}_{\rm III} &=&  ((s-m_{H^+}^2)(\mstop^2-t)
/\mw^2-(\mstop^2 + 2 m_{H^+}^2-u))
         /((s-m_h^2)(t-m_{\sbot_i}^2))  \nonumber \\
{\cal T}_{\rm II}\!\!\times\!\!{\cal T}_{\rm III} &=& 
((s-m_{H^+}^2)(\mstop^2-t)/\mw^2-(\mstop^2 + 2 m_{H^+}^2-u))
         /((s-m_H^2)(t-m_{\sbot_i}^2))  \nonumber \\
\tsq &=&   \f1^2 {\cal T}_{\rm I}\!\!\times\!\!{\cal T}_{\rm I}
+ \f2^2 {\cal T}_{\rm II}\!\!\times\!\!{\cal T}_{\rm II}
+ \sum_{i=1}^2 \sum_{j=1}^2 \f3^i \f3^j {\cal T}_{\rm III}\!\!\times\!\!{\cal
T}_{\rm III} 
+ 2 \f1 \f2 {\cal T}_{\rm I}\!\!\times\!\!{\cal T}_{\rm II}
+ \nl 2\sum_{i=1}^2 \f1 \f3^i  {\cal T}_{\rm I}\!\!\times\!\!{\cal
T}_{\rm III}
+ 2\sum_{i=1}^2 \f2 \f3^i  {\cal T}_{\rm II}\!\!\times\!\!{\cal
T}_{\rm III}
\end{eqnarray}

\subsection*{$\stop_1\stop_1^*\longrightarrow H H \quad [h h] \quad [h H]$}
 I. t-channel $\stop_{(1,2)}$ exchange \hfill\\
 II.  u-channel $\stop_{(1,2)}$ exchange \hfill\\
 III. point interaction \hfill\\
 IV. s-channel $h$ annihilation \hfill\\
 V. s-channel $H$ annihilation \hfill\\
\begin{eqnarray}
\f1^{i} &=& C_{\stop_1 - \stop_i - H}^2 \quad [C_{\stop_1 - \stop_i - h}^2]
\quad [ C_{\stop_1 - \stop_i - h} \; C_{\stop_1 - \stop_i - H}] 
\nonumber \\
\f2^{i} &=& C_{\stop_1 - \stop_i - H}^2  \quad [C_{\stop_1 - \stop_i - h}^2]
\quad [ C_{\stop_1 - \stop_i - h} \; C_{\stop_1 - \stop_i - H}]
\nonumber \\
\f3 &=& C_{\stop_1 - \stop_1 - H -H}  \quad [C_{\stop_1 - \stop_1 - h -h}]
\quad [C_{\stop_1 - \stop_1 - h -H}] \nonumber \\
\f4 &=& C_{\stop_1 - \stop_1 - h} \;  C_{h-H-H} \quad [C_{\stop_1 - \stop_1 - h}
\;  C_{h-h-h}] \quad [C_{\stop_1 - \stop_1 - h}
\;  C_{h-h-H}] \nonumber \\
\f5 &=& C_{\stop_1 - \stop_1 - H} \;  C_{H-H-H}  \quad [C_{\stop_1 - \stop_1 -
H} \;  C_{h-h-H}]  \quad [C_{\stop_1 - \stop_1 -
H} \;  C_{h-H-H}] \nonumber \\
{\cal T}_{\rm I}\!\!\times\!\!{\cal T}_{\rm I} &=&  1/((t -
m_{\stop_{i}}^2)(t-m_{\stop_{j}}^2))  \nonumber \\
{\cal T}_{\rm II}\!\!\times\!\!{\cal T}_{\rm II} &=&  1/((u -
m_{\stop_{i}}^2)(u-m_{\stop_{j}}^2))  \nonumber \\
{\cal T}_{\rm III}\!\!\times\!\!{\cal T}_{\rm III} &=&  1 \nonumber \\
{\cal T}_{\rm IV}\!\!\times\!\!{\cal T}_{\rm IV} &=& 1/(s-m_h^2)^2  \nonumber \\
{\cal T}_{\rm V}\!\!\times\!\!{\cal T}_{\rm V} &=&  1/(s-m_H^2)^2 \nonumber \\
{\cal T}_{\rm I}\!\!\times\!\!{\cal T}_{\rm II} &=& 1/((t - m_{\stop_{i}}^2)(u -
m_{\stop_{j}}^2))  \nonumber \\
{\cal T}_{\rm I}\!\!\times\!\!{\cal T}_{\rm III} &=& -1/(t - m_{\stop_{i}}^2) 
\nonumber \\
{\cal T}_{\rm I}\!\!\times\!\!{\cal T}_{\rm IV} &=&  1/((t - m_{\stop_{i}}^2)(s
- m_h^2))  \nonumber \\
{\cal T}_{\rm I}\!\!\times\!\!{\cal T}_{\rm V} &=&  1/((t -
m_{\stop_{i}}^2)(s-m_H^2)) \nonumber \\
{\cal T}_{\rm II}\!\!\times\!\!{\cal T}_{\rm III} &=&  -1/(u - m_{\stop_{i}}^2)
\nonumber \\
{\cal T}_{\rm II}\!\!\times\!\!{\cal T}_{\rm IV} &=& 1/((u -
m_{\stop_{i}}^2)(s-m_h^2))   \nonumber \\
{\cal T}_{\rm II}\!\!\times\!\!{\cal T}_{\rm V} &=&  1/((u -
m_{\stop_{i}}^2)(s-m_H^2)) \nonumber \\
{\cal T}_{\rm III}\!\!\times\!\!{\cal T}_{\rm IV} &=&  -1/(s- m_h^2)  \nonumber
\\
{\cal T}_{\rm III}\!\!\times\!\!{\cal T}_{\rm V} &=& -1/(s- m_H^2)  \nonumber \\
{\cal T}_{\rm IV}\!\!\times\!\!{\cal T}_{\rm V} &=&  1/((s-m_H^2)(s-m_h^2)) 
\nonumber \\
\tsq &=&   \sum_{i,j=1}^2 \left( \f1^{i} \f1^{j} {\cal T}_{\rm
I}\!\!\times\!\!{\cal T}_{\rm I} + \f2^{i} \f2^{j} {\cal T}_{\rm II}\!\!
\times\!\!{\cal T}_{\rm II} + 2 \f1^{i} \f2^{j} {\cal T}_{\rm I}\!\!\times\!
\!{\cal T}_{\rm II} \right)
+   2 \sum_{i=1}^2 \left( \f1^{i} \f3 {\cal T}_{\rm I}\!\!\times\!\!{\cal
T}_{\rm III} 
+  \right. \nl \left. \f1^{i} \f4 {\cal T}_{\rm I}\!\!\times\!\!{\cal T}_{\rm
IV}
+  \f1^{i} \f5 {\cal T}_{\rm I}\!\!\times\!\!{\cal T}_{\rm V}
+  \f2^{i}  \f3 {\cal T}_{\rm II}\!\!\times\!\!{\cal T}_{\rm III}
+  \f2^{i}  \f4 {\cal T}_{\rm II}\!\!\times\!\!{\cal
T}_{\rm IV}
+  \f2^{i}  \f5 {\cal T}_{\rm II}\!\!\times\!\!{\cal T}_{\rm V} \right)
+ \nl \f3^2 {\cal T}_{\rm III}\!\!\times\!\!{\cal T}_{\rm III}
+ \f4^2 {\cal T}_{\rm IV}\!\!\times\!\!{\cal T}_{\rm IV}
+  \f5^2 {\cal T}_{\rm V}\!\!\times\!\!{\cal T}_{\rm V}  
+  2 \f3 \f4 {\cal T}_{\rm III}\!\!\times\!\!{\cal T}_{\rm IV}
+ \nl 2 \f3 \f5 {\cal T}_{\rm III}\!\!\times\!\!{\cal T}_{\rm V}
+ 2 \f4 \f5 {\cal T}_{\rm IV}\!\!\times\!\!{\cal T}_{\rm V}
\end{eqnarray}

\subsection*{$\stop_1\stop_1^*\longrightarrow A  A$}
 I. t-channel $\stop_{2}$ exchange \hfill\\
 II.  u-channel $\stop_{2}$ exchange \hfill\\
 III. point interaction \hfill\\
 IV. s-channel $h$ annihilation \hfill\\
 V. s-channel $H$ annihilation \hfill\\
\begin{eqnarray}
\f1 &=& C_{\stop_1 - \stop_2 -A}^2 \nonumber \\
\f2 &=& C_{\stop_1 - \stop_2 -A}^2 \nonumber \\
\f3 &=& C_{\stop_1 - \stop_1 -A -A} \nonumber \\
\f4 &=& C_{\stop_1 - \stop_1 -h} \; C_{h-A-A} \nonumber \\
\f5 &=& C_{\stop_1 - \stop_1 -H} \; C_{H-A-A} \nonumber \\
{\cal T}_{\rm I}\!\!\times\!\!{\cal T}_{\rm I} &=&  1/(t - m_{\stop_2}^2)^2 
\nonumber \\
{\cal T}_{\rm II}\!\!\times\!\!{\cal T}_{\rm II} &=&  1/(u - m_{\stop_2}^2)^2 
\nonumber \\
{\cal T}_{\rm III}\!\!\times\!\!{\cal T}_{\rm III} &=& 1  \nonumber \\
{\cal T}_{\rm IV}\!\!\times\!\!{\cal T}_{\rm IV} &=& 1/(s-m_h^2)^2  \nonumber \\
{\cal T}_{\rm V}\!\!\times\!\!{\cal T}_{\rm V} &=&  1/(s-m_H^2)^2 \nonumber \\
{\cal T}_{\rm I}\!\!\times\!\!{\cal T}_{\rm II} &=&  1/((t - m_{\stop_2}^2)(u -
m_{\stop_2}^2)) \nonumber \\
{\cal T}_{\rm I}\!\!\times\!\!{\cal T}_{\rm III} &=&  1/(t - m_{\stop_2}^2)
\nonumber \\
{\cal T}_{\rm I}\!\!\times\!\!{\cal T}_{\rm IV} &=&  -1/((t -
m_{\stop_2}^2)(s-m_h^2))  \nonumber \\
{\cal T}_{\rm I}\!\!\times\!\!{\cal T}_{\rm V} &=&  -1/((t -
m_{\stop_2}^2)(s-m_H^2)) \nonumber \\
{\cal T}_{\rm II}\!\!\times\!\!{\cal T}_{\rm III} &=&  1/(u - m_{\stop_2}^2)
\nonumber \\
{\cal T}_{\rm II}\!\!\times\!\!{\cal T}_{\rm IV} &=&  -1/((u -
m_{\stop_2}^2)(s-m_h^2))  \nonumber \\
{\cal T}_{\rm II}\!\!\times\!\!{\cal T}_{\rm V} &=&  -1/((t -
m_{\stop_2}^2)(s-m_H^2)) \nonumber \\
{\cal T}_{\rm III}\!\!\times\!\!{\cal T}_{\rm IV} &=&  -1/(s-m_h^2)  \nonumber
\\
{\cal T}_{\rm III}\!\!\times\!\!{\cal T}_{\rm V} &=& -1/(s-m_H^2)  \nonumber \\
{\cal T}_{\rm IV}\!\!\times\!\!{\cal T}_{\rm V} &=&  1/((s-m_h^2)(s-m_H^2)) 
\nonumber \\
\tsq &=&   \f1^2 {\cal T}_{\rm I}\!\!\times\!\!{\cal T}_{\rm I} 
+ \f2^2 {\cal T}_{\rm II}\!\! \times\!\!{\cal T}_{\rm II} 
+ \f3^2 {\cal T}_{\rm III}\!\!\times\!\!{\cal T}_{\rm III}
+ \f4^2 {\cal T}_{\rm IV}\!\!\times\!\!{\cal T}_{\rm IV}
+  \f5^2 {\cal T}_{\rm V}\!\!\times\!\!{\cal T}_{\rm V}
+ \nl 2 \f1 \f2 {\cal T}_{\rm I}\!\!\times\! \!{\cal T}_{\rm II}   
+  2  \f1 \f3 {\cal T}_{\rm I}\!\!\times\!\!{\cal T}_{\rm III} 
+  2 \f1 \f4 {\cal T}_{\rm I}\!\!\times\!\!{\cal T}_{\rm IV}
+  2 \f1 \f5 {\cal T}_{\rm I}\!\!\times\!\!{\cal T}_{\rm V}
+ \nl  2 \f2  \f3 {\cal T}_{\rm II}\!\!\times\!\!{\cal T}_{\rm III}
+  2 \f2  \f4 {\cal T}_{\rm II}\!\!\times\!\!{\cal T}_{\rm IV}
+  2 \f2  \f5 {\cal T}_{\rm II}\!\!\times\!\!{\cal T}_{\rm V} 
+  2 \f3 \f4 {\cal T}_{\rm III}\!\!\times\!\!{\cal T}_{\rm IV}
+ \nl 2 \f3 \f5 {\cal T}_{\rm III}\!\!\times\!\!{\cal T}_{\rm V}
+ 2 \f4 \f5 {\cal T}_{\rm IV}\!\!\times\!\!{\cal T}_{\rm V}
\end{eqnarray}

\subsection*{$\stop_1\stop_1^*\longrightarrow h A \quad [H A]$}
 I. t-channel $\stop_{2}$ exchange \hfill\\
 II.  u-channel $\stop_{2}$ exchange \hfill\\
\begin{eqnarray}
\f1 &=& C_{\stop_1 - \stop_2 -h} \; C_{\stop_1 - \stop_2 -A} \quad [C_{\stop_1 -
\stop_2 -H} \; C_{\stop_1 - \stop_2 -A}]   \nonumber \\
\f2 &=&  C_{\stop_1 - \stop_2 -h} \; C_{\stop_1 - \stop_2 -A}  \quad
[C_{\stop_1 - \stop_2 -H} \; C_{\stop_1 - \stop_2 -A}]  \nonumber \\
{\cal T}_{\rm I}\!\!\times\!\!{\cal T}_{\rm I} &=&  1/(t - m_{\stop_2}^2)^2 
\nonumber \\
{\cal T}_{\rm II}\!\!\times\!\!{\cal T}_{\rm II} &=&  1/(u - m_{\stop_2}^2)^2 
\nonumber \\
{\cal T}_{\rm I}\!\!\times\!\!{\cal T}_{\rm II} &=&  1/((t - m_{\stop_2}^2)(u -
m_{\stop_2}^2)) \nonumber \\
\tsq &=&   \f1^2 {\cal T}_{\rm I}\!\!\times\!\!{\cal T}_{\rm I} 
+ \f2^2 {\cal T}_{\rm II}\!\! \times\!\!{\cal T}_{\rm II} 
+ 2 \f1 \f2 {\cal T}_{\rm I}\!\!\times\! \!{\cal T}_{\rm II}   
\end{eqnarray}

\subsection*{$\stop_1\stop_1^*\longrightarrow H^+ H^-$}
 I. t-channel $\sbot_{(1,2)}$ exchange \hfill\\
 II.  point interaction \hfill\\
 III. s-channel $h$ \hfill\\
 IV. s-channel $H$ \hfill\\
\begin{eqnarray}
\f1^{i} &=& (C_{\stop_1 - \sbot_i - H^+})^2 \nonumber \\
\f2 &=&  C_{\stop_1 - \stop_1 - H^+ - H^-} \nonumber \\
\f3 &=& C_{\stop_1 - \stop_1 - h} \, C_{h - H^+ - H^-} \nonumber \\
\f4 &=&  C_{\stop_1 - \stop_1 - H} \, C_{H - H^+ - H^-} \nonumber \\
{\cal T}_{\rm I}\!\!\times\!\!{\cal T}_{\rm I} &=&  1/((t -
m_{\sbot_i}^2)(t-m_{\sbot_j}^2))  \nonumber \\
{\cal T}_{\rm II}\!\!\times\!\!{\cal T}_{\rm II} &=&  1  \nonumber \\
{\cal T}_{\rm III}\!\!\times\!\!{\cal T}_{\rm III} &=& 1/(s-m_h^2)^2 \nonumber \\
{\cal T}_{\rm IV}\!\!\times\!\!{\cal T}_{\rm IV} &=& 1/(s-m_H^2)^2 \nonumber \\
{\cal T}_{\rm I}\!\!\times\!\!{\cal T}_{\rm II} &=&  -1/(t - m_{\sbot_i}^2)
\nonumber \\
{\cal T}_{\rm I}\!\!\times\!\!{\cal T}_{\rm III} &=& 1/((s-m_h^2)(t -
m_{\sbot_i}^2)) \nonumber \\
{\cal T}_{\rm I}\!\!\times\!\!{\cal T}_{\rm IV} &=& 1/((s-m_H^2)(t -
m_{\sbot_i}^2)) \nonumber \\
{\cal T}_{\rm II}\!\!\times\!\!{\cal T}_{\rm III} &=& -1/(s-m_h^2) \nonumber \\
{\cal T}_{\rm II}\!\!\times\!\!{\cal T}_{\rm IV} &=& -1/(s-m_H^2) \nonumber \\
{\cal T}_{\rm III}\!\!\times\!\!{\cal T}_{\rm IV} &=& 1/((s-m_h^2)(s-m_H^2))
\nonumber \\
\tsq &=&   \sum_{i,j=1}^2 \f1^{i} \f1^{j} {\cal T}_{\rm
I}\!\!\times\!\!{\cal T}_{\rm I} 
+ \f2^2 {\cal T}_{\rm II}\!\! \times\!\!{\cal T}_{\rm II} 
+ \f3^2 {\cal T}_{\rm III}\!\! \times\!\!{\cal T}_{\rm III}
+  \f4^2 {\cal T}_{\rm IV}\!\! \times\!\!{\cal T}_{\rm IV}
+ 2 \sum_{i=1}^2 \f1^{i} \f2 {\cal T}_{\rm I}\!\!\times\! \!{\cal T}_{\rm II}
+ \nl 2 \sum_{i=1}^2 \f1^{i} \f3 {\cal T}_{\rm I}\!\!\times\! \!{\cal T}_{\rm
III}
+ 2 \sum_{i=1}^2 \f1^{i} \f4 {\cal T}_{\rm I}\!\!\times\! \!{\cal T}_{\rm IV}
+ 2 \f2 \f3 {\cal T}_{\rm II}\!\! \times\!\!{\cal T}_{\rm III}
+ 2 \f2 \f4 {\cal T}_{\rm II}\!\! \times\!\!{\cal T}_{\rm IV}
+ \nl 2 \f3 \f4 {\cal T}_{\rm III}\!\! \times\!\!{\cal T}_{\rm IV}
\end{eqnarray}

\subsection*{$\stop_1\stop_1 \longrightarrow tt$}
 I. t-channel gluino exchange \hfill\\
 II.  u-channel gluino exchange \hfill\\
 III. t-channel $\chi^0_{(1,2,3,4)}$ exchange \hfill\\
 IV. u-channel $\chi^0_{(1,2,3,4)}$ exchange \hfill\\ 
\begin{eqnarray}
f_{1L} &=& f_{2L} \equiv f_L = (C_{\stop_1-t-\tilde{g}}^L)^2 \nonumber \\
f_{1R} &=& f_{2R} \equiv f_R = (C_{\stop_1-t-\tilde{g}}^R)^2 \nonumber \\
J_i &=& (C_{\stop_1-t-\tilde{\chi}_i^0}^{L} +
C_{\stop_1-t-\tilde{\chi}_i^0}^{R} )/2 \nonumber \\ 
K_i &=& (C_{\stop_1-t-\tilde{\chi}_i^0}^{L} -
C_{\stop_1-t-\tilde{\chi}_i^0}^{R} )/2  \nonumber \\
{\cal T}_{\rm I}\!\!\times\!\!{\cal T}_{\rm I} &=& (2/9) ((f_L^2+f_R^2)
\mgluino^2 (s - 2 m_t^2) 
         - f_L f_R (4 m_t^2 \mgluino^2+ 2 s t+ 2
        (\mstop^2-m_t^2-t)^2)
       \nl  + \sqrt{f_L f_R} \, (f_L+f_R) \, 4 \,  m_t \, \mgluino
       (t+m_t^2-\mstop^2))
          /(t-\mgluino^2)^2  \nonumber \\
{\cal T}_{\rm II}\!\!\times\!\!{\cal T}_{\rm II} &=&  (2/9) ((f_L^2+f_R^2)
\mgluino^2 (s - 2 m_t^2) 
      - f_L f_R (4 m_t^2 \mgluino^2+ 2 s u+ 2
          (\mstop^2-m_t^2-u)^2)
   \nl    + \sqrt{f_L f_R} \, (f_L+f_R) \, 4 \, m_t \, \mgluino
   (u+m_t^2-\mstop^2))
         /(u-\mgluino^2)^2  \nonumber \\
{\cal T}_{\rm III}\!\!\times\!\!{\cal T}_{\rm III} &=&    (-2 \, (4 \, J_i \,
J_j \, K_i \, K_j \,
\mxi \, \mxj \, s  + 
      K_i^2 (J_j^2 (-\mstop^4 - 2 \mstop^2 m_t^2 + 3 m_t^4 - 
                 4 m_t^3 \mxi \nl + 4 m_t^3 \mxj - 4 m_t^2 \mxi \mxj  + 
                 \mstop^2 s - m_t^2 s + m_t \mxi s - m_t \mxj s + 
                 \mxi \mxj s \nl + \mstop^2 t  + 3 m_t^2 t - m_t \mxi t  + 
                 m_t \mxj t + \mstop^2 u - m_t^2 u + m_t \mxi u  - 
                 m_t \mxj u \nl - t u)  + 
              K_j^2 (\mstop^4 + 2 \mstop^2 m_t^2 - 3 m_t^4 + 
                4 m_t^3 \mxi + 4 m_t^3 \mxj - 4 m_t^2 \mxi \mxj \nl - 
                 \mstop^2 s + m_t^2 s - m_t \mxi s - m_t \mxj s + 
                 \mxi \mxj s - \mstop^2 t  - 3 m_t^2 t  + m_t \mxi t \nl + 
                 m_t \mxj t - \mstop^2 u + m_t^2 u - m_t \mxi u - 
                 m_t \mxj u + t u))  + 
           J_i^2 (K_j^2 (-\mstop^4 \nl - 2 \mstop^2 m_t^2 + 3 m_t^4 + 
                 4 m_t^3 \mxi - 4 m_t^3 \mxj - 4 m_t^2 \mxi \mxj  + 
                 \mstop^2 s - m_t^2 s \nl - m_t \mxi s + m_t \mxj s + 
                 \mxi \mxj s + \mstop^2 t + 3 m_t^2 t + m_t \mxi t  - 
                 m_t \mxj t \nl + \mstop^2 u - m_t^2 u - m_t \mxi u + 
                 m_t \mxj u - t u)  + 
              J_j^2 (\mstop^4 + 2 \mstop^2 m_t^2 - 3 m_t^4 \nl - 
                 4 m_t^3 \mxi - 4 m_t^3 \mxj - 4 m_t^2 \mxi \mxj - 
                 \mstop^2 s  + m_t^2 s + m_t \mxi s + m_t \mxj s \nl + 
                 \mxi \mxj s - \mstop^2 t - 3 m_t^2 t - m_t \mxi t  - 
                 m_t \mxj t - \mstop^2 u + m_t^2 u \nl + m_t \mxi u + 
                 m_t \mxj u + t u))))
		  /((\mxi^2 - t) (-\mxj^2 + t))
  \nonumber \\
{\cal T}_{\rm IV}\!\!\times\!\!{\cal T}_{\rm IV} &=& (-2 (4 \, J_i \, J_j \, K_i
\, K_j \, \mxi \,  \mxj s  + 
           J_i^2 (K_j^2 (-\mstop^4 - 2 \mstop^2 m_t^2 + 3 m_t^4  + 
                 4 m_t^3 \mxi \nl - 4 m_t^3 \mxj  - 4 m_t^2 \mxi \mxj + 
                 \mstop^2 s - m_t^2 s - m_t \mxi s + m_t \mxj s + 
                 \mxi \mxj s \nl + \mstop^2 t - m_t^2 t  - m_t \mxi t + 
                 m_t \mxj t + \mstop^2 u + 3 m_t^2 u + m_t \mxi u - 
                 m_t \mxj u - t u) \nl + 
              J_j^2 (\mstop^4 + 2 \mstop^2 m_t^2 - 3 m_t^4 - 
                 4 m_t^3 \mxi - 4 m_t^3 \mxj - 4 m_t^2 \mxi \mxj - 
                 \mstop^2 s \nl + m_t^2 s + m_t \mxi s + m_t \mxj s + 
                 \mxi \mxj s - \mstop^2 t + m_t^2 t + m_t \mxi t + 
                 m_t \mxj t \nl - \mstop^2 u  - 3 m_t^2 u - m_t \mxi u - 
                 m_t \mxj u + t u))  + 
           K_i^2 (J_j^2 (-\mstop^4  - 2 \mstop^2 m_t^2 \nl + 3 m_t^4 - 
                 4 m_t^3 \mxi + 4 m_t^3 \mxj - 4 m_t^2 \mxi \mxj  + 
                 \mstop^2 s - m_t^2 s + m_t \mxi s \nl - m_t \mxj s + 
                 \mxi \mxj s + \mstop^2 t - m_t^2 t + m_t \mxi t  - 
                 m_t \mxj t + \mstop^2 u + 3 m_t^2 u \nl - m_t \mxi u + 
                 m_t \mxj u - t u)  + 
              K_j^2 (\mstop^4 + 2 \mstop^2 m_t^2 - 3 m_t^4 + 
                 4 m_t^3 \mxi \nl + 4 m_t^3 \mxj - 4 m_t^2 \mxi \mxj - 
                 \mstop^2 s  + m_t^2 s - m_t \mxi s - m_t \mxj s + 
                 \mxi \mxj s \nl - \mstop^2 t + m_t^2 t - m_t \mxi t  - 
                 m_t \mxj t - \mstop^2 u - 3 m_t^2 u + m_t \mxi u + 
                 m_t \mxj u \nl + t u))))/((\mxi^2 - u) (-\mxj^2 + u))
  \nonumber \\
{\cal T}_{\rm I}\!\!\times\!\!{\cal T}_{\rm II} &=&  (-2/27) (f_L f_R (2
(\mstop^2-m_t^2-u) (\mstop^2-
          m_t^2 - t) - 4 (s-2 m_t^2) (\mstop^2 - m_t^2))
        \nl  + \sqrt{f_L f_R} \, (f_L+f_R) m_t \, \mgluino (t-u)
          + (f_L^2+f_R^2) \mgluino^2 (s-2 m_t^2)
          - 4 m_t^2 t f_L f_R
        \nl  - m_t \mgluino \sqrt{f_L f_R} (f_L+f_R) (t+u)  
          - 4 m_t^2 \mgluino^2 f_L f_R)/((t-\mgluino^2) (u-\mgluino^2))
	  \nonumber \\
{\cal T}_{\rm I}\!\!\times\!\!{\cal T}_{\rm III} &=&  0 \nonumber \\
{\cal T}_{\rm I}\!\!\times\!\!{\cal T}_{\rm IV} &=&  0  \nonumber \\
{\cal T}_{\rm II}\!\!\times\!\!{\cal T}_{\rm III} &=&  0 \nonumber \\
{\cal T}_{\rm II}\!\!\times\!\!{\cal T}_{\rm IV} &=&  0  \nonumber \\
{\cal T}_{\rm III}\!\!\times\!\!{\cal T}_{\rm IV} &=&  (8 \, J_i \, J_j \, K_i
\, K_j \, \mxi \,
\mxj s + 
         J_i^2 (-(K_j^2 (2 \mstop^4 + 4 \mstop^2 m_t^2 - 6 m_t^4 - 
                 8 m_t^3 \mxi \nl + 8 m_t^3 \mxj  + 8 m_t^2 \mxi \mxj + 
                 2 \mstop^2 s + 2 m_t^2 s + 2 m_t \mxi s - 
                 2 m_t \mxj s \nl - 2 \mxi \mxj s  - s^2 - 2 \mstop^2 t - 
                 2 m_t^2 t + 2 m_t \mxi t + 2 m_t \mxj t + t^2 - 
                 2 \mstop^2 u \nl - 2 m_t^2 u  - 2 m_t \mxi u - 
                 2 m_t \mxj u + u^2))  + 
            J_j^2 (2 \mstop^4 + 4 \mstop^2 m_t^2  - 6 m_t^4 \nl - 
               8 m_t^3 \mxi - 8 m_t^3 \mxj - 8 m_t^2 \mxi \mxj  + 
               2 \mstop^2 s  + 2 m_t^2 s + 2 m_t \mxi s + 
               2 m_t \mxj s \nl + 2 \mxi \mxj s - s^2 - 2 \mstop^2 t  - 
               2 m_t^2 t + 2 m_t \mxi t - 2 m_t \mxj t + t^2 - 
               2 \mstop^2 u \nl - 2 m_t^2 u - 2 m_t \mxi u  + 
               2 m_t \mxj u + u^2))  + 
         K_i^2 (K_j^2 (2 \mstop^4 + 4 \mstop^2 m_t^2  - 6 m_t^4 \nl + 
               8 m_t^3 \mxi + 8 m_t^3 \mxj  - 8 m_t^2 \mxi \mxj + 
               2 \mstop^2 s + 2 m_t^2 s - 2 m_t \mxi s - 
               2 m_t \mxj s \nl + 2 \mxi \mxj s  - s^2 - 2 \mstop^2 t - 
               2 m_t^2 t - 2 m_t \mxi t + 2 m_t \mxj t + t^2 - 
               2 \mstop^2 u \nl - 2 m_t^2 u  + 2 m_t \mxi u - 
               2 m_t \mxj u + u^2)  - 
            J_j^2 (2 \mstop^4 + 4 \mstop^2 m_t^2 - 6 m_t^4 \nl + 
               8 m_t^3 \mxi - 8 m_t^3 \mxj + 8 m_t^2 \mxi \mxj  + 
               2 \mstop^2 s + 2 m_t^2 s - 2 m_t \mxi s + 
               2 m_t \mxj s \nl - 2 \mxi \mxj s - s^2 - 2 \mstop^2 t  - 
               2 m_t^2 t - 2 m_t \mxi t - 2 m_t \mxj t + t^2 - 
               2 \mstop^2 u \nl - 2 m_t^2 u + 2 m_t \mxi u  + 
               2 m_t \mxj u + u^2)))/((\mxi^2 - t) (\mxj^2 - u))
  \nonumber \\
\tsq &=&   {\cal T}_{\rm I}\!\!\times\!\!{\cal T}_{\rm I} 
+  {\cal T}_{\rm II}\!\! \times\!\!{\cal T}_{\rm II}
+ \sum_{i,j=1}^4 \left( {\cal T}_{\rm III}\!\! \times\!\!{\cal T}_{\rm III}
 +  {\cal T}_{\rm IV}\!\! \times\!\!{\cal T}_{\rm IV} \right) 
+  2 {\cal T}_{\rm I}\!\!\times\! \!{\cal T}_{\rm II}
+ \nl 2 \sum_{i=1}^4 \left( {\cal T}_{\rm I}\!\!\times\! \!{\cal T}_{\rm III}
+   {\cal T}_{\rm I}\!\!\times\! \!{\cal T}_{\rm IV}
+  {\cal T}_{\rm II}\!\!\times\! \!{\cal T}_{\rm III}
+  {\cal T}_{\rm II}\!\!\times\! \!{\cal T}_{\rm IV} \right)
+  2 \sum_{i,j=1}^4 {\cal T}_{\rm III}\!\!\times\! \!{\cal T}_{\rm IV}
\end{eqnarray}

\subsection*{$\ch \stop_1 \longrightarrow t g$}
 I. s-channel $t$ annihilation \hfill\\
 II.  t-channel $\stop_1$ exchange \hfill\\
\begin{eqnarray}
f_{1L} &=& C_{\stop_1 - t -\chi^0_1}^{L} \; C_{t-t-g} \nonumber \\
f_{1R} &=& C_{\stop_1 - t -\chi^0_1}^{R} \; C_{t-t-g}  \nonumber \\
f_{2L} &=&  C_{\stop_1 - t -\chi^0_1}^{L} \; C_{\stop_1-\stop_1-g} \nonumber
\\
f_{2R} &=&  C_{\stop_1 - t -\chi^0_1}^{R} \; C_{\stop_1-\stop_1-g} 
\nonumber \\
{\cal T}_{\rm I}\!\!\times\!\!{\cal T}_{\rm I} &=&   (-4/6)((f_{1L}^2+f_{1R}^2)
2 ((3 m_t^2 - s) (s + \mchi^2 - m_{\stop_1}^2) + (s - m_t^2) (\mchi^2 + m_t^2 -
t)) \nl - 8 f_{1L} f_{1R} \, m_t \, \mchi (s + m_t^2))/(s-m_t^2)^2 \nonumber \\
{\cal T}_{\rm II}\!\!\times\!\!{\cal T}_{\rm II} &=&  (-4/6)(2(t +
m_{\stop_1}^2)((f_{2L}^2 + f_{2R}^2)(\mchi^2 + m_t^2 - t)
- 4 \mchi m_t f_{2L} f_{2R}))/(t - m_{\stop_1}^2)^2  \nonumber \\
{\cal T}_{\rm I}\!\!\times\!\!{\cal T}_{\rm II} &=& 
(-4/6)((f_{1L}f_{2L}+f_{1R}f_{2R})/2 ((\mchi^2 + m_t^2 - t)
(s - 2 \mchi^2 + 2 m_{\stop_1}^2 + m_t^2) \nl - (s + m_t^2)
(2 s - 3 \mchi^2 - 2 m_{\stop_1}^2 + u) 
+ (2 m_{\stop_1}^2 + 3 m_t^2 - 2 u - s)
(s + \mchi^2 - m_{\stop_1}^2) 
\nl + 2 m_t^2 (2 s - 3 \mchi^2 - 2 m_{\stop_1}^2 + u)) 
-(f_{1L}f_{2R} + f_{1R} f_{2L}) \, m_t \, \mchi
 (4 m_{\stop_1}^2 + 4 m_t^2 \nl - 2 \mchi^2- 2u))
/((s-m_t^2)(t-m_{\stop_1}^2))    \nonumber \\
\tsq &=&   {\cal T}_{\rm I}\!\!\times\!\!{\cal T}_{\rm I} 
+  {\cal T}_{\rm II}\!\! \times\!\!{\cal T}_{\rm II} 
+ 2 {\cal T}_{\rm I}\!\!\times\! \!{\cal T}_{\rm II}
\end{eqnarray}

\subsection*{$\ch \stop_1 \longrightarrow t Z$}
 I.  u-channel $\tilde{\chi}^0_{(1,2,3,4)}$ exchange \hfill\\
\begin{eqnarray}
f_{LL}^i &=& C_{\stop_1-t-\tilde{\chi}^0_i}^{L} \; C_{\tilde{\chi}_1^0 -
\tilde{\chi}_i^0 - Z}^{L}  \nonumber \\
f_{LR}^i &=&  C_{\stop_1-t-\tilde{\chi}^0_i}^{L} \; C_{\tilde{\chi}_1^0 -
\tilde{\chi}_i^0 - Z}^{R} \nonumber \\
f_{RL}^i &=& C_{\stop_1-t-\tilde{\chi}^0_i}^{R} \; C_{\tilde{\chi}_1^0 -
\tilde{\chi}_i^0 - Z}^{L}   \nonumber \\
f_{RR}^i &=& C_{\stop_1-t-\tilde{\chi}^0_i}^{R} \; C_{\tilde{\chi}_1^0 -
\tilde{\chi}_i^0 - Z}^{R}   \nonumber \\
{\cal T}_{\rm I}\!\!\times\!\!{\cal T}_{\rm I} &=&  (1/2)((f_{LL}^i \, f_{LL}^j
+ f_{RR}^i \, f_{RR}^j)(((u - \mstop^2
           + m_t^2) (u + \mchi^2 - \mz^2) - (\mchi^2 + m_t^2-t)u) 
           \nl + ((\mchi^2 + \mz^2-u)/\mz^2)((\mchi^2-\mz^2-u)
             (u- \mstop^2+ m_t^2)
          -(s- m_t^2- \mz^2)u ) )
     \nl +(f_{LR}^i \, f_{RR}^j + f_{RL}^i \, f_{LL}^j) m_t \mxi ((u + \mchi^2
           - \mz^2) + (1/\mz^2)((\mchi^2+ \mz^2-u) \nl \times \, 
           (\mchi^2- \mz^2-u)))
          +(f_{LL}^i \, f_{RL}^j + f_{RR}^i \, f_{LR}^j ) m_t \mxj
     ((u + \mchi^2
           - \mz^2) \nl +(1/\mz^2)((\mchi^2+ \mz^2-u)
     (\mchi^2- \mz^2-u)))
       \nl   +(f_{LR}^i \, f_{LR}^j +f_{RL}^i \, f_{RL}^j) \mxi \mxj
           ((t+ \mchi^2- m_t^2) \nl + (1/\mz^2) ((\mchi^2+ \mz^2-u)
     (s- m_t^2- \mz^2) ))
      \nl  + (f_{LL}^i \, f_{RR}^j + f_{RR}^i \, f_{LL}^j) 6 m_t \mchi u
          + (f_{LL}^i \, f_{LR}^j + f_{RR}^i \, f_{RL}^j) 3 \mxj (u+ m_t^2-
	  \mstop^2)
          \mchi \nl + (f_{LR}^i \, f_{LL}^j + f_{RL}^i \, f_{RR}^j) 3 \mxi
           (u + m_t^2 - \mstop^2) \mchi \nl + (f_{LR}^i \, f_{RL}^j + f_{RL}^i
	   \, f_{LR}^j) 6
           \mxi \mxj m_t \mchi)/((t-\mxi^2) (t-\mxj^2))  \nonumber \\
\tsq &=&   \sum_{i,j=1}^4 {\cal T}_{\rm I}\!\! \times\!\!{\cal T}_{\rm I}
\end{eqnarray}

\subsection*{$\ch \stop_1 \longrightarrow b W^+$}
 I.  u-channel $\tilde{\chi}^+_{(1,2)}$ exchange \hfill\\
\begin{eqnarray}
f_{LL}^i &=& C_{\stop_1-b-\tilde{\chi}^+_i}^{L} \; C_{\tilde{\chi}_1^0 -
\tilde{\chi}_i^- - W^+}^{L}  \nonumber \\
f_{LR}^i &=&  C_{\stop_1-b-\tilde{\chi}^+_i}^{L} \; C_{\tilde{\chi}_1^0 -
\tilde{\chi}_i^- - W^+}^{R} \nonumber \\
f_{RL}^i &=& C_{\stop_1-b-\tilde{\chi}^+_i}^{R} \; C_{\tilde{\chi}_1^0 -
\tilde{\chi}_i^- - W^+}^{L}   \nonumber \\
f_{RR}^i &=& C_{\stop_1-t-\tilde{\chi}^+_i}^{R} \; C_{\tilde{\chi}_1^0 -
\tilde{\chi}_i^- - W^+}^{R}   \nonumber \\
{\cal T}_{\rm I}\!\!\times\!\!{\cal T}_{\rm I} &=&  (1/2) ((f_{LL}^i \, f_{LL}^j
+ f_{RR}^i \, f_{RR}^j)      (((u-\mstop^2
         + m_b^2) (u + \mchi^2- \mw^2) - (\mchi^2 + m_b^2 - t)u) 
         \nl + ((\mchi^2 + \mw^2- u)/\mw^2)((\mchi^2 - \mw^2-u)
             (u-\mstop^2 + m_b^2)
          -(s-m_b^2- \mw^2)u ) )
      \nl   +(f_{LR}^i \, f_{RR}^j + f_{RL}^i \, f_{LL}^j)  m_b \mchari 
      (( u + \mchi^2
           -\mw^2) \nl +(1/\mw^2)((\mchi^2 + \mw^2-u)
           (\mchi^2- \mw^2-u)))
        \nl    +(f_{LL}^i \, f_{RL}^j + f_{RR}^i \, f_{LR}^j) m_b \mcharj
           ((u+\mchi^2
           -\mw^2) \nl +(1/\mw^2)((\mchi^2+ \mw^2-u)
           (\mchi^2- \mw^2-u)))
        \nl    +(f_{LR}^i \, f_{LR}^j + f_{RL}^i \, f_{RL}^j) \mchari \mcharj
           ((t+ \mchi^2- m_b^2) \nl + (1/\mw^2)((\mchi^2 + \mw^2-u)
           (s-m_b^2-\mw^2) ))
         \nl  + (f_{LL}^i \, f_{RR}^j + f_{RR}^i \, f_{LL}^j) 6 m_b  \mchi u
        + (f_{LL}^i \, f_{LR}^j + f_{RR}^i \, f_{RL}^j) 3 \mcharj (u+ m_b^2-
\mstop^2) \mchi \nl + (f_{LR}^i \, f_{LL}^j + f_{RL}^i \, f_{RR}^j) 3 \mchari
         (u + m_b^2- \mstop^2) \mchi \nl + (f_{LR}^i \, f_{RL}^j + f_{RL}^i \,
	   f_{LR}^j) 6 
        \mchari \mcharj m_b  \mchi)/((t-\mchari^2)(t-\mcharj^2))  \nonumber \\
\tsq &=&   \sum_{i,j=1}^2 {\cal T}_{\rm I}\!\! \times\!\!{\cal T}_{\rm I}
\end{eqnarray}

\subsection*{$\ch \stop_1 \longrightarrow t H \quad [t h]$}
 I. s-channel $t$ annihilation \hfill\\
 II.  t-channel $\stop_{(1,2)}$ exchange \hfill\\
 III. u-channel $\chi^0_{(1,2,3,4)}$ exchange \hfill\\ 
\begin{eqnarray}
f_{1L} &=& C_{\stop_1-t- \tilde{\chi}_1^0}^{L} \; C_{t-t-H} \quad
[C_{\stop_1-t- \tilde{\chi}_1^0}^{L} \; C_{t-t-h}] \nonumber \\
f_{1R} &=& C_{\stop_1-t- \tilde{\chi}_1^0}^{R} \; C_{t-t-H} \quad
[C_{\stop_1-t- \tilde{\chi}_1^0}^{R} \; C_{t-t-h}] \nonumber \\
f_{2L}^i &=& C_{\stop_1-t- \tilde{\chi}_1^0}^{L} \;  C_{\stop_1 - \stop_1 -H}
\quad [C_{\stop_1-t- \tilde{\chi}_1^0}^{L} \;  C_{\stop_1 - \stop_1 -h}]
\nonumber \\
f_{2R}^i &=&  C_{\stop_1-t- \tilde{\chi}_1^0}^{R} \;  C_{\stop_1 - \stop_1
-H} \quad [C_{\stop_1-t- \tilde{\chi}_1^0}^{R} \;  C_{\stop_1 - \stop_1-h}]
\nonumber \\
f_{3LL}^i &=&  C_{\stop_1-t- \tilde{\chi}_i^0}^{L} \;  C_{\tilde{\chi}_1^0 -
\tilde{\chi}_i^0 - H}^{L} \quad [C_{\stop_1-t- \tilde{\chi}_i^0}^{L} \; 
C_{\tilde{\chi}_1^0 - \tilde{\chi}_i^0 - h}^{L}] \nonumber \\
f_{3LR}^i &=&  C_{\stop_1-t- \tilde{\chi}_i^0}^{L} \;  C_{\tilde{\chi}_1^0 -
\tilde{\chi}_i^0 - H}^{R} \quad [C_{\stop_1-t- \tilde{\chi}_i^0}^{L} \; 
C_{\tilde{\chi}_1^0 -
\tilde{\chi}_i^0 - h}^{R}]\nonumber \\
f_{3RL}^i &=&  C_{\stop_1-t- \tilde{\chi}_i^0}^{R} \;  C_{\tilde{\chi}_1^0 -
\tilde{\chi}_i^0 - H}^{L} \quad [C_{\stop_1-t- \tilde{\chi}_i^0}^{R} \; 
C_{\tilde{\chi}_1^0 -
\tilde{\chi}_i^0 - h}^{L}]  \nonumber \\
f_{3RR}^i &=&  C_{\stop_1-t- \tilde{\chi}_i^0}^{R} \;  C_{\tilde{\chi}_1^0 -
\tilde{\chi}_i^0 - H}^{R} \quad [C_{\stop_1-t- \tilde{\chi}_i^0}^{R} \; 
C_{\tilde{\chi}_1^0 -
\tilde{\chi}_i^0 - h}^{R}] \nonumber \\
{\cal T}_{\rm I}\!\!\times\!\!{\cal T}_{\rm I} &=&  (1/2)((f_{1L}^2+f_{1R}^2)
((s + m_t^2 - m_{H[h]}^2) (s + \mchi^2 - m_{\stop_1}^2)
 - (s - m_t^2) (\mchi^2 + m_t^2 \nl - t) 
+ 2 m_t^2 (s + \mchi^2 - m_{\stop_1}^2))
 - f_{1L} f_{1R} 4 \mchi m_t (2 s+2 m_t^2 - m_{H[h]}^2))/(s-m_t^2)^2 
\nonumber  \\
{\cal T}_{\rm II}\!\!\times\!\!{\cal T}_{\rm II} &=&  (1/2)((f_{2L}^i f_{2L}^j
+ f_{2R}^i f_{2R}^j) (\mchi^2 + m_t^2 - t)
 - 2 m_t \mchi (f_{2L}^i f_{2R}^j + f_{2R}^i f_{2L}^j))
\nl /((t - m_{\stop_i}^2)(t - m_{\stop_j}^2))  \nonumber \\
{\cal T}_{\rm III}\!\!\times\!\!{\cal T}_{\rm III} &=& (1/2)((f_{3LR}^i
f_{3LR}^j + f_{3RL}^i f_{3RL}^j) ((m_t^2 - m_{\stop_1}^2 + u)  (\mchi^2 -
m_{H[h]}^2 + u) \nl - u (\mchi^2 + m_t^2 - t))
 + (f_{3LL}^i f_{3LL}^j + f_{3RR}^i f_{3RR}^j) \mxi
\mxj (\mchi^2 + m_t^2 - t) \nl
- (f_{3LR}^i f_{3LL}^j + f_{3RL}^i f_{3RR}^j) \, \mxj
(m_t^2 - m_{\stop_1}^2 + u) 
- (f_{3LL}^i f_{3LR}^j + f_{3RR}^i f_{3RL}^j) \nl \times \,  \mxi \, \mchi
(m_t^2 - m_{\stop_1}^2 + u)
 + (f_{3RL}^i f_{3LL}^j + f_{3LR}^i f_{3RR}^j) \, m_t \, \mxj
 (\mchi^2 - m_{H[h]}^2 \nl + u)
 + (f_{3RR}^i f_{3LR}^j + f_{3LL}^i f_{3RL}^j)  \, m_t \, \mxi
(\mchi^2 - m_{H[h]}^2 + u) 
- (f_{3RL}^i f_{3LR}^j \nl + f_{3LR}^i f_{3RL}^j) 2 \, m_t \, \mchi \, u
 - (f_{3RR}^i f_{3LL}^j + f_{3LL}^i f_{3RR}^j) \, 2 \, \mxi \, \mxj \, m_t \,
 \mchi)
\nl /((u-\mxi^2)(u-\mxj^2))  \nonumber \\
{\cal T}_{\rm I}\!\!\times\!\!{\cal T}_{\rm II} &=&  (1/2)((f_{1L} f_{2L}^i +
f_{1R} f_{2R}^i) m_t (s+2 \mchi^2+m_t^2 - m_{\stop_1}^2 - t)
 -(f_{1R} f_{2L}^i \nl +f_{1L} f_{2R}^i) 
\mchi (s+3 m_t^2 - m_{H[h]}^2))
/((s- m_t^2)(t-m_{\stop_i}^2)) \nonumber \\
{\cal T}_{\rm I}\!\!\times\!\!{\cal T}_{\rm III} &=&  (1/2)((f_{1L} f_{3RL}^i +
f_{1R} f_{3LR}^i)((s+\mchi^2- m_{\stop_1}^2)
(u + m_t^2- \mstop^2)  - (\mchi^2+m_t^2 \nl -t)(m_t^2+\mchi^2
- \mstop^2- m_{H[h]}^2)  + (u+\mchi^2-m_{H[h]}^2)(s+m_t^2-m_{H[h]}^2) \nl 
+ 2 m_t (u+\mchi^2- m_{H[h]}^2))  + (f_{1L} f_{3LL}^i+f_{1R} f_{3RR}^i) 
\, \mxi \, m_t \, (s+2 \mchi^2+ m_t^2-\mstop^2 \nl -t)
  - (f_{1L} f_{3LR}^i+f_{1R} f_{3RL}^i) 
\, \mchi \, m_t \, (u+2 m_t^2+ \mchi^2- 2 \mstop^2- m_{H[h]}^2) 
\nl - (f_{1L} f_{3RR}^i+f_{1R} f_{3LL}^i)\,  
\mchi \, \mxi\, ( s+3 m_t^2-m_{H[h]}^2))
/((s-m_t^2)(u-\mxi^2)) \nonumber \\
{\cal T}_{\rm II}\!\!\times\!\!{\cal T}_{\rm III} &=& (1/2)((f_{2R}^j f_{3LR}^i
+f_{2L}^j f_{3RL}^i) \, m_t \, (u+\mchi^2-m_{H[h]}^2)
 + (f_{2L}^j f_{3LL}^i +f_{2R}^j f_{3RR}^i) \nl \times \, \mxi \,
 (\mchi^2+m_t^2-t)
 - (f_{2L}^j f_{3LR}^i + f_{2R}^j f_{3RL}^i) 
\, \mchi \, (u+m_t^2-\mstop^2)
\nl - (f_{2R}^j f_{3LL}^i+f_{2L}^j f_{3RR}^i) 2 \, \mchi \, m_t \, \mxi)
/((t-m_{\stop_j}^2)(u-\mxi^2))  \nonumber \\
\tsq &=&   {\cal T}_{\rm I}\!\!\times\!\!{\cal T}_{\rm I} 
+  \sum_{i,j=1}^2 {\cal T}_{\rm II}\!\! \times\!\!{\cal T}_{\rm II}
+  \sum_{i,j=1}^4 {\cal T}_{\rm III}\!\! \times\!\!{\cal T}_{\rm III} 
+ 2 \sum_{i=1}^2 {\cal T}_{\rm I}\!\!\times\! \!{\cal T}_{\rm II}
+ 2 \sum_{i=1}^4 {\cal T}_{\rm I}\!\!\times\! \!{\cal T}_{\rm III}
+ \nl 2 \sum_{j=1}^2 \sum_{i=1}^4 {\cal T}_{\rm II}\!\!\times\! \!{\cal T}_{\rm
III}
\end{eqnarray}

\subsection*{$\ch \stop_1 \longrightarrow t A$}
 I. s-channel $t$ annihilation \hfill\\
 II.  t-channel $\stop_{2}$ exchange \hfill\\
 III. u-channel $\chi^0_{(1,2,3,4)}$ exchange \hfill\\ 
\begin{eqnarray}
f_{1L} &=& C_{\stop_1 - t -\tilde{\chi}_1^0}^{L} \; C_{t-t-A} \nonumber \\
f_{1R} &=& C_{\stop_1 - t -\tilde{\chi}_1^0}^{R} \; C_{t-t-A} \nonumber \\
f_{2L} &=&  C_{\stop_2 - t -\tilde{\chi}_1^0}^{L} \; C_{\stop_1-\stop_2-A}
\nonumber \\
f_{2R} &=&  C_{\stop_2 - t -\tilde{\chi}_1^0}^{R} \; C_{\stop_1-\stop_2-A}
\nonumber \\
f_{3LL}^i &=&  C_{\stop_1 - t -\tilde{\chi}_i^0}^{L} \; C_{\tilde{\chi}_1^0 -
\tilde{\chi}_i^0 - A}^{L} \nonumber \\
f_{3LR}^i &=&  C_{\stop_1 - t -\tilde{\chi}_i^0}^{L} \; C_{\tilde{\chi}_1^0 -
\tilde{\chi}_i^0 - A}^{R} \nonumber \\
f_{3RL}^i &=&  C_{\stop_1 - t -\tilde{\chi}_i^0}^{R} \; C_{\tilde{\chi}_1^0 -
\tilde{\chi}_i^0 - A}^{L} \nonumber \\
f_{3RR}^i &=&  C_{\stop_1 - t -\tilde{\chi}_i^0}^{R} \; C_{\tilde{\chi}_1^0 -
\tilde{\chi}_i^0 - A}^{R} \nonumber \\
{\cal T}_{\rm I}\!\!\times\!\!{\cal T}_{\rm I} &=& (1/2)((f_{1L}^2+f_{1R}^2)((s-
m_t^2-m_A^2) (s+\mchi^2-\mstop^2) - (s-m_t^2)(\mchi^2+m_t^2-t))
\nl - f_{1L}f_{1R}\, 4 \, \mchi \, m_t \, m_A^2)/(s-m_t^2)^2   \nonumber \\
{\cal T}_{\rm II}\!\!\times\!\!{\cal T}_{\rm II} &=&  
(1/2)((f_{2L}f_{2L}+f_{2R}f_{2R})(\mchi^2+m_t^2-t)
- 2 \, m_t \, \mchi (f_{2L} f_{2R}+f_{2R} f_{2L}))
\nl /((t-m_{\stop_2}^2)^2) \nonumber \\
{\cal T}_{\rm III}\!\!\times\!\!{\cal T}_{\rm III} &=&
(1/2)((f_{3LR}^i f_{3LR}^j +f_{3RL}^i f_{3RL}^j)
((m_t^2-\mstop^2+u)(\mchi^2-m_A^2+u)  -u
(\mchi^2+m_t^2 \nl -t))
 + (f_{3LL}^i f_{3LL}^j  + f_{3RR}^i f_{3RR}^j) \mxi \, \mxj
(\mchi^2+m_t^2-t)
 - (f_{3LR}^i f_{3LL}^j \nl + f_{3RL}^i f_{3RR}^j) \mchi\, \mxj  
(m_t^2-\mstop^2+u)
 - (f_{3LL}^i f_{3LR}^j + f_{3RR}^i f_{3RL}^j) \mxi \, \mchi
 (m_t^2 \nl -\mstop^2+u)
 + (f_{3RL}^i f_{3LL}^j + f_{3LR}^i f_{3RR}^j) m_t \, \mxj
(\mchi^2-m_A^2+u)
 + (f_{3RR}^i f_{3LR}^j \nl + f_{3LL}^i f_{3RL}^j) m_t \, \mxi
(\mchi^2-m_A^2+u)
 - (f_{3RL}^i f_{3LR}^j + f_{3LR}^i f_{3RL}^j) \, 2 \, m_t \, \mchi \, u
 \nl - (f_{3RR}^i f_{3LL}^j + f_{3LL}^i f_{3RR}^j) \, 2 \, \mxi \, \mxj \, m_t
 \, \mchi)
 /((u-\mxi^2)(u-\mxj^2))  \nonumber \\
{\cal T}_{\rm I}\!\!\times\!\!{\cal T}_{\rm II} &=&
(1/2)((f_{2R}f_{1R}-f_{2L}f_{1L})m_t(u-\mchi^2-m_A^2)
 + (f_{2L}f_{1R}-f_{2R}f_{1L})\, \mchi (s-m_t^2 \nl -m_A^2))
/((s-m_t^2)(t-m_{\stop_2}2))  \nonumber \\
{\cal T}_{\rm I}\!\!\times\!\!{\cal T}_{\rm III} &=& (1/2)((f_{3LR}^i f_{1R}-
f_{3RL}^i f_{1L})((\mchi^2+m_t^2-t)
(m_t^2+\mchi^2-\mstop^2-m_A^2) \nl -(s-m_t^2-m_A^2)
(u+\mchi^2-m_A^2)-(u+m_t^2-\mstop^2)
(s+\mchi^2-\mstop^2))/2 
\nl+ (f_{3RR}^i f_{1R}-f_{3LL}^i f_{1L}) \, 
\mxi \, m_t \, (u-m_A^2-\mchi^2) 
 + (f_{3LR}^i f_{1L}-f_{3RL}^i f_{1R}) \nl \times \, m_t \, \mchi
 (u-\mchi^2+m_A^2)
 + (f_{3RR}^i f_{1L}-f_{3LL}^i f_{1R})\mxi\mchi(m_t^2+m_A^2-s))
\nl /((s-m_t^2)(u-\mxi^2))  \nonumber \\
{\cal T}_{\rm II}\!\!\times\!\!{\cal T}_{\rm III} &=& (1/2)((f_{2R} f_{3LR}^i +
f_{2L} f_{3RL}^i) \, m_t \, 
(u+\mchi^2-m_A^2)
  + (f_{2L} f_{3LL}^i + f_{2R} f_{3RR}^i) \nl \times \, \mxi (\mchi^2+m_t^2-t)
           - (f_{2L} f_{3LR}^i+f_{2R} f_{3RL}^i) \mchi (u+m_t^2-\mstop^2)
     \nl    - (f_{2R} f_{3LL}^i+f_{2L} f_{3RR}^i) \, 2 \, \mchi \, m_t \, \mxi)
         /((t-m_{\stop_2}^2) (u-\mxi^2))  \nonumber \\
\tsq &=&   {\cal T}_{\rm I}\!\!\times\!\!{\cal T}_{\rm I} 
+  {\cal T}_{\rm II}\!\! \times\!\!{\cal T}_{\rm II}
+  \sum_{i,j=1}^4 {\cal T}_{\rm III}\!\! \times\!\!{\cal T}_{\rm III} 
+ 2  {\cal T}_{\rm I}\!\!\times\! \!{\cal T}_{\rm II}
+ 2 \sum_{i=1}^4 {\cal T}_{\rm I}\!\!\times\! \!{\cal T}_{\rm III}
+  2  \sum_{i=1}^4 {\cal T}_{\rm II}\!\!\times\! \!{\cal T}_{\rm
III}
\end{eqnarray}

\subsection*{$\ch \stop_1 \longrightarrow b H^+$}
 I. s-channel $t$ annihilation \hfill\\
 II.  t-channel $\tilde{\chi}^+_{(1,2)}$ exchange \hfill\\
 III. u-channel $\sbot_{(1,2)}$ exchange \hfill\\ 
\begin{eqnarray}
f_{1LL} &=& C_{t-b-H^+}^{L} \; C_{\stop_1 - t - \tilde{\chi}_1^0}^{L}
\nonumber \\
f_{1LR} &=& C_{t-b-H^+}^{L} \; C_{\stop_1 - t - \tilde{\chi}_1^0}^{R}
\nonumber \\
f_{1RL} &=& C_{t-b-H^+}^{R} \; C_{\stop_1 - t - \tilde{\chi}_1^0}^{L}
\nonumber \\
f_{1RR} &=& C_{t-b-H^+}^{R} \; C_{\stop_1 - t - \tilde{\chi}_1^0}^{R}
\nonumber \\
f_{2LL}^i &=&  C_{\stop_1 - b - \tilde{\chi}^+_i}^{L} \;
C_{\tilde{\chi}_1^0-\tilde{\chi}^+_i-H^+}^{L} \nonumber \\
f_{2LR}^i &=&  C_{\stop_1 - b - \tilde{\chi}^+_i}^{L} \;
C_{\tilde{\chi}_1^0-\tilde{\chi}^+_i-H^+}^{R} \nonumber \\
f_{2RL}^i &=&  C_{\stop_1 - b - \tilde{\chi}^+_i}^{R} \;
C_{\tilde{\chi}_1^0-\tilde{\chi}^+_i-H^+}^{L} \nonumber \\
f_{2RR}^i &=&  C_{\stop_1 - b - \tilde{\chi}^+_i}^{R} \;
C_{\tilde{\chi}_1^0-\tilde{\chi}^+_i-H^+}^{R} \nonumber \\
f_{3L}^i &=&  C_{\sbot_i-b-\tilde{\chi}_1^0}^{L} \; C_{\stop_1 - \sbot_i -
H^+} \nonumber \\
f_{3R}^i &=&  C_{\sbot_i-b-\tilde{\chi}_1^0}^{R} \; C_{\stop_1 - \sbot_i -
H^+} \nonumber \\
{\cal T}_{\rm I}\!\!\times\!\!{\cal T}_{\rm I} &=&  (1/2)((f_{1LR}^2+f_{1RL}^2)
      ((s+m_b^2-m_{H^+}^2) (s+\mchi^2-\mstop^2)
         - s (\mchi^2 + m_b^2 -u)) \nl + (f_{1LL}^2 + f_{1RR}^2) m_t^2 
         (\mchi^2 + m_b^2-u) 
	  - 2 (f_{1LL} f_{1LR}+f_{1RR} f_{1RL}) \, \mchi \, m_t \nl \times \,  
         (s+m_b^2-m_{H^+}^2) 
	  + 2 (f_{1LR} f_{1RR} + f_{1LL} f_{1RL}) m_b \, m_t 
         (s+\mchi^2-\mstop^2) 
	 \nl -4 (f_{1LR} f_{1RL}) m_b \, \mchi \,  s
         - 4 f_{1LL} f_{1RR} \, m_t^2 \, \mchi \, m_b)/(s-m_t^2)^2  \nonumber \\
{\cal T}_{\rm II}\!\!\times\!\!{\cal T}_{\rm II} &=&  (1/2) ((f_{2RR}^i
f_{2RR}^j+f_{2LL}^i f_{2LL}^j) 
            ((t+m_b^2-\mstop^2) (t+\mchi^2-m_{H^+}^2) 
           - t (\mchi^2+m_b^2 \nl -u))
          + (f_{2RL}^i f_{2RL}^j+f_{2LR}^i f_{2LR}^j) \, \mchari \, \mcharj \, 
            (\mchi^2+m_b^2-u)
          - (f_{2RR}^i f_{2RL}^j \nl +f_{2LL}^i f_{2LR}^j) \, \mchi \, \mcharj
	  \, 
             (t+m_b^2-\mstop^2)
          - (f_{2RL}^i f_{2RR}^j+f_{2LR}^i f_{2LL}^j) \, \mchari \, \mchi \nl
	  \times \, 
             (t-m_b^2-\mstop^2)
          + (f_{2LL}^i f_{2RL}^j+f_{2RR}^i f_{2LR}^j) \, m_b \, \mcharj \, 
             (t+\mchi^2-m_{H^+}^2)
         \nl + (f_{2LR}^i f_{2RR}^j  +f_{2RL}^i f_{2LL}^j) \, m_b \, \mchari \,
             (t+\mchi^2-m_{H^+}^2)
         - (f_{2LL}^i f_{2RR}^j+f_{2RR}^i f_{2LL}^j) \nl \times \, 2 \, m_b \,
	 \mchi \, t
           - (f_{2LR}^i f_{2RL}^j+f_{2RL}^i f_{2LR}^j) \, 2 \, m_b \, \mchi \, 
     \mchari \, 
         \mcharj)
	 \nl /((t - \mchari^2) (t - \mcharj^2))  \nonumber \\
{\cal T}_{\rm III}\!\!\times\!\!{\cal T}_{\rm III} &=& (1/2) ((f_{3L}^i
f_{3L}^j+f_{3R}^i f_{3R}^j)
                   (\mchi^2+m_b^2-u)
              - 2 \, \mchi \, m_b (f_{3L}^i f_{3R}^j+f_{3R}^i f_{3L}^j))
               \nl    /((u-m_{\sbot_i}^2) (u-m_{\sbot_j}^2))  \nonumber \\
{\cal T}_{\rm I}\!\!\times\!\!{\cal T}_{\rm II} &=&  (1/2) ((f_{2RR}^i
f_{1LR}+f_{2LL}^i f_{1RL}) (1/2)
          ((t+m_b^2-\mstop^2) (s+\mchi^2-\mstop^2) 
          -(\mchi^2 \nl +m_b^2-u) (\mchi^2+m_b^2-\mstop^2-m_{H^+}^2) 
            +(s+m_b^2-m_{H^+}^2) (t+\mchi^2-m_{H^+}^2))
    \nl     + (f_{2RL}^i f_{1LL}+f_{2LR}^i f_{1RR}) \, \mchari \, m_t \, 
    (\mchi^2+m_b^2-u)
         - (f_{2RR}^i f_{1LL}+f_{2LL}^i f_{1RR}) \nl \times \, \mchi \, m_t \, 
    (t+m_b^2-\mstop^2)
         - (f_{2RL}^i f_{1LR}+f_{2LR}^i f_{1RL}) \, \mchari \, \mchi \, 
    (s+m_b^2-m_{H^+}^2)
    \nl     + (f_{2LL}^i f_{1LL}+f_{2RR}^i f_{1RR}) \, m_b \, m_t \, 
    (t+\mchi^2-m_{H^+}^2)
       + (f_{2LR}^i f_{1LR}+f_{2RL}^i f_{1RL}) \nl \times \, m_b \, \mchari \,  
    (s+\mchi^2-\mstop^2)
          - (f_{2LL}^i f_{1LR}+f_{2RR}^i f_{1RL}) \, m_b \, \mchi \, 
             (\mchi^2+m_b^2-\mstop^2 \nl -m_{H^+}^2)
          - (f_{2LR}^i f_{1LL}+f_{2RL}^i f_{1RR}) \, 2 \, m_b \, \mchi \, m_t \,
	  \mchari)
          /((s-m_t^2) (t-\mchari^2)) \nonumber \\
{\cal T}_{\rm I}\!\!\times\!\!{\cal T}_{\rm III} &=&  (1/2) ((f_{1LR} f_{3R}^i
+ f_{1RL} f_{3L}^i) \, m_b \, 
                   (s+\mchi^2-\mstop^2)
                 + (f_{1LL} f_{3L}^i+f_{1RR} f_{3R}^i) \, m_t \nl \times \, 
                    (\mchi^2+m_b^2-u)
                 - (f_{1LR} f_{3L}^i+f_{1RL} f_{3R}^i) \, \mchi \, 
                    (s+m_b^2-m_{H^+}^2)
                 - (f_{1LL} f_{3R}^i \nl +f_{1RR} f_{3L}^i) \, 2 \, m_t \, m_b
		 \, \mchi)
              /((u-m_{\sbot_i}^2) (s-m_t^2)) \nonumber \\
{\cal T}_{\rm II}\!\!\times\!\!{\cal T}_{\rm III} &=&  (1/2) ((f_{2LR}^j
f_{3R}^i+f_{2RL}^j f_{3L}^i) \, m_b \, 
                    (t+\mchi^2-m_{H^+}^2)
                 + (f_{2LL}^j f_{3L}^i+f_{2RR}^j f_{3R}^i) \nl \times \, \mcharj
                    (\mchi^2+m_b^2-u)
                 - (f_{2LR}^j f_{3L}^i+f_{2RL}^j f_{3R}^i) \, \mchi \, 
                    (t+m_b^2-\mstop^2)
    \nl            - (f_{2LL}^j f_{3R}^i+f_{2RR}^j f_{3L}^i) \, 2 \, \mcharj \,
    m_b \, \mchi)
                /((u-m_{\sbot_i}^2) (t-\mcharj^2)) \nonumber \\
\tsq &=&   {\cal T}_{\rm I}\!\!\times\!\!{\cal T}_{\rm I} 
+  \sum_{i,j=1}^2 {\cal T}_{\rm II}\!\! \times\!\!{\cal T}_{\rm II}
+  \sum_{i,j=1}^2 {\cal T}_{\rm III}\!\! \times\!\!{\cal T}_{\rm III} 
+ 2  \sum_{i=1}^2 {\cal T}_{\rm I}\!\!\times\! \!{\cal T}_{\rm II}
+ 2 \sum_{i=1}^2 {\cal T}_{\rm I}\!\!\times\! \!{\cal T}_{\rm III}
+ \nl 2  \sum_{i,j=1}^2 {\cal T}_{\rm II}\!\!\times\! \!{\cal T}_{\rm
III}
\end{eqnarray}

\subsection*{$\stop_1 \tilde{\ell} \longrightarrow t \ell$}
 I. t-channel $\tilde{\chi}^0_{(1,2,3,4)}$ exchange \hfill\\ 
\begin{eqnarray}
f_{LL}^i &=& C_{\stop_1 - t - \tilde{\chi}_i^0}^{L} \; C_{\tilde{\ell}_1 -
\ell - \tilde{\chi}_i^0}^{L} \nonumber \\
f_{LR}^i &=& C_{\stop_1 - t - \tilde{\chi}_i^0}^{L} \; C_{\tilde{\ell}_1 -
\ell - \tilde{\chi}_i^0}^{R} \nonumber \\
f_{RL}^i &=& C_{\stop_1 - t - \tilde{\chi}_i^0}^{R} \; C_{\tilde{\ell}_1 -
\ell - \tilde{\chi}_i^0}^{L} \nonumber \\
f_{RR}^i &=& C_{\stop_1 - t - \tilde{\chi}_i^0}^{R} \; C_{\tilde{\ell}_1 -
\ell - \tilde{\chi}_i^0}^{R} \nonumber \\
{\cal T}_{\rm I}\!\!\times\!\!{\cal T}_{\rm I} &=&  ((f_{LR}^i
f_{RL}^j+f_{RL}^i f_{LR}^j) ((\mstop^2-m_t^2-t)
              (t+m_{\ell}^2-\msl^2)-t (s-m_t^2-m_{\ell}^2))
    \nl      + (f_{LL}^i f_{LL}^j+f_{RR}^i f_{RR}^j) \, \mxi \, \mxj \,
     (s-m_t^2-m_{\ell}^2)
    \nl      - (f_{LR}^i f_{LL}^j+f_{RL}^i f_{RR}^j) \, m_{\ell} \, \mxj \,
    (\mstop^2 -m_t^2-t)
    \nl      - (f_{LL}^i f_{RL}^j+f_{RR}^i f_{LR}^j) \, m_{\ell} \, \mxi \,
    (\mstop^2 -m_t^2-t)
    \nl       + (f_{RL}^i f_{LL}^j+f_{LR}^i f_{RR}^j) \, m_t \, \mxj \, 
    (t+m_{\ell}^2-\msl^2)
    \nl       + (f_{RR}^i f_{RL}^j+f_{LL}^i f_{LR}^j) \, m_t \, \mxi \,
    (t+m_{\ell}^2-\msl^2)
  \nl      - (f_{RL}^i f_{LR}^j+f_{LR}^i f_{LR}^j) \, m_t \, m_{\ell} \, 2 \, t
        - (f_{RR}^i f_{LL}^j+f_{LL}^i f_{RR}^j) \, 2 \, m_t \, m_{\ell} \, \mxi
	\, \mxj)
     \nl        /((t-\mxi^2) (t-\mxj^2))  \nonumber \\
\tsq &=&   \sum_{i,j=1}^4 {\cal T}_{\rm I}\!\!\times\!\!{\cal T}_{\rm I} 
\end{eqnarray}

\subsection*{$\stop_1 \tilde{\ell} \longrightarrow b \nu$}
 I. t-channel $\tilde{\chi}^+_{(1,2)}$ exchange \hfill\\
\begin{eqnarray}
f_{LL}^i &=& C_{\stop_1-b-\tilde{\chi}^+_i}^{L} \;
C_{\tilde{\ell}_1-\nu-\tilde{\chi}^+_i}^{L}  \nonumber \\
f_{LR}^i &=& C_{\stop_1-b-\tilde{\chi}^+_i}^{L} \;
C_{\tilde{\ell}_1-\nu-\tilde{\chi}^+_i}^{R}  \nonumber \\
f_{RL}^i &=& C_{\stop_1-b-\tilde{\chi}^+_i}^{R} \;
C_{\tilde{\ell}_1-\nu-\tilde{\chi}^+_i}^{L}  \nonumber \\
f_{RR}^i &=& C_{\stop_1-b-\tilde{\chi}^+_i}^{R} \;
C_{\tilde{\ell}_1-\nu-\tilde{\chi}^+_i}^{R}  \nonumber \\
{\cal T}_{\rm I}\!\!\times\!\!{\cal T}_{\rm I} &=&  ((f_{LR}^i
f_{RL}^j+f_{RL}^i f_{LR}^j) ((\mstop^2-m_b^2-t)
    (t-m_{\tilde{\ell}_1}^2)-t (s-m_b^2))
 \nl  + (f_{LL}^i f_{LL}^j+f_{RR}^i f_{RR}^j) \, \mchari \, \mcharj \, 
          (s-m_b^2)
\nl     + (f_{RL}^i f_{LL}^j+f_{LR}^i f_{RR}^j) \, m_b \, \mcharj
            \,  (t-\msl^2)
  \nl       + (f_{RR}^i f_{RL}^j+f_{LL}^i f_{LR}^j) \, m_b \, \mchari \, 
               (t-\msl^2))
   \nl        /((t-\mchari^2) (t-\mcharj^2))   \nonumber \\
\tsq &=&   \sum_{i,j=1}^2 {\cal T}_{\rm I}\!\!\times\!\!{\cal T}_{\rm I}
\end{eqnarray} 

\subsection*{$\stop_1 \tilde{\ell}^* \longrightarrow t \bar{\ell}$}
 I. t-channel $\tilde{\chi}^0_{(1,2,3,4)}$ exchange \hfill\\ 
\begin{eqnarray}
f_{LL}^i &=& C_{\stop_1 - t - \tilde{\chi}_i^0}^{L} \; C_{\tilde{\ell}_1 -
\ell - \tilde{\chi}_i^0}^{L} \nonumber \\
f_{LR}^i &=& C_{\stop_1 - t - \tilde{\chi}_i^0}^{L} \; C_{\tilde{\ell}_1 -
\ell - \tilde{\chi}_i^0}^{R} \nonumber \\
f_{RL}^i &=& C_{\stop_1 - t - \tilde{\chi}_i^0}^{R} \; C_{\tilde{\ell}_1 -
\ell - \tilde{\chi}_i^0}^{L} \nonumber \\
f_{RR}^i &=& C_{\stop_1 - t - \tilde{\chi}_i^0}^{R} \; C_{\tilde{\ell}_1 -
\ell - \tilde{\chi}_i^0}^{R} \nonumber \\
{\cal T}_{\rm I}\!\!\times\!\!{\cal T}_{\rm I} &=&  ((f_{LL}^i
f_{LL}^j+f_{RR}^i f_{RR}^j) ((\mstop^2-m_t^2-t)
               (t+m_{\ell}^2-\msl^2)-t (s-m_t^2-m_{\ell}^2))
  \nl     + (f_{LR}^i f_{RL}^j+f_{RL}^i f_{LR}^j) \, \mxi \, \mxj \, 
  (s-m_t^2-m_{\ell}^2)
    \nl       - (f_{LL}^i f_{RL}^j+f_{RR}^i f_{LR}^j) \, m_{\ell} \, \mxj \, 
    (\mstop^2-m_t^2-t)
    \nl       - (f_{LR}^i f_{LL}^j+f_{RL}^i f_{RR}^j) \, m_{\ell} \, \mxi \, 
    (\mstop^2-m_t^2-t)
    \nl       + (f_{RR}^i f_{RL}^j+f_{LL}^i f_{LR}^j) \, m_t \, \mxj \, 
    (t+m_{\ell}^2-\msl^2)
     \nl       + (f_{RL}^i f_{LL}^j+f_{LR}^i f_{RR}^j) \, m_t \, \mxi \, 
     (t+m_{\ell}^2-\msl^2)
     \nl       - (f_{RR}^i f_{LL}^j+f_{LL}^i f_{RR}^j) \, m_t \, m_{\ell} \, 2
     \, t
        - (f_{RL}^i f_{RL}^j+f_{LR}^i f_{LR}^j) \, 2 \, m_t \, m_{\ell} \, \mxi
    \,  \mxj)
     \nl       /((t-\mxi^2) (t-\mxj^2))  \nonumber \\
\tsq &=&   \sum_{i,j=1}^4 {\cal T}_{\rm I}\!\!\times\!\!{\cal T}_{\rm I} 
\end{eqnarray}



\begin{thebibliography}{99}

\bibitem{EHNOS}
J. Ellis, J.S. Hagelin, D.V. Nanopoulos, K.A. Olive
and M. Srednicki, Nucl. Phys. B {\bf 238}, 453 (1984); see also
H. Goldberg, Phys. Rev. Lett. {\bf 50}, 1419 (1983).

\bibitem{stopco}
C.~Boehm, A.~Djouadi and M.~Drees,
Phys.\ Rev.\ D {\bf 62}, 035012 (2000) 
[arXiv:hep-ph/9911496].

\bibitem{omegah2} N.~Bahcall, J.~P.~Ostriker, S.~Perlmutter and
P.~J.~Steinhardt, Science {\bf 284}, 1481 (1999). 

\bibitem{gs}
K. Griest and D. Seckel, Phys. Rev. D {\bf 43}, 3191 (1991).

\bibitem{oldcoann}
S.~Mizuta and M.~Yamaguchi,
Phys.\ Lett.\ B {\bf 298}, 120 (1993)
[arXiv:hep-ph/9208251].

\bibitem{eg}
J.~Edsjo and P.~Gondolo,
Phys.\ Rev.\ D {\bf 56}, 1879 (1997)
[arXiv:hep-ph/9704361].

\bibitem{efo}
J.~Ellis, T.~Falk and K.A.~Olive,
Phys.\ Lett.\ B {\bf 444}, 367 (1998).

\bibitem{efosi}
J.~Ellis, T.~Falk, K.A.~Olive and M.~Srednicki, Astropart. Phys. {\bf 13}
181 (2000).

\bibitem{coann}
M.~E.~G\'omez,
G.~Lazarides and C.~Pallis,
Phys.\ Rev.\ D {\bf 61}, 123512 (2000)
[arXiv:hep-ph/9907261]
and
Phys.\ Lett.\ B {\bf 487}, 313 (2000) [arXiv:hep-ph/0004028];
see also:
A.~B.~Lahanas, D.~V.~Nanopoulos and V.~C.~Spanos, Phys.\ Rev.\ D {\bf 62},
023515 (2000), [arXive:hep-ph/9909497].
\bibitem{ADS}
R.~Arnowitt, B.~Dutta and Y.~Santoso, 
Nucl.\ Phys.\ B {\bf 606}, 59 (2001)
[arXiv:hep-ph/0102181].

\bibitem{focus}
J.~L.~Feng, K.~T.~Matchev and T.~Moroi,
Phys.\ Rev.\ Lett.\  {\bf 84}, 2322 (2000)
[arXiv:hep-ph/9908309];
J.~L.~Feng, K.~T.~Matchev and T.~Moroi,
Phys.\ Rev.\ D {\bf 61}, 075005 (2000)
[arXiv:hep-ph/9909334];
J.~L.~Feng, K.~T.~Matchev and F.~Wilczek,
Phys.\ Lett.\ B {\bf 482}, 388 (2000)
[arXiv:hep-ph/0004043].

\bibitem{EFGOSi}
J.~R.~Ellis, T.~Falk, G.~Ganis, K.~A.~Olive and M.~Srednicki,
Phys.\ Lett.\ B {\bf 510}, 236 (2001)
[arXiv:hep-ph/0102098].   


\bibitem{funnel}
M.~Drees and M.~M.~Nojiri,
Phys.\ Rev.\ D {\bf 47}, 376 (1993);
H.~Baer and M.~Brhlik,
Phys.\ Rev.\ D {\bf 53}, 597 (1996) and Phys.\ Rev.\ D {\bf 57}, 567 
(1998);
H.~Baer, M.~Brhlik, M.~A.~Diaz, J.~Ferrandis, P.~Mercadante, P.~Quintana
and X.~Tata,
Phys.\ Rev.\ D {\bf 63}, 015007 (2001);
A.~B.~Lahanas and V.~C.~Spanos, hep-ph/0106345.


\bibitem{swo}
M.~Srednicki, R.~Watkins and K.~A.~Olive,
Nucl.\ Phys.\ B {\bf 310}, 693 (1988).

\bibitem{gg}
P.~Gondolo and G.~Gelmini,
Nucl.\ Phys.\ B {\bf 360}, 145 (1991).

\bibitem{fkosi}
T.~Falk, K.~A.~Olive and M.~Srednicki,
Phys.\ Lett.\ B {\bf 339}, 248 (1994)
[arXiv:hep-ph/9409270].

\bibitem{g-2}
H.~N.~Brown {\it et al.}  [Muon g-2 Collaboration],
Phys.\ Rev.\ Lett.\  {\bf 86}, 2227 (2001)
[hep-ex/0102017].

\bibitem{FeynHiggs}
S.~Heinemeyer, W.~Hollik and G.~Weiglein,
Comput.\ Phys.\ Commun.\  {\bf 124}, 76 (2000)
[arXiv:hep-ph/9812320];
S.~Heinemeyer, W.~Hollik and G.~Weiglein,
Eur.\ Phys.\ J.\ C {\bf 9}, 343 (1999)
[arXiv:hep-ph/9812472].

\bibitem{AF}
H. Baer, M. Brhlik and D. Casta\~no, Phys.\ Rev.\ D {\bf 54} (1996) 6944;
S. Abel and T. Falk, Phys.\ Lett.\ B {\bf 444} (1998) 427.


\end{thebibliography}
\end{document}